\documentclass[12pt,english]{article}
\usepackage{amsmath}
\usepackage{amsthm}
\usepackage{mathptmx}
\usepackage{newtxmath}
\usepackage[T1]{fontenc}
\usepackage[latin9]{inputenc}
\usepackage{geometry}
\usepackage{color}
\usepackage{babel}
\usepackage{float}
\usepackage{url}
\usepackage{graphicx}
\usepackage{booktabs}
\usepackage{setspace}
\usepackage{needspace}
\usepackage[authoryear,round]{natbib}
\let\citepbase\citep
\let\citetbase\citet
\let\citebase\cite
\let\citealtbase\citealt
\let\citealpbase\citealp
\renewcommand{\citep}{\citepbase*}
\renewcommand{\citet}{\citetbase*}
\renewcommand{\cite}{\citebase*}
\renewcommand{\citealt}{\citealtbase*}
\renewcommand{\citealp}{\citealpbase*}
\usepackage{lineno}

\usepackage[unicode=true,pdfusetitle,
 bookmarks=true,bookmarksnumbered=false,bookmarksopen=false,
 breaklinks=false,pdfborder={0 0 0},pdfborderstyle={},backref=false,colorlinks=true]
 {hyperref}
\hypersetup{
 linkcolor=red,urlcolor=red,citecolor=blue}

\usepackage{eurosym}

\theoremstyle{plain}
\newtheorem{prop}{Proposition}
\newtheorem{cor}{Corollary}

\makeatletter

\providecommand{\tabularnewline}{\\}

\usepackage{babel}
\usepackage[bottom]{footmisc}
\usepackage{indentfirst}
\usepackage{lscape}
\usepackage{xcolor}
\usepackage{titlesec}
\usepackage{caption}

\date{October 19, 2026}

\@ifundefined{showcaptionsetup}{}{%
 \PassOptionsToPackage{caption=false}{subfig}}
\usepackage{subfig}
\makeatother

\begin{document}
\title{Trading Frictions in Dynamic Cap-and-Trade Markets\footnote{Nicola Borri, LUISS University, email: nborri@luiss.it. Yukun Liu, University of Rochester, email: yliu229@ur.rochester.edu. Aleh Tsyvinski, Department of Economics at Yale University, email: a.tsyvinski@yale.edu. Xi Wu, University of California, Berkeley, email: xiwu@berkeley.edu. The empirical analysis in this paper subsumes that of ``Inefficiencies of Carbon Trading Markets'' first posted on arXiv on August 12, 2024. We thank Bruno Biais, Felix Bierbrauer, Lasse Pedersen and Christian Schlag for comments.}}
\author{Nicola Borri, Yukun Liu, Aleh Tsyvinski, and Xi Wu}

\maketitle
\begin{abstract}
\noindent Using 2.8 million EU ETS transactions and compliance records, 2005--2022, we document three facts about the world's largest carbon market: each year, 37\% of regulated operators do not trade, and non-trading is persistent. Purchases concentrate in the April surrender month, when returns are systematically high. Operator order flow predicts returns, concentrated among frequent traders. A model with costly market access, limited intermediation, and heterogeneous information rationalizes all three. Because access is endogenous, the frictions interact non-additively, shaping how efficiently the market corrects the externality. Staggering deadlines across four groups lowers the surrender-month premium by 58\% without changing the cap. 
\end{abstract}

\thispagestyle{empty}

\newpage

\setcounter{page}{1}

\section{Introduction}

Carbon emissions are the classic example of a climate externality. The leading real-world implementation of a market-based corrective instrument is the cap-and-trade system, in which agents can trade allowances to emit carbon \citep{dales1968pollution,montgomery1972markets,Rubin1996}. By making emission rights tradable and bankable (that is, transferable across time), a cap-and-trade system creates an asset market: the cap determines aggregate scarcity and the secondary allowance market must reallocate compliance obligations across firms and over time, and aggregate information about scarcity. The central question of this paper is how multiple trading frictions interact to shape prices and trading behavior in a market designed to correct an externality. We address it empirically, theoretically, and quantitatively, using the universe of registry transactions and compliance records of the world's largest carbon market. We document who trades, when, and what their trades reveal about future prices. We then develop a model in which trading frictions shape prices and interact in equilibrium. Finally, we calibrate the model to the data to quantify the interactions and to evaluate market design.

In canonical Pigouvian settings, of which climate is a leading example, an optimal corrective instrument can target the externality directly \citep{Weitzman1974,nordhaus1992optimal}. \cite{GolosovHasslerKrusellTsyvinski2014} show that in a wide class of climate models the optimal correction admits a simple closed-form solution which can be implemented either as a carbon tax or as a cap-and-trade system. \cite{Barrage2020} and \cite{Kubler2025} characterize how this unconstrained benchmark is modified by distortionary fiscal policy and by incomplete financial markets, respectively. When multiple frictions are present, as in real-world implementations of cap-and-trade markets, designing and analyzing a corrective instrument is substantially harder. The main challenge is that the joint effect of the corrective instrument on prices and quantities need not equal the sum of the separate effects of each friction, as in the classic work by \citet{LipseyLancaster1956}, and the interactions of the frictions may amplify or dampen the size of the response.

Recent climate-finance research treats climate risk, carbon pricing, and green finance as important objects for asset pricing and market design \citep{EngleGiglioKellyLeeStroebel2020,GiglioKellyStroebel2021}. An important paper by \cite{Pedersen2026CarbonPricing} compares carbon pricing and green finance as alternative tools for the climate transition and emphasizes the central role of price-based instruments. \cite{kanzig2023unequal} shows that restrictive carbon policy shocks in Europe reduce emissions and affect aggregate economic activity, which is evidence that carbon prices have real effects. \cite{HongWangYang2023} model climate disasters as Poisson arrivals and show that adaptation and capital taxes complement carbon pricing in the optimal climate policy mix. Other recent work studies financial dimensions of cap-and-trade from complementary angles: \cite{BiaisHombertSchmidtWeill2026} examine the spot-futures wedge under transition risk with imperfect hedging and financial constraints, and \cite{BiaisLandier2026} study how emission caps interact with firms' green-technology investment under government commitment. We develop a dynamic stochastic model of cap-and-trade with three trading frictions---slow participation, limited intermediation and heterogeneous information, which interact in equilibrium. We bring it to the data of the European Union Emissions Trading System (EU ETS), the leading cap-and-trade market, which covers about 40\% of the EU's greenhouse-gas emissions, using the universe of registry transactions and compliance records from the European Union Transaction Log (EUTL): about 2.8 million registry records from 2005 through September 2022.\footnote{The EUTL publishes transaction-level records with a three-year delay, so the transaction sample ends with the latest month available in the current public release we use, September 2022.} The paper's main contribution is to characterize and quantify, in an actual market, how these frictions interact.

The EUTL transaction data cover the first three phases of the EU ETS, as well as part of the ongoing fourth phase, which is scheduled to conclude in 2030. There are three main groups of participants in this market: regulated firms, intermediaries, and regulators. They are largely represented in the EU ETS by operator holding accounts, person holding accounts, and administrative accounts, respectively. Our analyses focus on regulated firms, or operators, and study their transactions with other regulated firms and intermediaries.\footnote{Throughout the paper, ``regulated'' refers to the EU ETS surrender obligation, not to financial regulation. Financial intermediaries are supervised under other frameworks, and emission allowances have been financial instruments under MiFID~II since January 2018.} The data allow us to observe how regulated firms interact with the secondary allowance market: whether they participate, when they trade, with whom they trade, and whether their order flow predicts future returns.

We first document three stylized facts about participation, surrender-month trading, and operator flow. First, approximately 37\% of regulated operators do not trade in a given year, falling to about 23\% in Phase IV through September 2022. Extending evidence of low participation in the early EU ETS \citep{martino2013back,jaraite2010transaction}, we show that non-trading is persistent throughout the market's history.

Second, operator purchases concentrate in April, the month in which firms must surrender allowances covering their verified emissions. Net purchases in April are nearly eighteen times larger than in typical non-April months, coinciding with systematically high April returns averaging about 9\% in the post-2008 price sample. This concentration of trading near the surrender deadline persists into Phase IV.

Third, operator order flow positively and significantly predicts future allowance returns in the full sample. The relation also holds after excluding the pilot phase, with smaller coefficients. Trading is highly heterogeneous: about 10\% of operators trade more than four times their annual surrender obligations. These patterns are consistent with some active operators trading on information about future scarcity that is not yet fully reflected in prices.

To interpret these facts, we first consider the frictionless benchmark underlying the classic case for cap-and-trade. In a market with bankable allowances and no trading frictions, firms trade until marginal abatement costs are equalized across firms and over time. This benchmark delivers three sharp implications: non-trading should arise only when a firm's initial allowance holdings already coincide with its cost-minimizing compliance and banking needs. Predictable compliance-related demand should be reflected in prices before the surrender deadline. Operator order flow based only on public compliance needs should not forecast future returns.

Motivated by these facts, we develop a dynamic stochastic model with costly market access, limited intermediation, and heterogeneous information. Costly market access makes participation endogenous. Access choices determine how much compliance demand remains unresolved near the surrender deadline. When intermediary capacity is limited, this residual demand raises allowance prices above their frictionless benchmark. We call this expected price difference the \textbf{surrender-month premium}. The expected premium, in turn, feeds back into firms' incentives to acquire access before the deadline. The same access margin determines which operators can act on private information, linking access to informative flow. The interaction runs in two directions. Dampening occurs when agents' endogenous access responses partially offset the direct effect of a policy that reduces one friction, so the surrender-month premium falls by less than it would if market access remained unchanged. Amplification occurs when one friction makes the equilibrium premium more responsive to another, so the price effect of a given friction is larger when other frictions are more severe. A static analysis would predict price responses that reflect each friction's direct effect alone. In our dynamic model, in contrast, the dominant effect is dampening: a reduction in intermediary return impact lowers the equilibrium premium less than one for one, because endogenous access responses raise residual terminal demand. When access costs are proportional to expected shortfalls, we prove that higher access costs weaken the access response and raise the equilibrium premium's elasticity to intermediary return impact.

We first illustrate the interaction of these frictions in a two-period model and then develop a continuous-time framework with a closed-form access rule and a unique equilibrium surrender-month premium. In this specification, we prove that endogenous access dampens the effect of intermediary frictions: locally, a 1\% reduction in intermediary return impact lowers the premium by approximately 0.5\% to 1\%, rather than the 1\% implied by fixed market access. A reduced-form information component links market access to informative operator flow. Beyond rationalizing the three stylized facts, the model generates additional empirical implications that we evaluate in the data.

We then test these additional implications and examine alternative explanations. We assess whether matched free allocations or banking can account for persistent non-trading, test whether operators with less pre-surrender market access purchase more in April conditional on compliance shortfalls, and examine whether return predictability is concentrated in flow from frequent traders. The predictive content comes from frequent traders, defined as operators with an above-median number of trades: with month fixed effects, their flow predicts returns at every horizon while flow from infrequent traders does not, as the model's information channel predicts.

The model also allows us to quantify the access and intermediation frictions and to evaluate market-design counterfactuals. Trading-frequency moments discipline firms' access intensities, and the relation between April return premia and model-implied terminal demand disciplines surrender-window return impact. The resulting model-implied terminal demand has a correlation of 0.82 with observed April compliance purchases at the year level. Under the one-shot access specification used in the quantitative exercise, our main policy counterfactual in which surrender deadlines are staggered across four equally sized groups reduces the surrender-month premium by about 58\% and the implied April premium paid by delayed buyers (the additional cost of purchasing allowances in April rather than earlier in the cycle) by about 41\%, with no change in aggregate emissions or the cap.

The calibrated model further quantifies how the access and intermediary frictions interact. Halving intermediary return impact alone lowers the equilibrium surrender-month premium by about 38\%, rather than the 50\% a static demand-pressure model would predict, because endogenous access responses raise residual terminal demand by about 24\%. Halving the access cost alone lowers the premium by about 21\%, but halving both frictions jointly lowers it by exactly 50\%: the joint effect is not the sum of the separate effects. When the access-cost friction is doubled relative to the baseline calibration, the elasticity of the equilibrium surrender-month premium with respect to intermediary return impact rises from 0.68 to 0.70, illustrating how the interaction can amplify the response to a single friction. In euro terms, the baseline implied April premium paid by delayed buyers is about \euro237 million per cycle and falls to roughly \euro118 million under the joint counterfactual and \euro140 million under the staggered-deadline policy.

The paper contributes to three strands of the literature. The first studies climate finance, sustainable investing, and the pricing of climate-related risks. Recent work shows that climate exposures and carbon-transition risks are reflected in asset prices and portfolio choices \citep{SautnerVanLentVilkovZhang2023,BoltonKacperczyk2023,croce2025green}. Related work studies how sustainability preferences and ESG characteristics enter portfolio choice and equilibrium expected returns \citep{PastorStambaughTaylor2021,PedersenFitzgibbonsPomorski2021}. Our paper differs by studying the secondary allowance market at the core of cap-and-trade---the leading real-world setting in which carbon prices are formed through trading. Rather than asking how climate risk is priced in equities or portfolios, we study how trading frictions and their interaction shape participation, intermediation, and return predictability in the secondary market for emission allowances.

The second strand studies climate economics, carbon pricing, and market-based environmental policy. Modern climate-macro models characterize optimal carbon taxation and the general-equilibrium effects of fossil-fuel use \citep{nordhaus1992optimal,GolosovHasslerKrusellTsyvinski2014,Barrage2020,Kubler2025,BierbrauerPolbornRitterrathWeizsacker2026}. Recent empirical work studies the real effects of carbon pricing: \cite{colmer2025does} show that the EU ETS reduced regulated firms' emissions without detectable contractions in economic activity. \citet{RamadoraiZeni2024} combine firms' disclosures with a dynamic model to show how beliefs about climate regulation shape emissions abatement and how reputational interactions across firms amplify regulatory effects. Recent theoretical and quantitative work studies climate damages, social costs of carbon, macroeconomic climate modeling, and directed innovation in the energy transition \citep{AcemogluAghionBarrageHemous2024,BarrageNordhaus2024,FernandezVillaverdeGillinghamScheidegger2025,FoliniFriedlKueblerScheidegger2025}. In cap-and-trade, \cite{BiaisHombertSchmidtWeill2026} study imperfect hedging, financial constraints, delayed spot purchases, and the spot-futures wedge under transition risk, while \cite{BiaisLandier2026} study how emission caps interact with firms' investment in green technologies and government commitment. Several papers use the registry data: \citet{BertaGautheratGun2017} document that most volume is transacted by financial accounts while regulated firms trade little and mostly for compliance, \citet{CludiusBetz2020} show that banks intermediate operators' compliance trading, \citet{AbrellCludiusLehmannSchleichBetz2022} characterize corporate trading behaviour over the first decade, and \citet{QueminPahle2023} document the growing role of financial participants. \citet{HitzemannUhrigHomburg2018} model equilibrium EUA price dynamics with banking and abatement. Relative to this work, we use the full 2005--2022 universe, document the persistence of non-trading at the firm level, test the access mechanism in the cross-section, and show that operator flow predicts returns. We take the cap-and-trade system as given and study the secondary market mechanism through which permits are traded, compliance demand is reallocated, and scarcity information is incorporated into prices.

The third strand studies trading frictions and information in asset markets. We build on models of slow-moving capital and search frictions \citep{DuffieGarleanuPedersen2007,Duffie2010}, constrained intermediation and demand pressure \citep{GrossmanMiller1988,garleanu2009demand,acharya2013limits,hendershott2014price}, and heterogeneous information and informative order flow \citep{GrossmanStiglitz1980,Kyle1985,GlostenMilgrom1985,subrahmanyam1991risk}. The surrender-month pattern we study belongs to a family of settings in which large, anticipated, uninformed demand meets limited risk-bearing capacity on a known date: Treasury auction cycles \citep{LouYanZhang2013}, quarter-end balance-sheet constraints \citep{DuTepperVerdelhan2018}, and dividend-payment cash flows \citep{HartzmarkSolomon2025}. The cap-and-trade deadline is a particularly clean instance because it is set by regulation, uniform across regulated firms, and movable by the regulator. Our representation of infrequent market access as a chosen Poisson intensity at a convex cost follows the optimal-inattention literature, in which agents observe or trade at endogenously chosen intervals \citep{DuffieSun1990,AbelEberlyPanageas2007,AbelEberlyPanageas2013,AlvarezGuisoLippi2012}. The individual mechanisms are well established: liquidity suppliers absorb temporary order imbalances, costly information acquisition prevents prices from being fully revealing, and inattentive agents rebalance infrequently. Our contribution lies not in any of these mechanisms in isolation, but in their equilibrium interaction with endogenous market access around a predictable regulatory deadline. Endogenous access determines participation and residual compliance demand, limited intermediary capacity turns residual demand into surrender-month price pressure, and heterogeneous information makes operator order flow predictive of future allowance returns. The solved equilibrium generates feedback between access choices and surrender-month pressure. The reduced-form information component then links these access choices to the speed of information incorporation. Our intermediation channel also parallels the funding-liquidity mechanism of \cite{BrunnermeierPedersen2009}, in which intermediary capacity and asset prices are jointly determined in equilibrium.

\section{Institutional Setting and Data\label{sec:setting_data}}
\label{subsec:emp_data}

The EU ETS is the world's first major carbon market, covering over 13,000 facilities across multiple industrial sectors. It operates as a cap-and-trade system, allocating emission permits, known as European Union Allowances (EUAs), to regulated firms, which we refer to as operators. The system is organized into ``phases.'' Phase I (2005--2007) was a pilot period that established trading rules and procedures. Phase II (2008--2012) expanded the scope of the system and standardized allocation rules. Phase III ran from 2013 to 2020, and Phase IV began in 2021 and runs through 2030.

Compliance follows a fixed annual calendar. Operators must report their annual emissions, and an accredited verifier must certify those emissions by the end of March. By the end of April, each operator must surrender allowances matching its verified emissions for the previous year. April is therefore the uniform surrender deadline for every regulated operator in the system. Missing the deadline carries a penalty of \euro100 for each tonne not covered by a surrendered allowance, in addition to the obligation to surrender the missing allowances, and the names of penalized operators are made public. Compliance is close to $100\%$ in our sample, so the margin of interest is not whether operators comply but when they acquire the allowances they need.

Allowances enter the market through primary issuance before they are traded. Within a trading phase, free allocations for a given year are credited to operator accounts by the end of February of that year, before the April deadline for the previous year's emissions, so operators can use newly issued allowances to meet the current obligation. Since 2013, a large share of the cap is sold at regular auctions throughout the year, so intermediaries and operators can also acquire allowances from primary supply. Primary issuance is recorded in the registry as transfers from administrative accounts, and our analysis takes it as given and studies secondary-market trading between operators and other participants.

Three account types appear in the registry~\citep{abrell_database}. Operator holding accounts belong to regulated installations and are used to receive, transfer, and surrender allowances. Person holding accounts belong to non-regulated participants, including financial intermediaries, which trade allowances but face no surrender obligation of their own. Administrative accounts belong to regulators and are used to allocate allowances and to receive surrendered ones. This asymmetry matters for what follows: only operators face the April deadline, so the intermediary sector is the natural counterparty when compliance demand concentrates near surrender. Person holding accounts are far less numerous than operator holding accounts but account for the majority of traded value, and a single company may hold accounts of more than one type.

Our working sample includes the universe of EUTL transactions from February 2005 through September 2022 and the associated compliance records, covering the first three EU ETS phases and part of the fourth. The registry records transfers between any two accounts, including reallocations within a firm and transfers to and from administrative accounts. We exclude transactions involving administrative accounts, because allocation and surrender are regulatory flows rather than market activity, while retaining surrender quantities and the identities of operators that interact only with administrative accounts. We aggregate transactions to the company level, identified by the company registration number attached to each account, so transfers recorded within the same company net to zero. Because the registry does not identify ultimate parents, aggregation stops at the legal entity: a group that centralizes trading in one subsidiary appears as one active company and one or more non-trading companies, so the participation measures below are defined at the level of the registered company. The Online Appendix provides a detailed description of the market and the data.

\section{Stylized Facts\label{sec:facts}}

We begin with three facts that motivate the model: many regulated operators trade infrequently, operator purchases and allowance returns concentrate in the April surrender month, and operator flow predicts future returns. This section presents descriptive stylized facts. Section~\ref{sec:empirical} reports the full empirical specifications and evaluates alternative explanations and additional cross-sectional predictions.

\subsection{Infrequent and Persistent Trading\label{subsec:fact_participation}}

\begin{figure}[p]
\caption{Distribution of Operator Trading Amounts\label{fig:trading-frequency}}

\medskip{}

{\footnotesize~The top panels plot the fraction of regulated companies with different trading volumes in each year, excluding transactions involving administrative accounts. Panels A and B correspond to the overall sample and Phases I through IV, respectively. For each figure, we group the operators into the following
categories: no trade; 0 $\leq$ trades $<$ 20k; 20k $\leq$ trades
$<$ 40k; 40k $\leq$ trades $<$ 60k; 60k $\leq$ trades $<$ 80k;
and 80k $\leq$ trades. The bottom panels plot the fraction of regulated companies with different trading ratios in each year, where the trading ratio is defined as the amount traded, excluding transactions with administrative accounts, divided by the company's surrendering amount. Panels C and D correspond to the overall sample and Phases I through IV, respectively. For each figure, we group the operators
into the following categories: no trade; trade ratios < 1; 1 $\leq$
trade ratios $<$ 2; 2 $\leq$ trade ratios $<$ 3; 3 $\leq$ trade
ratios $<$ 4; 4 $\leq$ trade ratios. Data is for the period from
February 2005 to September 2022. The 2022 observations cover January through September.}

\medskip{}

\subfloat[Overall]{\begin{centering}
\includegraphics[width=8cm]{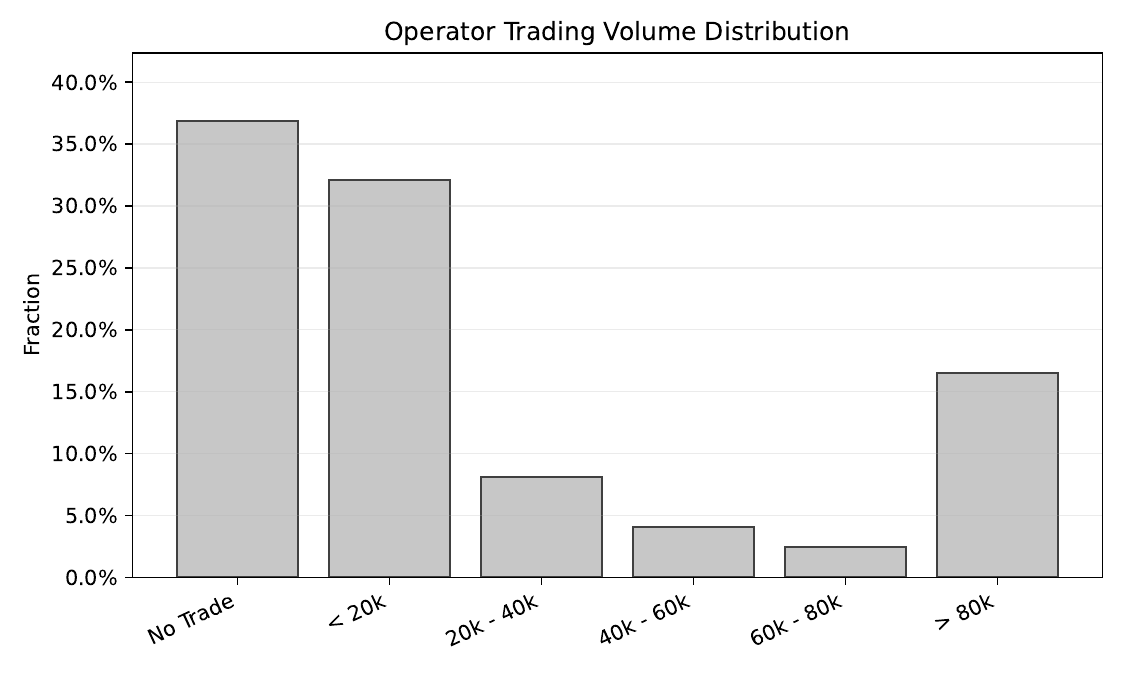}
\par\end{centering}
}
\subfloat[EU-ETS Phases]{\begin{centering}
\includegraphics[width=8cm]{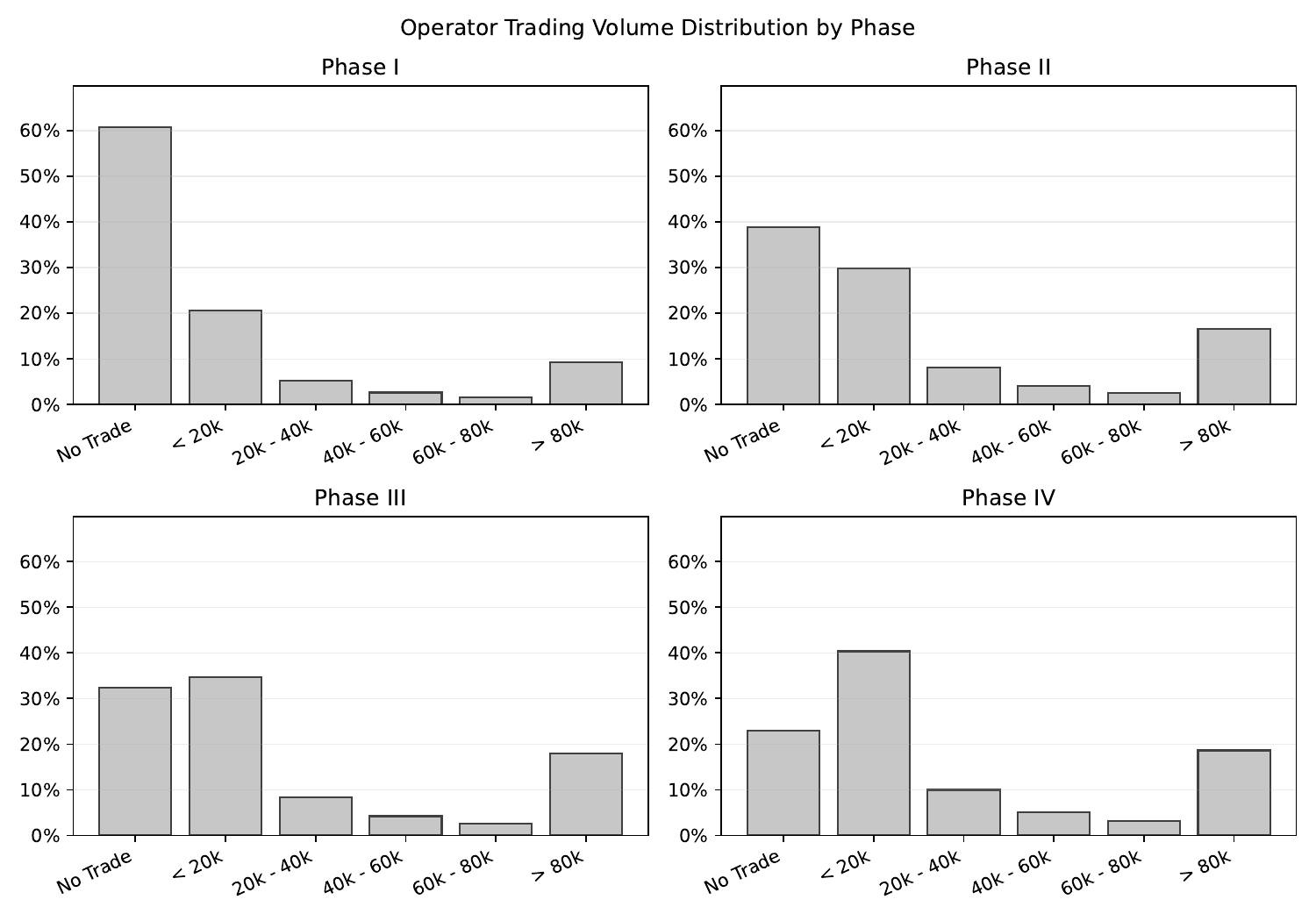}
\par\end{centering}
}

\subfloat[Overall]{\begin{centering}
\includegraphics[width=8cm]{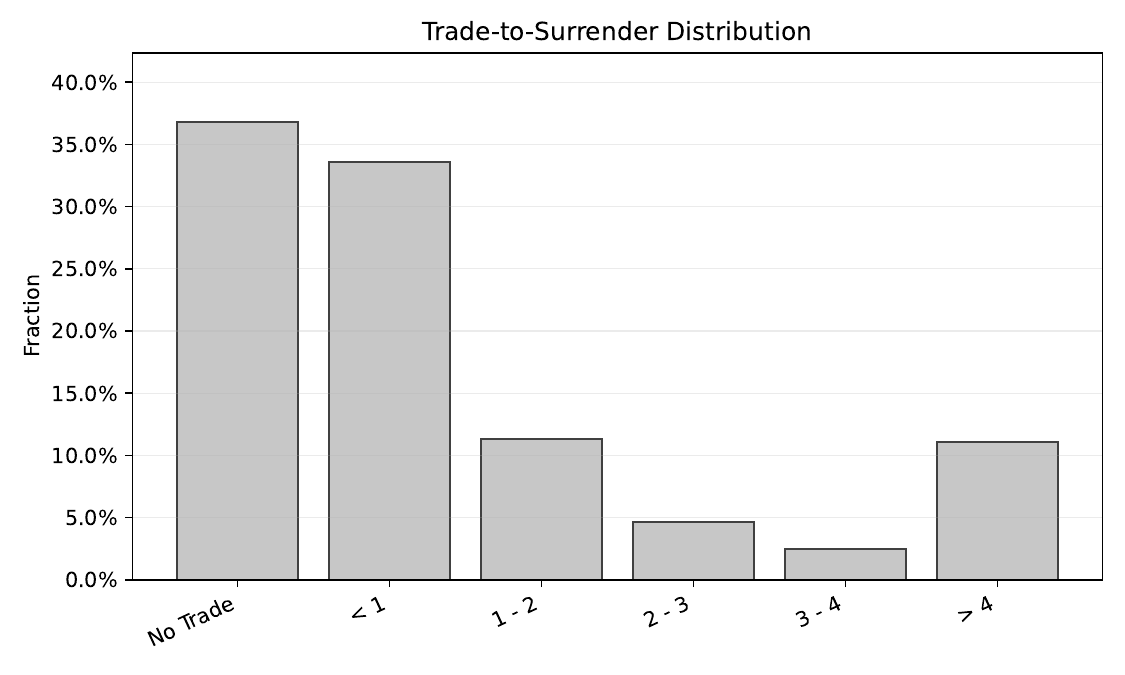}
\par\end{centering}
}
\subfloat[EU-ETS Phases]{\begin{centering}
\includegraphics[width=8cm]{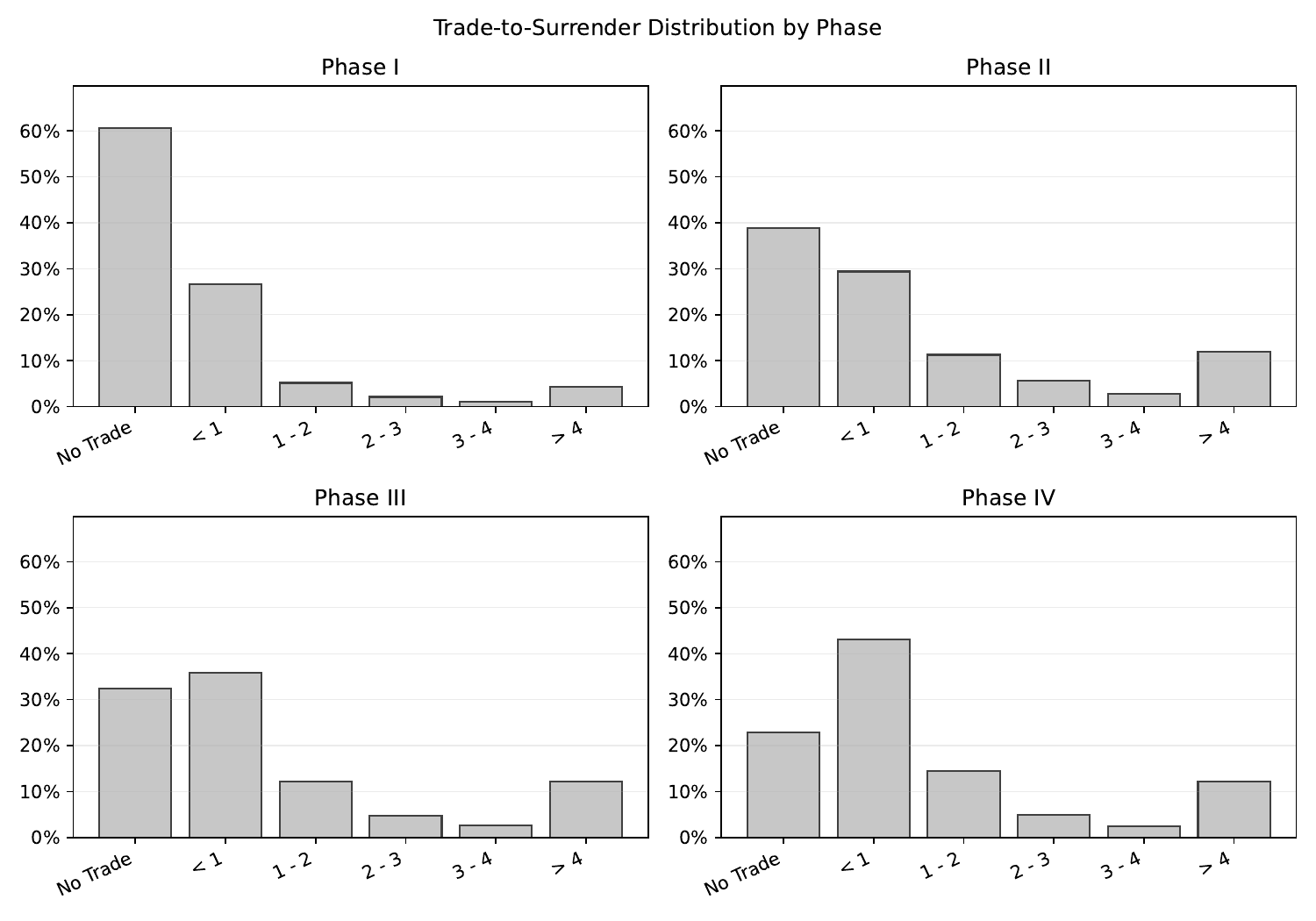}
\par\end{centering}
}
\end{figure}

We analyze the trading behavior of firms with operator holding accounts in the EU ETS. The top panels of Figure~\ref{fig:trading-frequency} show the yearly share of firms by trading volume---Panel A for the full sample, Panel B by EU ETS phases. A large fraction of firms do not trade in a given year: about 37\% overall, nearly 60\% in Phase I, and over 35\% in Phases II and III. Through September 2022, the Phase IV non-trading share is about 23\%.

Figure~\ref{fig:trading-frequency} also reports the trading-volume bins and trading-ratio categories used in the analysis.

Among firms that do trade, average annual volume is approximately 95,000 allowances, but the distribution is highly skewed. The median is about 3,700. 10\% of firms trade fewer than 100 allowances, while another 10\% trade more than 86,000. To control for firm size, we normalize trading volume by the surrendering amount. The bottom panels of Figure~\ref{fig:trading-frequency} show that about 34\% of firms trade less than they surrender (ratio $<1$), while over 10\% trade more than four times their surrender obligation. This pattern holds across EU ETS phases.

Non-trading is also persistent. In the full sample, an operator that does not trade in year $t$ is about 40 percentage points more likely not to trade in year $t+1$ than an operator that does trade. Section~\ref{subsec:emp_slow} reports the full persistence specifications and examines whether matched free allocations or allowance banking can account for the pattern.

\subsection{Surrender-Month Trading and Returns\label{subsec:fact_pressure}}

\begin{figure}[!t]

\caption{Trading and Prices in the Surrendering Month\label{fig:Trading-Concentrate}}

\medskip{}

{\footnotesize This figure plots the monthly net purchase of emission allowances
by operators and the average emission allowance returns for each month.
Panel A depicts monthly net purchases of allowances by operators over time, with April months in red,
while Panel B shows the average EUA log return by calendar month over January 2009 through September 2022, computed from daily log returns of the ICE EUA December futures settlement price, winsorized at the 1st and 99th percentiles and summed within each month. Transactions involving admin accounts are excluded. At least one party
in a transaction must be an operator for the transaction to be included
in the analysis.\par}

\centering

\medskip{}

\subfloat[Net Transaction Amount]{
\includegraphics[width=10cm]{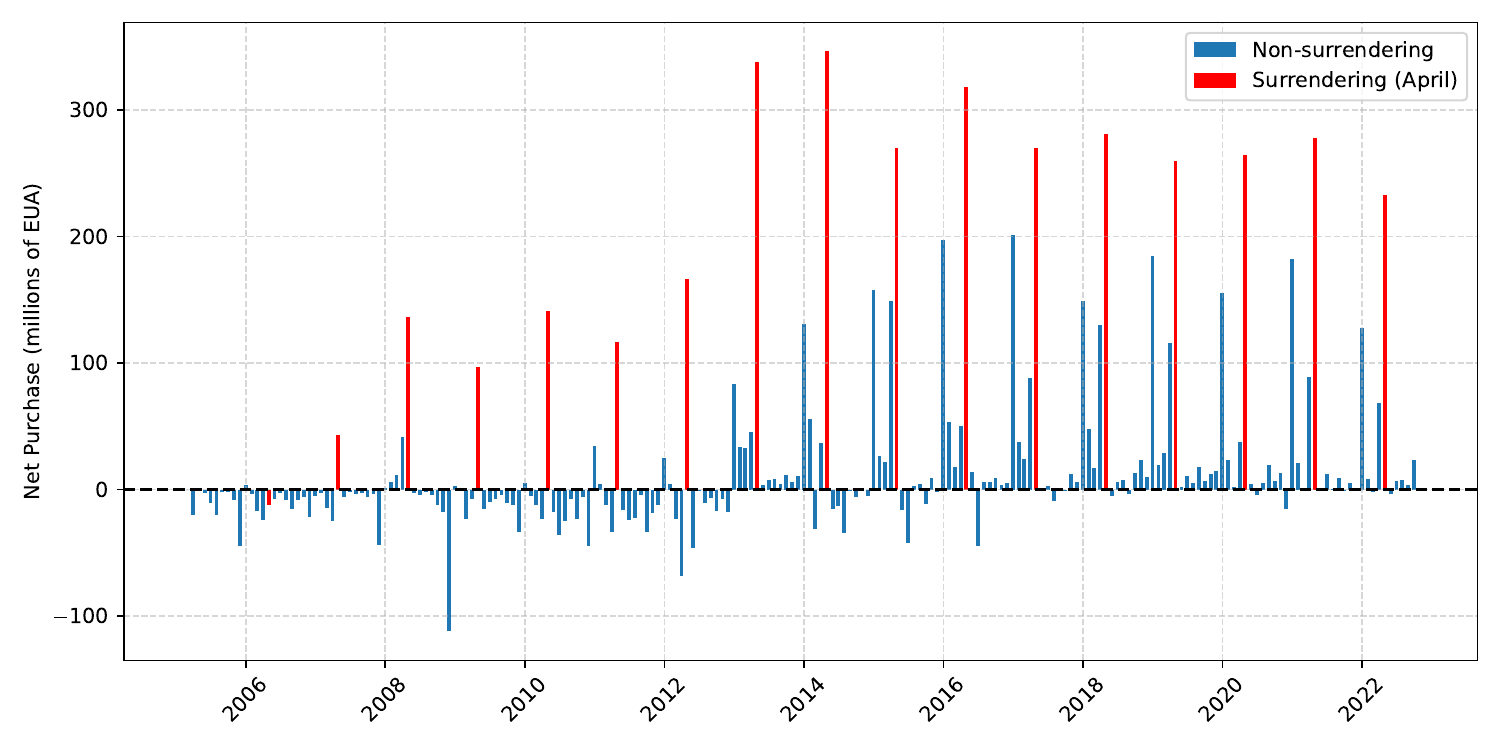}
\par
}

\subfloat[Average EUA Return]{
\includegraphics[width=10cm]{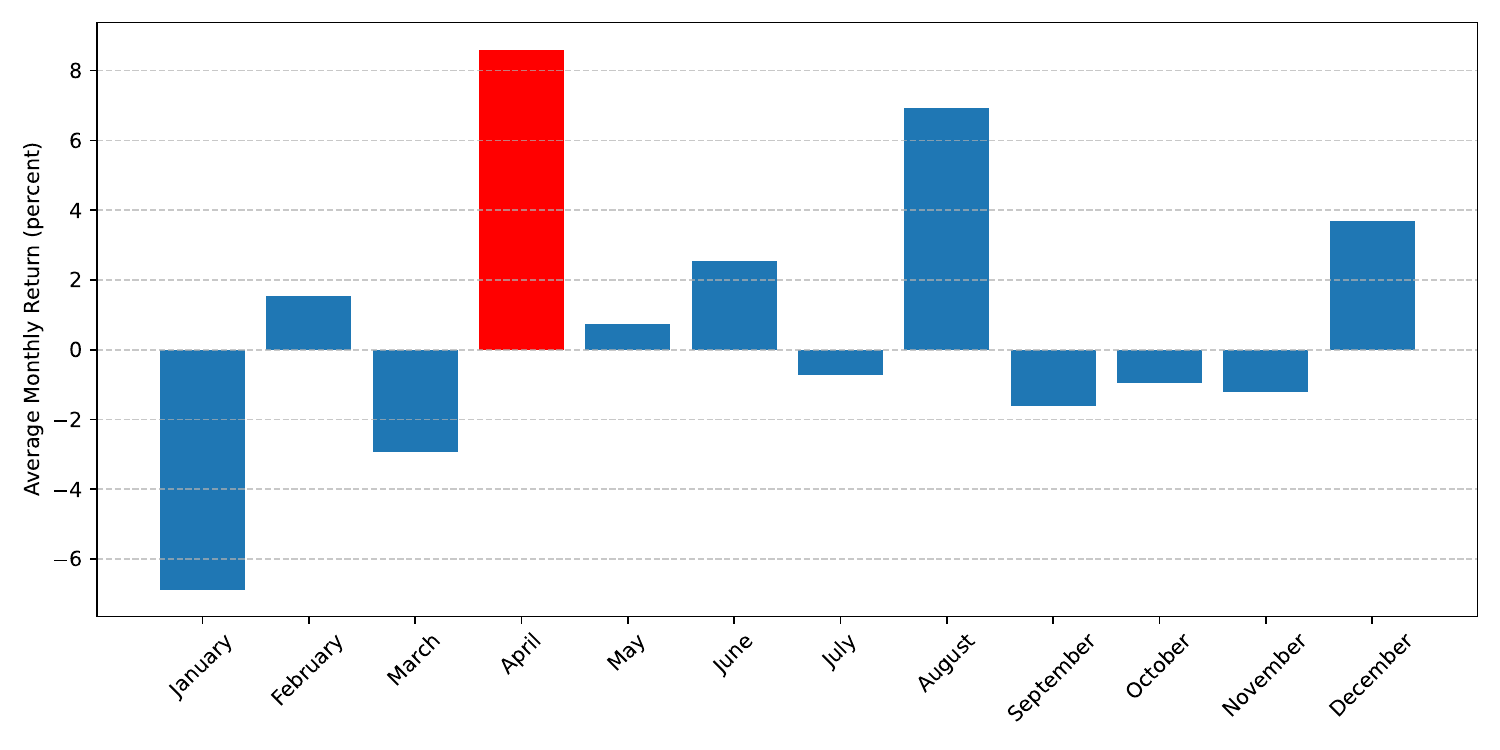}
\par
}

\end{figure}

We compute the monthly net purchase of emission allowances by firms, excluding transactions involving administrative accounts. Panel A of Figure~\ref{fig:Trading-Concentrate} shows a striking pattern: April consistently has the largest net operator purchases throughout the sample. Average April net purchases are about 197 million allowances, roughly eighteen times the median absolute net purchase in non-April months. The second-largest net purchase month is typically December, possibly because of tax considerations or the expiration of the most common EUA futures contract. Although some April buying may reflect uncertainty resolved by the end-March verification process, verification timing alone is unlikely to account for the size and persistence of the April spike.

Panel B reports average monthly log allowance returns by calendar month over January 2009 through September 2022, computed from daily settlement prices of the ICE EUA December futures contract, the price series used throughout the paper.\footnote{An earlier draft used a daily EUA spot series from Datastream with quantitatively similar results.} The average return is highest in April, at 8.6\%. August, at 6.9\%, is the only other month above 4\%. Panel B starts in 2009, the first surrender year with bankable allowances. The pilot-phase Aprils of 2006 and 2007, when Phase I allowances collapsed, are reported in the Online Appendix. Concentrated operator buying in the surrender month thus coincides with systematically high April returns. Daily data sharpen the timing: the return accrues mainly in the first half of April, as verified emissions become known and compliance purchases reach the market, rather than in a last-day squeeze, and it does not reverse in the ten trading days after the deadline (Appendix~\ref{subsec:daily_april_evidence}).

\subsection{Operator Flow and Future Returns\label{subsec:fact_predictability}}

To capture regulated firms' net purchases in a month, we construct \emph{Operator Flow Imbalance}: the net flow from person holding accounts to operator holding accounts, scaled by total bilateral flows between the two groups. In the full sample, Operator Flow Imbalance positively predicts cumulative allowance returns from one to twelve months ahead, and the relation remains statistically significant after controlling for calendar-month effects. The measure has a standard deviation of 0.55. With calendar-month effects, a one-standard-deviation increase is associated with allowance returns about 6 percentage points higher over the following month and about 19 percentage points higher over the following three months, relative to a monthly return standard deviation of 25 percentage points. Pure temporary liquidity pressure would predict short-run impact followed by reversal. The positive longer-horizon coefficients instead indicate that operator flow contains information about future scarcity or persistent demand not yet fully reflected in prices. The relation holds in the full sample and after excluding the pilot phase (Section~\ref{subsec:emp_predictability}). Section~\ref{subsec:emp_frequent} examines which operators account for this predictability.

\section{The Mechanism in a Two-Period Benchmark\label{sec:model}}

This section develops a simple two-period model of a cap-and-trade market with trading frictions.\footnote{\cite{goulder2013markets,schmalensee2013so2,cantillon2023market} provide excellent reviews of cap-and-trade models.} Its role is to isolate the mechanism that we then embed in a continuous-time model in Section~\ref{sec:ctmodel}. The building blocks are a permit-market model following \cite{harstad2010trading}, combined with established finance mechanisms of liquidity provision, constrained intermediation, and informed trading \citep{GrossmanStiglitz1980,GrossmanMiller1988,subrahmanyam1991risk,garleanu2009demand,acharya2013limits,hendershott2014price}, and with the evidence that anticipated uninformed demand moves prices when risk-bearing capacity is limited \citep{LouYanZhang2013,HartzmarkSolomon2025}. The model introduces the smallest set of frictions needed to rationalize the empirical facts documented in Section~\ref{sec:facts}. Full derivations are provided in the Online Appendix.

Consider a competitive market for bankable emission allowances in two
periods, $t=0,1$. Firm $i$ is a permit-compliant operator which chooses emissions
$x_{it}$, net purchases of allowances $m_{it}$, and the quantity of
allowances banked from period 0 to period 1, denoted by $b_{i1}$.
The firm's flow profit in each period is
\begin{align}
\Pi_{it}=-\frac{1}{2}\left(\theta_{it}-x_{it}\right)^2-p_t m_{it},
\end{align}
where a firm's type is $\theta_{it}$, which denotes business-as-usual emissions. Allowance
holdings satisfy
\begin{align}
b_{i1}=b_{i0}+q_{i0}+m_{i0}-x_{i0}, \qquad 
0=b_{i1}+q_{i1}+m_{i1}-x_{i1}, \qquad b_{i1}\geq 0,
\end{align}
where $q_{it}$ denotes the quantity of allowances allocated to the
firm. Let $M_t=\sum_i m_{it}$ denote aggregate operator net order flow.
Thus, $M_t>0$ denotes net operator buying pressure, equivalently net allowance flow from intermediaries or person holding accounts to operator holding accounts. $M_t<0$ denotes net operator selling.
Intermediaries provide immediacy by absorbing this order flow and incur a quadratic cost when post-trade inventories deviate from target, where $\phi>0$ indexes limited intermediation capacity.

In the frictionless benchmark, $\phi=0$, participation costs are zero,
and no trader possesses private information\footnote{Throughout the paper, the frictionless benchmark is used to organize the analysis of secondary-market functioning: cost-effective reallocation of allowances, the timing with which predictable compliance demand is incorporated into prices, and the informational content of order flow. We do not use this benchmark to claim that observed market outcomes are Pareto-inefficient once trading frictions and institutional constraints are taken as given.}. Let $\xi_{i0}$ and
$\xi_{i1}$ denote the shadow values of one additional allowance in
periods 0 and 1, and let $p_t^*$ denote the corresponding
frictionless price. For an interior solution, the firm's optimality
conditions imply
\begin{align}
\theta_{i0}-x_{i0} &= \xi_{i0}, \qquad \theta_{i1}-x_{i1} = \xi_{i1},
\label{eq:model_mac_main}\\
p_0^* &= \xi_{i0}, \qquad p_1^* = \xi_{i1},
\label{eq:model_price_main}\\
p_0^* &= \delta E_0[p_1^*]. \label{eq:model_euler_main}
\end{align}
Equations \eqref{eq:model_mac_main}--\eqref{eq:model_euler_main}
imply that, absent frictions, marginal abatement costs are equalized
across firms and frictionless allowance prices satisfy the standard
no-arbitrage condition for a bankable asset.
The banking constraint $b_{i1}\geq 0$ is asymmetric: firms can store surplus allowances across compliance years but cannot run a negative bank, so borrowing future allocations to cover current obligations is ruled out. The first-order conditions \eqref{eq:model_mac_main}--\eqref{eq:model_euler_main} hold for an interior banking solution. When the no-borrowing corner $b_{i1}=0$ binds, the Euler equation \eqref{eq:model_euler_main} holds with inequality, $p_0^*\geq\delta E_0[p_1^*]$, because the firm would like to shift additional allowance resources from the future to the present but cannot run a negative bank. With frictionless trading the multiplier on $b_{i1}\geq 0$ equals $p_0^*-\delta E_0[p_1^*]$ for every firm, so the corner binds for all firms at once, when the aggregate bank is exhausted, and current allowances are then more valuable at the margin than their discounted expected future value. The no-borrowing constraint can reinforce the surrender-month demand pressure formalized below: firms can use banked allowances to smooth compliance but cannot borrow future allocations to avoid deadline-month purchases.
This benchmark yields three immediate
implications. First, non-trading over the compliance horizon is a knife-edge outcome: a firm trades in neither period only if its initial allowance position already coincides with its cost-minimizing compliance and banking plan under the frictionless benchmark. Second, predictable
seasonal compliance demand, such as the April surrender deadline,
should already be reflected in prices before the deadline arrives.
Third, order flow based only on public compliance needs should not
predict future returns. We interpret the documented stylized facts of the EU ETS as departures from this benchmark.

When intermediation is limited, $\phi>0$, actual prices satisfy
\begin{align}
p_t = p_t^* + \phi M_t. \label{eq:model_intermediary_main}
\end{align}
Equation \eqref{eq:model_intermediary_main} says that prices can
depart from frictionless benchmark prices when intermediaries must
absorb net order imbalances and charge a premium for immediacy provision.\footnote{The Online Appendix derives this relation from a short-window inventory problem in the tradition of the classic inventory models of dealer pricing \citep{Stoll1978,HoStoll1981,GrossmanMiller1988}: intermediaries enter the window at target inventory and value retained allowances at $p_t^*$, so their first-order condition and market clearing yield \eqref{eq:model_intermediary_main}. Away from the target, the price would also load on the pre-existing inventory deviation.} Operators and intermediaries transact at the same actual price $p_t$, and operators' optimality conditions \eqref{eq:model_mac_main}--\eqref{eq:model_price_main} then hold with $p_t$ in place of $p_t^*$. $p_t^*$ is a counterfactual benchmark, not a second transaction price. For example, if many operators delay purchases until the surrender month, intermediaries must absorb a large wave of compliance demand, which can push April prices above their frictionless benchmark level.

\begin{figure}[tbp]
\caption{Timeline of the Two-Period Model with Intermediation.\label{fig:model_timeline}}

\medskip{}

{\footnotesize This figure illustrates the timing of the two-period model used in the
main text. Allowances are allocated at the beginning of periods 0 and
1, compliance occurs at the end of periods 0 and 1, unused allowances
can be banked from period 0 to period 1, and intermediaries absorb net
order flow in each period. In the EU ETS, the compliance date corresponds to the April compliance deadline. The figure also highlights the price-impact
relation $p_t=p_t^*+\phi M_t$.}

\medskip{}

\centering{}\includegraphics[width=0.9\textwidth]{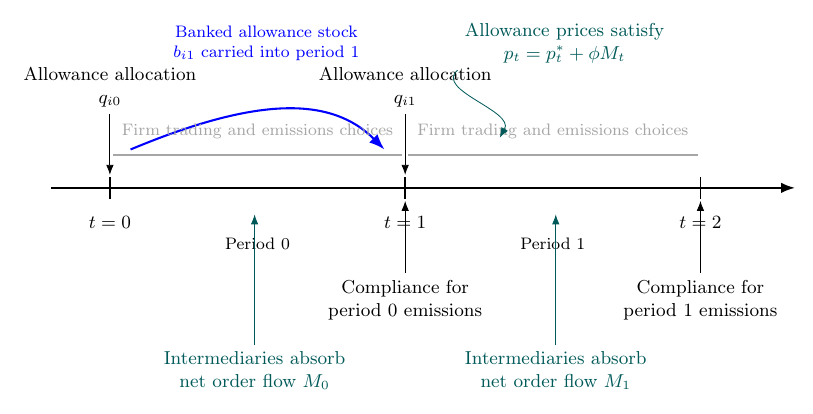}
\end{figure}

The frictionless benchmark above describes the cost-minimizing allocation with bankable allowances. To study the timing of compliance trades within a yearly compliance cycle, we now use a simplified within-cycle representation with an early trading date $e$ and a surrender-month date $A$. This representation isolates the participation and timing mechanism that generates delayed trading and surrender-month price pressure. Figure~\ref{fig:model_timeline} provides a high-level summary of the timing of the two-period setup and the role of intermediation.

\subsection{Participation Costs and Surrender-Month Trading}

To interpret the large fraction of firms that do not trade and the
concentration of purchases in April, suppose that firms face a fixed
participation cost $K_i>0$ whenever they trade actively before the
surrender deadline. This cost can be interpreted broadly as the
internal cost of attention, treasury management, internal approval,
compliance monitoring, or establishing and maintaining a trading
relationship.

Consider one compliance cycle with two trading dates: an early date
$e$ and the surrender month $A$, which corresponds to the April compliance deadline in the EU ETS. Figure~\ref{fig:model_within_period}
in the Online Appendix provides a more detailed within-cycle
timeline, including early trading, verification, and surrender. Let
$d_i\equiv\max\{\text{expected allowance shortfall}_i,0\}$ denote firm
$i$'s expected positive shortfall at the surrender date, that is, the expected quantity of allowances it still needs to acquire to meet compliance after accounting for allocations and any banked permits. To keep the
mechanism parsimonious, we abstract from verification uncertainty and
treat firms' compliance needs as known in expectation before the
surrender date. A firm that
becomes active at date $e$ can purchase allowances early at price
$p_e$ rather than wait and buy in April at price $p_A$. Ignoring
within-year discounting, the firm trades early if and only if
\begin{align}
d_i \left(E_0[p_A]-p_e\right) \geq K_i. \label{eq:model_participation_main}
\end{align}
Condition \eqref{eq:model_participation_main} governs the timing of purchases among firms with positive expected shortfalls: firms for which the condition fails delay purchases until the surrender month, while firms for which the condition holds buy early.\footnote{Equivalently, one can write the threshold as $d_i(E_0[p_A]-p_e)\geq K_i^e-K_i^A$, where $K_i^e$ is the cost of early market participation and $K_i^A$ is the lower incremental cost of deadline-month compliance trading. We normalize $K_i^A=0$ so that $K_i\equiv K_i^e$ is the incremental cost of early relative to deadline-month compliance trading. Condition \eqref{eq:model_participation_main} should be interpreted as a reduced-form participation condition rather than as the solution to a fully specified equilibrium problem. The expected April premium is taken as given by the individual firm, even though in equilibrium it is shaped by delayed compliance demand and intermediary immediacy-provision costs.} Complete non-trading arises in a different regime: it applies to firms whose expected shortfall is zero or can be covered with initial or banked allowances, or to firms whose effective market-access costs prevent discretionary rebalancing despite nonzero imbalances. Because intermediaries
absorb net order flow subject to immediacy-provision costs, prices at each
within-cycle date $\ell\in\{e,A\}$ satisfy
\begin{align}
p_{\ell}=p_{\ell}^*+\phi M_{\ell}. \label{eq:model_april_price_main}
\end{align}
When many firms delay participation, compliance demand is concentrated
at the surrender date, so $M_A$ becomes large and the April price can
rise above the early price even though the compliance wave is
predictable. This mechanism is closely related to models in which limited arbitrage and intermediary balance-sheet constraints affect asset prices
\citep{garleanu2009demand,hendershott2014price,gabaix2015international}.

In the continuous-time model below, the fixed cost $K_i$ becomes a chosen Poisson market-access intensity with a convex attention or treasury-management cost: the two-period condition captures the early-versus-surrender timing tradeoff, while the continuous-time model endogenizes the speed at which firms access the market during the compliance cycle.

\subsection{Private Information and Speculative Trading}

Participation costs alone cannot explain why operator order flow
predicts future allowance returns. To capture this fact, suppose that
a subset of active firms receives a private signal about future permit
scarcity or future permit values. Let $\mathcal{F}_t$ denote the
public information set at time $t$, and let
$\mathcal{I}_{it}\supseteq\mathcal{F}_t$ denote firm $i$'s
information set, which may also contain private information. Define
$\mu_{it}$ as the private component of expected benchmark returns,
\begin{align}
\mu_{it}
= E_t\!\left[p_{t+1}^*-p_t^* \mid \mathcal{I}_{it}\right]
- E_t\!\left[p_{t+1}^*-p_t^* \mid \mathcal{F}_t\right],
\end{align}
where $p_t^*$ is the frictionless benchmark price. This convention isolates information not yet reflected in the public state and avoids treating predictable public April price pressure as private information. Taking prices as given, an active firm can, in addition to its compliance-motivated position, take a speculative inventory position $n_{it}$ and solves
\begin{align}
\max_{n_{it}} \left\{ \mu_{it} n_{it} - \frac{\gamma_i}{2} n_{it}^2 \right\},
\end{align}
where $\gamma_i>0$ captures inventory, balance-sheet, or risk
management costs. The optimal speculative position is therefore
\begin{align}
n_{it}=\frac{\mu_{it}}{\gamma_i}. \label{eq:model_spec_main}
\end{align}
Equation \eqref{eq:model_spec_main} implies that aggregate operator order flow can predict future benchmark-price changes $p_{t+1}^*-p_t^*$ when a subset of firms trades on private information. Predictability of actual returns additionally requires that contemporaneous price impact under-reveal the signal, a condition stated in Section~\ref{subsec:ct_info}. The model suggests that return predictability
should be strongest for frequent traders, since these are the firms
most likely to have low participation costs and to trade on private
information. This is the pattern we test below in the
return-predictability tests, where forecasting power is concentrated
among frequent traders and the implied ex post trading performance is economically
large.

\subsection{April Premium Paid by Delayed Buyers\label{subsec:private_costs}}

The two-period model clarifies the April premium paid by regulated firms that delay compliance purchases. If firm $i$ expects to be short by $d_i$ allowances at the surrender date, its expected additional payment from delaying purchases until April is
\[
E_0[\text{Premium}^{\text{delay}}_i] = d_i(E_0[p_A]-p_e).
\]
Aggregating across delayed buyers yields the back-of-the-envelope measure computed in Section~\ref{subsec:emp_pressure}. This object is not a direct measure of aggregate resource costs, because allowance payments are partly transfers across market participants. It indicates the size of the participation and adjustment frictions needed to rationalize delayed trading.

\section{Continuous-Time Model with Endogenous Access\label{sec:ctmodel}}

The two-period model is useful for intuition, but it compresses the recurring compliance cycle into a single early-versus-late trading decision. In practice, regulated firms manage allowance positions over repeated annual surrender cycles, access the market gradually rather than continuously, and trade in a market where prices adjust as intermediaries absorb order imbalances. This section generalizes the same mechanism in a continuous-time framework designed to discipline the timing and return-predictability implications. It preserves the same three economic forces as the two-period model---participation frictions, limited intermediation, and informational advantages among active operators---and yields closed-form results in a seasonal benchmark and transparent comparative statics beyond that benchmark.\footnote{Full derivations and proofs are in the Online Appendix.} 

The relationship between the two-period model and the continuous-time model is direct. The high participation cost of the two-period model becomes an endogenous Poisson market-access intensity: in simple terms, firms do not rebalance continuously, but instead choose how often to receive trading opportunities, balancing the value of earlier adjustment against attention or treasury costs. A lower chosen arrival rate means slower effective access to the market. Limited intermediary capacity becomes affine price impact: when intermediaries must absorb more net buying or selling pressure at a given moment, prices move more, and in the model this effect is summarized by a linear relation between contemporaneous net order flow and the price deviation from the frictionless benchmark. Private information becomes private signals received by a subset of operators when they gain market access, implying that information enters prices gradually rather than all at once.

\subsection{Dynamic Environment and Frictionless Benchmark}

Time is continuous, and compliance obligations recur on a yearly
calendar. Let $s_t$ denote the time remaining until the next surrender
date. The public state at date $t$ is
\begin{align}
X_t = (B_t, z_t, s_t),
\label{eq:ct_state}
\end{align}
where $B_t$ is the aggregate bank of allowances and $z_t$ is a public
scarcity state that captures tightness of the cap and aggregate
compliance demand. We use an affine benchmark price,
\begin{align}
P_t^* = A_0(s_t) + A_B(s_t) B_t + A_z(s_t) z_t,
\label{eq:ct_benchmark}
\end{align}
which obtains under affine state dynamics and linear marginal compliance value, as implied by the quadratic-cost benchmark in Section~\ref{sec:model}, and provides a tractable approximation more generally.
The coefficient functions $A_0(s_t)$, $A_B(s_t)$, and $A_z(s_t)$
capture the smoothing of predictable compliance demand across the
compliance cycle. In the frictionless benchmark, predictable
compliance demand is reflected in prices well before the surrender
date, surrender-month returns are not systematically elevated, and
public operator order flow does not forecast future returns.

\subsection{Slow Participation as Endogenous Poisson Market Access\label{subsec:ct_access}}

The key departure from the frictionless benchmark is that operators
adjust compliance positions only intermittently. Rather than treating market-access intensity as exogenous, we represent slow participation endogenously by letting operators choose, at a convex attention or treasury-management cost, how often they receive discretionary trading opportunities during the compliance cycle, in the spirit of optimal-inattention models in which agents choose how often to observe the market or rebalance \citep{DuffieSun1990,AbelEberlyPanageas2013,AlvarezGuisoLippi2012}. Let $c_{it}$ denote operator $i$'s
\emph{compliance allowance holdings} and $h_{it}^*$ the operator's
target compliance holdings, which reflect expected end-of-cycle
obligations and any optimal banking motive.

\paragraph{Endogenous access choice.}
At the beginning of each compliance cycle, operator $i$ chooses a market-access intensity $\lambda_i$ for the cycle. Conditional on this choice, discretionary trading opportunities arrive according to a Poisson process with intensity $\lambda_i$. Let $\tau$ denote the time between the access-choice date and the surrender deadline, $d_i\geq 0$ the operator's expected compliance shortfall, and let $\pi_A$ denote the expected surrender-month return premium relative to the frictionless benchmark. For expositional simplicity, we normalize the expected frictionless return between the early trading date and the surrender month to zero and work in return units relative to the early-trading price level:
\begin{align}
\pi_A \equiv E_0\!\left[\frac{P_A - P^*(X_A)}{P_e}\right]
= E_0\!\left[\frac{P_A - P_e}{P_e}\right]
\approx E_0[\log P_A - \log P_e],
\label{eq:ct_DeltaA}
\end{align}
where $X_A=(B_A,z_A,s_A)$ denotes the public state in the surrender month, $P^*(X_A)$ is the corresponding frictionless benchmark price, $P_e$ is the relevant early-trading price, and $P_A$ is the surrender-month price. Two simplifications underlie \eqref{eq:ct_DeltaA}: the early-trading price equals its frictionless benchmark, $P_e=P_e^*$, so that price impact at the early date is abstracted from, and expected informative flow in the surrender window is zero, $E_0[M_A^{\text{info}}]=0$, so that the fixed point below involves compliance demand only. To keep the access-choice problem on the same scale, let $\tilde{\psi}_i\equiv \psi_i/P_e$ denote the normalized access-cost parameter. In the solved access-choice problem, the private value of obtaining at least one discretionary access opportunity before surrender comes from avoiding terminal compliance buying for the expected shortfall:
\begin{align}
S_i = d_i\,\pi_A.
\label{eq:ct_S}
\end{align}
The expression $S_i=d_i\pi_A$ includes compliance savings only. Expected speculative profits are not part of the access-choice objective. It treats any access opportunity during $[0,\tau]$ as enabling early purchase at the frictionless benchmark price level.
The operator chooses $\lambda_i$ to solve
\begin{align}
\max_{\lambda_i\geq 0}
\left\{ S_i\!\left(1-e^{-\lambda_i\tau}\right) - \frac{\tilde{\psi}_i}{2}\lambda_i^2 \right\},
\label{eq:ct_access_problem}
\end{align}
where $\tilde{\psi}_i>0$ indexes the operator's normalized cost of market access, attention, and treasury management. The first-order condition is
\begin{align}
S_i\,\tau\, e^{-\lambda_i\tau} = \tilde{\psi}_i\,\lambda_i,
\label{eq:ct_FOC}
\end{align}
and the unique solution is, using the Lambert $W$ function\footnote{The Lambert $W$ function is the inverse of $x\mapsto x e^x$ on $\mathbb{R}_+$, so if $y e^y=z$, then $y=W(z)$.},
\begin{align}
\lambda_i^* = \frac{1}{\tau}\,W\!\left(\frac{S_i\,\tau^{2}}{\tilde{\psi}_i}\right) = \frac{1}{\tau}\,W\!\left(\frac{d_i\,\pi_A\,\tau^{2}}{\tilde{\psi}_i}\right).
\label{eq:ct_lambdaStar}
\end{align}
Since $W$ is increasing on $\mathbb{R}_+$, the chosen intensity $\lambda_i^*$ is increasing in $d_i$ and $\pi_A$, and decreasing in $\tilde{\psi}_i$.

The solution makes slow participation endogenous. Firms with larger expected shortfalls or higher expected surrender-month premia choose higher access intensity, while firms facing a higher normalized access-cost parameter choose lower access intensity. Low-access firms are more likely to leave compliance needs unresolved until the surrender month. The chosen intensity $\lambda_i^*$ is the dynamic counterpart of the fixed participation cost $K_i$ in the two-period model: a higher $\tilde{\psi}_i$ raises, and a higher value of access $S_i=d_i\pi_A$ lowers, the probability $e^{-\lambda_i^*\tau}$ of reaching the surrender month without having traded, the analogue of failing the participation condition \eqref{eq:model_participation_main}.

\paragraph{Compliance-holdings dynamics under endogenous access.}
Compliance positions can be rebalanced only when operator $i$'s Poisson counting process for access opportunities, $N_{it}$, jumps. Upon access, the operator resets compliance holdings to the desired level:
\begin{align}
dc_{it} = \bigl( h_{it}^* - c_{it^-} \bigr)\, dN_{it},
\qquad
E_t[dN_{it}] = \lambda_i^*\, dt.
\label{eq:ct_access}
\end{align}
Here $c_{it^-}$ denotes compliance holdings immediately before the access jump at time $t$.
The intensity $\lambda_i^*$ is constant within a compliance cycle and is reset at the beginning of each cycle through the access-choice problem in \eqref{eq:ct_access_problem}, as $(d_i,\pi_A,\tilde{\psi}_i)$ realize for that cycle.\footnote{Allowing operators to close only a fraction $\alpha_i\in(0,1]$ of the gap upon access replaces $\lambda_i^*$ by $\alpha_i\lambda_i^*$ in the expected gap dynamics and residual-demand formula. We use full reset in the main text to keep the endogenous-access mechanism transparent.}

Define the operator-level compliance gap as
\begin{align}
g_{it} = h_{it}^* - c_{it}.
\label{eq:ct_gap_def}
\end{align}
We treat $h_{it}^*$ as a stochastic target: it has a deterministic drift component capturing the calendar build-up of compliance demand toward surrender, an idiosyncratic component reflecting firm-level innovations to expected obligations between verification cycles, and a public-state component that loads on the aggregate scarcity state $z_t$. Specifying $h_{it}^*$ as stochastic ensures that operator-level compliance gaps are not fully predictable away from access shocks and gives private signals about future scarcity a non-trivial object to be informative about.
In expectation, away from the terminal surrender date, unresolved compliance demand at the operator level obeys
the linear adjustment equation
\begin{align}
E_t[dg_{it}] = E_t[dh_{it}^*] - \lambda_i^*\, g_{it} \, dt.
\label{eq:ct_gap}
\end{align}
Aggregating across operators, aggregate unresolved compliance demand $G_t = \sum_i g_{it}$ satisfies
\begin{align}
E_t[dG_t] = A(s_t,X_t)\, dt - \sum_i \lambda_i^*\, g_{it}\, dt = A(s_t,X_t)\, dt - \bar\lambda_t^*\, G_t\, dt,
\qquad
\bar\lambda_t^* \equiv \frac{\sum_i \lambda_i^*\, g_{it}}{G_t},
\label{eq:ct_Gagg}
\end{align}
where $A(s_t,X_t)=\sum_i a_i(s_t,X_t)$ captures how desired compliance demand rises as the surrender date approaches and $\bar\lambda_t^*$ is the gap-weighted average chosen access intensity over operators with positive gaps. Aggregate adjustment is slower when many operators have high normalized access costs $\tilde{\psi}_i$, small expected shortfalls $d_i$, or low perceived value of early trading $\pi_A$. When desired compliance demand rises toward surrender and operators adjust only gradually, unresolved
demand accumulates and is largest near surrender.

\paragraph{Terminal compliance and residual demand.}
Conditional on the cycle-level choice $\lambda_i^*$, the realized compliance gap evolves in continuous time during the cycle. Because surrender is mandatory, any remaining positive compliance shortfall must be closed at the surrender date through a terminal compliance trade. Under full reset upon access (that is, $\alpha_i=1$), the probability that operator $i$ receives no discretionary access opportunity before surrender is $e^{-\lambda_i^*\tau}$, and the corresponding expected residual terminal demand for operator $i$ is
\begin{align}
R_{iA} = d_i\, e^{-\lambda_i^*\,\tau}.
\label{eq:ct_RiA}
\end{align}
For the closed-form fixed-point specification and the quantitative calibration below, we specialize the richer dynamic environment to a one-shot known-shortfall object at the cycle level: each operator enters the cycle with a single expected shortfall $d_i$, and terminal demand depends on whether at least one discretionary access opportunity arrives before surrender. Aggregating across operators then yields the expected residual terminal demand
\begin{align}
D_A(\pi_A) = \sum_i d_i\, e^{-\lambda_i^*(\pi_A)\,\tau},
\label{eq:ct_DA_main}
\end{align}
where $\lambda_i^*(\pi_A)$ is the function defined by \eqref{eq:ct_lambdaStar}. In the closed-form fixed point, $D_A$ is expected \emph{net deadline buying pressure}: the residual positive-shortfall demand that must be met by net supply from the intermediary sector. The net object is the relevant one for the surrender-window price premium. If one tracks the realized residual gap $g_{iA-}=h^*_{iA}-c_{iA-}$ immediately before surrender, the terminal compliance trade is $\Delta c_{iA}=g_{iA-}^{+}$ and aggregate terminal order flow is
\begin{align}
M_A^{\text{term}} = \sum_i g_{iA-}^{+},
\label{eq:ct_Mterm}
\end{align}
where $g_{iA-}^{+}=\max\{g_{iA-},0\}$, since only positive shortfalls generate mandatory surrender-month purchases. Under the same reduced-form immediacy-supply relation introduced below, the surrender-month price impact is
\begin{align}
P_A - P^*(X_A) = \frac{\phi}{h_A}\left(M_A^{\text{term}} + M_A^{\text{info}}\right).
\label{eq:ct_PAterm}
\end{align}
We treat the surrender date as a windowed event of length $h_A$: $M_A^{\text{term}}$ and $M_A^{\text{info}}$ in \eqref{eq:ct_PAterm} are stock-equivalent integrated flows over the surrender window, and $\phi$ in \eqref{eq:ct_PAterm} is the same instantaneous price-level impact parameter as in \eqref{eq:ct_price}. Dividing by $h_A$ converts the integrated terminal flow into the average flow rate over the window, against which $\phi$ acts as in \eqref{eq:ct_price}.
Endogenous access determines how much demand remains unresolved at the surrender date: lower normalized access costs $\tilde{\psi}_i$ or larger benefits of early trading $S_i$ raise $\lambda_i^*$, reduce expected residual terminal demand $R_{iA}$, and weaken surrender-month price pressure. To connect this residual demand to the return-based object used in the quantitative section, let $\varphi>0$ denote the reduced-form surrender-window return-impact parameter, distinct from the instantaneous price-level impact parameter $\phi$ in \eqref{eq:ct_price}. Specifically, $\varphi=\phi/(P_e h_A)$, where $P_e$ is the early-cycle price level used to define $\pi_A$ and $h_A$ is the surrender-window length introduced after \eqref{eq:ct_PAterm}. The equilibrium surrender-month return premium then satisfies $\pi_A=\varphi D_A(\pi_A)$.
The fixed point is a buyer-side specification, for operators with expected positive shortfalls $d_i>0$. Operators with expected surpluses are outside this specification. They would value April access as an opportunity to sell into the premium, and their sales would add supply at the deadline, lowering net residual demand and the premium. Because $\hat{\varphi}$ in Section~\ref{sec:quant} is calibrated to observed April premia given buyer-side terminal demand, this offset is absorbed in the calibrated return impact rather than modeled.

\paragraph{Endogenous access and surrender-month pressure.}
The endogenous-access mechanism delivers the following closed-form characterization, which states the optimal access intensity, its comparative statics, and the existence and uniqueness of the equilibrium surrender-month premium in the one-shot known-shortfall specification.

\begin{prop}\label{prop:endog_access}
In the one-shot known-shortfall specification, suppose operator $i$ has expected compliance shortfall $d_i\geq 0$, normalized access-cost parameter $\tilde{\psi}_i>0$, and chooses a cycle-level Poisson market-access intensity $\lambda_i$ over horizon $\tau>0$. If the expected surrender-month return premium is $\pi_A\geq 0$, the optimal access intensity is
\begin{align}
\lambda_i^*(\pi_A) = \frac{1}{\tau}\,W\!\left(\frac{d_i\,\pi_A\,\tau^{2}}{\tilde{\psi}_i}\right),
\label{eq:prop_lambdaStar}
\end{align}
where $W(\cdot)$ is the Lambert $W$ function. The intensity $\lambda_i^*(\pi_A)$ is nondecreasing in $d_i$ and $\pi_A$, strictly increasing when $d_i>0$ and $\pi_A>0$, and decreasing in $\tilde{\psi}_i$. Under full reset upon access (that is, $\alpha_i=1$), expected aggregate residual terminal demand is
\begin{align}
D_A(\pi_A) = \sum_i d_i\, e^{-\lambda_i^*(\pi_A)\,\tau}.
\label{eq:prop_DA}
\end{align}
If the surrender-month return premium satisfies the reduced-form return-impact relation
\begin{align}
\pi_A = \varphi\,D_A(\pi_A),
\label{eq:prop_fixed_point}
\end{align}
with $\varphi>0$, then whenever $\sum_i d_i>0$ the map $\pi_A\mapsto \varphi D_A(\pi_A)$ is continuous and strictly decreasing, satisfies $\varphi D_A(0)=\varphi\sum_i d_i>0$, and converges to zero as $\pi_A\to\infty$. Hence there exists a unique positive equilibrium return premium solving \eqref{eq:prop_fixed_point}.
\end{prop}

\begin{cor}\label{cor:phi_elasticity}
At the equilibrium of Proposition~\ref{prop:endog_access}, whenever $\sum_i d_i>0$ the elasticity of the surrender-month return premium with respect to the surrender-window return-impact parameter is bounded strictly below unity:
\begin{align}
\frac{d\log \pi_A}{d\log \varphi} = \frac{1}{1+\varphi\,|D_A'(\pi_A)|} \in (0,\,1),
\label{eq:cor_elasticity}
\end{align}
where $D_A'(\pi_A)<0$ because endogenous access $\lambda_i^*(\pi_A)$ is strictly increasing in $\pi_A$.
\end{cor}

Corollary~\ref{cor:phi_elasticity} formalizes the \emph{dampening} feature emphasized in the introduction: endogenous access reabsorbs a fraction of any policy change in $\varphi$ through adjustment of access intensities and hence of residual terminal demand, so the equilibrium premium responds less than one-for-one to a change in intermediary return impact. A static price-impact model with exogenous $\lambda_i$ would instead deliver $d\log\pi_A/d\log\varphi=1$. The shortfall from unity is the elasticity gap absorbed by endogenous access.

\begin{cor}\label{cor:friction_interaction}
In the limit $\tilde\psi_i\to\infty$ for all $i$, $\lambda_i^*(\pi_A)\to 0$, $|D_A'(\pi_A)|\to 0$, and the elasticity \eqref{eq:cor_elasticity} converges to unity. The endogenous-access dampening of the equilibrium premium therefore vanishes in the static-access limit, and the equilibrium response to a change in $\varphi$ depends on the level of normalized access costs.
\end{cor}

Corollary~\ref{cor:friction_interaction} characterizes the static-access limit. Corollary~\ref{cor:proportional_amplification} below establishes monotone amplification under a proportional-cost benchmark.

\begin{cor}\label{cor:elasticity_bound}
At the equilibrium of Proposition~\ref{prop:endog_access}, whenever $\sum_i d_i>0$, the elasticity in \eqref{eq:cor_elasticity} admits the closed form
\begin{align}
\frac{d\log \pi_A}{d\log \varphi} = \frac{1}{1+\omega}, \qquad \omega \equiv \sum_i \varrho_i\,\frac{W_i}{1+W_i},
\label{eq:cor_omega}
\end{align}
where $W_i\equiv W\!\left(d_i\pi_A\tau^2/\tilde\psi_i\right)$, $\varrho_i\equiv d_i e^{-W_i}/\sum_j d_j e^{-W_j}$ is operator $i$'s share of aggregate residual terminal demand, and $W(\cdot)$ is the Lambert $W$ function. Because $W_i/(1+W_i)\in(0,1)$ for every operator with $d_i>0$ and the shares $\varrho_i$ sum to one over those operators, $\omega\in(0,1)$ and the elasticity is bounded strictly in $\left(\tfrac{1}{2},\,1\right)$. The elasticity converges to $1/2$ in the limit $\tilde\psi_i\to 0$ for all $i$.
\end{cor}

Corollary~\ref{cor:elasticity_bound} sharpens the dampening result of Corollary~\ref{cor:phi_elasticity}: the equilibrium-premium elasticity to intermediary return impact is bounded strictly above $1/2$ as well as strictly below $1$. The lower limit is approached when access frictions are negligible, in which case the equilibrium premium scales as $\sqrt{\varphi}$. The upper limit was characterized in Corollary~\ref{cor:friction_interaction}.

\begin{cor}\label{cor:scale_invariance}
For any $k>0$, the joint rescaling $(\varphi,\tilde\psi_i)\mapsto(k\varphi,\,k\tilde\psi_i)$ for all $i$ leaves the Lambert-$W$ argument $d_i\pi_A\tau^2/\tilde\psi_i$ in \eqref{eq:prop_lambdaStar} invariant and scales the equilibrium of Proposition~\ref{prop:endog_access} by the same factor: $\pi_A\mapsto k\pi_A$, with each $\lambda_i^*$ and $D_A(\pi_A)$ unchanged. In particular, halving both frictions ($k=1/2$) halves the equilibrium premium exactly.
\end{cor}

Corollary~\ref{cor:scale_invariance} explains why the joint half-$\varphi$ and half-$\tilde\psi$ row in Table~\ref{tab:quant_cf} lowers the equilibrium premium by exactly $50\%$: under the joint rescaling, the Lambert-$W$ argument is invariant, so access intensities and residual demand do not respond, and the equilibrium response coincides with the prediction of a static price-impact model in which $\lambda_i$ is held fixed and only $\varphi$ varies.

\Needspace{6\baselineskip}
\begin{cor}[Amplification with proportional access costs]\label{cor:proportional_amplification}
Under the conditions of Proposition~\ref{prop:endog_access}, suppose $\sum_i d_i>0$ and $\tilde\psi_i=a d_i$ for every operator with $d_i>0$, where $a>0$ is common across operators. Let $D_0\equiv\sum_i d_i$ and $w\equiv\tfrac{1}{2}W(2\varphi D_0\tau^2/a)>0$. Positive-shortfall operators choose the common intensity $\lambda_i^*=w/\tau$, and the equilibrium-premium elasticity satisfies
\begin{align}
\varepsilon_\varphi \equiv \frac{\partial\log\pi_A}{\partial\log\varphi}
&=\frac{1+w}{1+2w}\in\left(\tfrac{1}{2},1\right),
\label{eq:cor_proportional_elasticity}\\
\frac{\partial\varepsilon_\varphi}{\partial\log a}
&=\frac{w}{(1+2w)^3}>0,
\label{eq:cor_proportional_amplification}
\end{align}
where the derivative with respect to $\log a$ holds $\varphi$, the shortfalls, and $\tau$ fixed. Higher access costs therefore amplify the premium's response to intermediary return impact relative to a lower-access-cost economy, while that response remains dampened relative to exogenous access.
\end{cor}

The proportional-cost restriction allows heterogeneous shortfalls but imposes a common normalized access-cost parameter per allowance of expected shortfall. It yields common access intensities, so higher $a$ weakens access adjustment without changing operators' relative shares of residual demand. With unrestricted heterogeneity, changes in those shares can offset this effect, and global monotonicity need not hold. We do not impose the restriction in the quantitative exercise, which retains heterogeneous access costs and illustrates amplification numerically. Proofs of Proposition~\ref{prop:endog_access} and Corollaries~\ref{cor:phi_elasticity}--\ref{cor:proportional_amplification} are in the Online Appendix.

\subsection{Intermediation and Affine Price Impact}

We give intermediary supply a short-window inventory interpretation. The Online Appendix shows that a quadratic cost of deviating from target inventory yields $P_t-v_t=\phi(M_t^s+\bar I_t-I_{t^-})$. In the fast-reset benchmark, $I_{t^-}=\bar I_t$ and $v_t=P_t^*(X_t)$: retained allowances can be traded at the benchmark outside the congested window, and discounting and benchmark drift are second order over the short interval. Thus $P_t-P_t^*(X_t)=\phi M_t^s$. Operators and intermediaries transact at the same actual price $P_t$. $P_t^*(X_t)$ is a counterfactual benchmark, not a second transaction price. Market clearing at operator demand $M_t^s=M_t$ delivers
\begin{align}
P_t = P_t^*(X_t) + \phi M_t,
\label{eq:ct_price}
\end{align}
where $M_t$ denotes the contemporaneous net operator order-flow \emph{rate}---positive when operators are net buyers, equivalently when allowances flow from intermediaries or person holding accounts to operator holding accounts---and $\phi > 0$ is the reduced-form price impact of that contemporaneous net flow. It is useful to decompose the expected net flow rate into a compliance-related
component $M_t^{\text{comp}}$, driven by the aggregate unresolved gap
$G_t$, and an informational component $M_t^{\text{info}}$, defined below as the expected informative flow rate from
speculative orders submitted when operators gain access:
\begin{align}
E_t[M_t] = E_t[M_t^{\text{comp}}] + M_t^{\text{info}}.
\label{eq:ct_decomp}
\end{align}
Within the positive-shortfall buyer specification used for the fixed point, compliance flow is nonnegative, $M_t^{\text{comp}}\geq 0$, whereas informative flow is signed, $M_t^{\text{info}}\in\mathbb{R}$. Hence total net operator flow $M_t$ can be positive or negative even though terminal compliance demand is nonnegative.
For the compliance component, under full reset upon access, the expected contemporaneous compliance-related order-flow rate is
\begin{align}
E_t[M_t^{\text{comp}}]
= \sum_i \lambda_i^*\, g_{it}
= \bar\lambda_t^*\, G_t,
\qquad
\bar\lambda_t^* \equiv \frac{\sum_i \lambda_i^*\, g_{it}}{G_t},
\label{eq:ct_Mcomp}
\end{align}
whenever $G_t>0$, where $\lambda_i^*$ is the cycle-level access intensity from \eqref{eq:ct_lambdaStar}. So contemporaneous compliance order flow is proportional to the aggregate unresolved compliance gap, with an endogenous gap-weighted access intensity $\bar\lambda_t^*$ that inherits the comparative statics of the underlying $\lambda_i^*$ in the value of access $S_i=d_i\pi_A$ and the normalized access-cost parameter $\tilde{\psi}_i$. The sign of $\phi$ is the direct counterpart of the slope of the dealers' inverse supply curve for immediacy in the two-period model: $\phi > 0$ implies that price pressure is stronger when unresolved demand is larger, and the pressure becomes more pronounced as the surrender date approaches or when the market is tighter.

\subsection{Equilibrium Surrender-Month Premium\label{subsec:ct_eqm}}

The expected surrender-month return premium $\pi_A$ enters individual access choices through the value of access $S_i=d_i\pi_A$ in \eqref{eq:ct_lambdaStar}, and it is in turn determined by aggregate residual terminal demand $D_A(\pi_A)$ in \eqref{eq:ct_DA_main}. We close the model's quantitative specification with the reduced-form return-impact condition $\pi_A=\varphi\,D_A(\pi_A)$, a price-pressure relation rather than a no-arbitrage condition. It presumes that neither active operators nor intermediaries carry inventory from the early date to the surrender window to exploit the publicly predictable premium: speculative positions respond only to the private component $\mu_{it}$ by construction, and intermediaries enter the surrender window at target inventory. Each operator takes $\pi_A$ as given when choosing $\lambda_i$ and does not internalize its own effect on aggregate terminal demand, a price-taking approximation for a large number of operators. Proposition~\ref{prop:endog_access} establishes that this fixed-point equation has a unique positive solution whenever aggregate expected shortfall $\sum_i d_i$ is positive, so that the surrender-month return premium is an equilibrium object rather than an assumed primitive: it reflects the joint outcome of operators' endogenous access choices and residual terminal demand. A higher premium raises the private value of early access, while costly access leaves residual terminal demand that generates the premium through the reduced-form surrender-window return-impact relation.

\subsection{Informational Advantages and Operator Flow\label{subsec:ct_info}}

We introduce a parsimonious
private-signal component. A subset $\mathcal{I}$ of operators receives a
private signal about short-run innovations to the frictionless
benchmark price. The private-information component is kept separate from
the compliance stock $c_{it}$: informed operators submit speculative
orders only conditional on gaining market access, and these orders do
not enter the compliance-gap law of motion
\eqref{eq:ct_gap}. Define operator $i$'s informational advantage
in expected-price-change units (euros per allowance) as
\begin{align}
\mu_{it} =
E_t\!\left[P_{t+\Delta}^* - P_t^* \mid \mathcal{I}_{it}\right]
-
E_t\!\left[P_{t+\Delta}^* - P_t^* \mid \mathcal{F}_t\right],
\label{eq:ct_mu}
\end{align}
where $\mathcal{F}_t=\sigma(X_t)$ denotes the public information set
generated by the public state $X_t$, and
$\mathcal{I}_{it}\supseteq\mathcal{F}_t$ denotes operator $i$'s
information set, including any private signal. Upon
access, taking prices as given, operator $i$ submits a speculative order $q_{it}^{\text{info}}$
that solves
\begin{align}
\max_{q_{it}^{\text{info}}}
\left\{
\mu_{it}\, q_{it}^{\text{info}}
- \frac{\gamma_i}{2}\bigl(q_{it}^{\text{info}}\bigr)^2
\right\},
\qquad
q_{it}^{\text{info}} = \frac{\mu_{it}}{\gamma_i},
\label{eq:ct_q}
\end{align}
where $\gamma_i > 0$ captures inventory, risk-management, or
balance-sheet costs. Because access arrivals are Poisson with chosen
intensity $\lambda_i^*$, the contemporaneous informative order-flow
rate contributed by operator $i$ is $\lambda_i^*\, \mu_{it}/\gamma_i$ in
expectation (each access re-establishes the speculative position from zero, so the order equals the position). Aggregating in expected-flow form across informed
operators,
\begin{align}
M_t^{\text{info}}
= \sum_{i \in \mathcal{I}} \lambda_i^*\,
\frac{\mu_{it}}{\gamma_i}.
\label{eq:ct_Minfo}
\end{align}
Because
access is intermittent, information enters prices gradually through
operator order flow, and a higher chosen $\lambda_i^*$ implies faster and more
informative flow. Operator flow imbalance predicts future
returns when informed operators trade on signals about future scarcity
that are not yet fully embedded in the price. Moreover, the expected magnitude of informed
trading is larger for operators with higher access intensity, holding signal quality and $\gamma_i$ fixed. Lower $\gamma_i$ increases the magnitude of the speculative order submitted upon access, but does not raise the compliance-driven $\lambda_i^*$ in the solved model. The frequent-versus-infrequent trader split provides a reduced-form empirical comparison motivated by this access channel.

Because actual prices include both the benchmark price and contemporaneous price impact, $P_t=P_t^*(X_t)+\phi M_t$, informative order flow $M_t^{\text{info}}$ predicts actual future returns only if contemporaneous price impact does not fully reveal the private signal. We assume that the signal content of $M_t^{\text{info}}$ is incorporated into the benchmark price gradually, so that subsequent adjustment in $P_t^*$ exceeds any temporary price-impact reversal. Concretely, if
\begin{align}
E_t\!\left[P_{t+\Delta}^*-P_t^* \mid \mathcal{F}_t, M_t^{\text{info}}\right] - E_t\!\left[P_{t+\Delta}^*-P_t^* \mid \mathcal{F}_t\right] = \beta\, M_t^{\text{info}},
\label{eq:ct_signal_revelation}
\end{align}
then expected price changes, and hence simple returns conditional on the current price, are positively related to informative flow when the signal-loading coefficient $\beta$ exceeds the expected reversal of contemporaneous price impact. We do not solve an underlying filtering or Kyle-style equilibrium.

\subsection{Model Predictions}

The continuous-time model delivers the same three core empirical predictions as the two-period model and one additional structural prediction about how the trading frictions interact, in a richer endogenous setting that also generates quantitative objects for the analysis below.

First, the model predicts endogenous slow participation. By Proposition~\ref{prop:endog_access}, operators choose lower cycle-level market-access intensity $\lambda_i^*$ when expected shortfalls $d_i$ are small, expected surrender-month return premia $\pi_A$ are low, or normalized access costs $\tilde{\psi}_i$ are high. These firms are less likely to receive discretionary trading opportunities during the compliance cycle and trade less often. As a result, unresolved compliance gaps $g_{it}$ are more persistent and adjust more slowly, concentrating a larger share of purchases near the surrender date. This prediction is the dynamic analog of the participation-cost mechanism in the two-period model: a high $\tilde{\psi}_i$ raises, and a high $S_i=d_i\pi_A$ lowers, the probability $e^{-\lambda_i^*\tau}$ of reaching the surrender month without having traded. Empirically, this prediction maps into the absence of discretionary trading before the surrender date and the concentration of purchases near surrender among operators with high attention or treasury costs. Complete non-trading in a year arises for operators whose expected shortfall is zero or covered by initial or banked allowances or, as in the two-period model, whose effective access costs prevent discretionary rebalancing altogether.

Second, the model predicts surrender-month price pressure. Endogenous access choices determine how much compliance demand remains unresolved before the deadline. During the compliance cycle, the unresolved gap $G_t$ generates compliance-related order flow with expected rate $E_t[M_t^{\text{comp}}]=\bar\lambda_t^* G_t$, where $\bar\lambda_t^*$ is the gap-weighted access intensity defined in \eqref{eq:ct_Mcomp}. At the surrender date, any remaining positive gap is converted into terminal compliance order flow $M_A^{\text{term}}$ as in \eqref{eq:ct_Mterm}, with aggregate expected residual demand $D_A(\pi_A)=\sum_i d_i e^{-\lambda_i^*(\pi_A)\tau}$. The resulting surrender-month return premium is the unique solution to the reduced-form fixed-point condition $\pi_A=\varphi D_A(\pi_A)$ in Proposition~\ref{prop:endog_access}. Within this closed-form specification, the premium is stronger when normalized access costs $\tilde{\psi}_i$ are higher, the surrender-window return impact $\varphi$ is larger, or expected shortfalls $d_i$ are larger. More generally, higher scarcity or a smaller allowance bank can also raise residual terminal demand and hence the premium. This price pressure is an equilibrium object, not an expression of market irrationality: it arises because slow participation leaves compliance demand unresolved until intermediaries must absorb a predictable, calendar-driven wave of demand at the surrender deadline. Empirically, this prediction maps into concentrated April purchases by operators and predictably high April allowance returns.

Third, under the maintained information-incorporation condition, the model predicts informative operator flow. Conditional on receiving market access, a subset $\mathcal{I}$ of operators trades on private signals about short-run innovations to the frictionless benchmark price, and the informational component of operator order flow, $M_t^{\text{info}}$, forecasts future allowance returns. Because access intensity is endogenous, operators with higher chosen access intensity $\lambda_i^*$ contribute more frequently to informative order flow: the expected contribution of operator $i$ to $M_t^{\text{info}}$ scales with $\lambda_i^*$. Higher chosen access intensities make informative order flow appear more quickly and more strongly in the market, so the predictive content of operator flow should be stronger among operators who trade more actively. Empirically, this prediction maps into Operator Flow Imbalance forecasting future allowance returns, especially for frequent traders.

Fourth, the model implies that the two trading frictions interact non-additively in their effect on the equilibrium premium. By Corollary~\ref{cor:phi_elasticity}, the elasticity of $\pi_A$ with respect to $\varphi$ is bounded strictly below unity: endogenous access reabsorbs a fraction of any policy change in $\varphi$ through adjustment of $\lambda_i^*$ and hence of residual terminal demand $D_A$, so a static demand-pressure model overstates the price response to changes in intermediary capacity. By Corollary~\ref{cor:friction_interaction}, this dampening vanishes in the static-access limit. Corollary~\ref{cor:proportional_amplification} establishes that higher access costs monotonically weaken dampening when access costs are proportional to shortfalls. The quantitative counterfactuals of Section~\ref{sec:quant} retain heterogeneity: halving $\varphi$ alone lowers the premium by less than $50\%$ because endogenous access raises $D_A$. Halving both $\varphi$ and $\tilde\psi$ jointly lowers it by exactly $50\%$ by Corollary~\ref{cor:scale_invariance}. Starting from doubled access costs yields a larger finite-change elasticity in the calibration, illustrating the amplification direction established in the proportional-cost benchmark.

\section{Quantification and Counterfactuals\label{sec:quant}}

This section uses the model to benchmark the magnitude of the trading frictions and to evaluate market-design counterfactuals. We use the closed-form access-choice condition \eqref{eq:ct_lambdaStar}, the residual terminal-demand expression \eqref{eq:ct_DA_main}, and the reduced-form surrender-window return-impact relation $\pi_A=\varphi\,D_A(\pi_A)$ to calibrate normalized access frictions and a return-impact sensitivity parameter, and then evaluate counterfactual changes to $\varphi$, $\tilde{\psi}_i$, and the timing of surrender obligations. We treat the recovered objects as model-implied magnitudes rather than as fully structural estimates of EU ETS primitives. Detailed construction choices, banking-adjusted shortfall measures, alternative premium definitions, and leave-one-out diagnostics are reported in Appendix~\ref{subsec:quant_appendix}.

\subsection{Construction of the Quantitative Objects\label{subsec:quant_construction}}

We define each compliance cycle as running from May of year $t-1$ through April of year $t$, with April as the surrender month. The obligation settled in that April is the one for emissions in calendar year $t-1$, so cycle $t$ is paired with the EUTL compliance record for year $t-1$ (its free allocation and the allowances surrendered against its verified emissions). For each operator $i$ and cycle $t$, we use the transparent allocation-based shortfall proxy
\begin{align}
d^{\mathrm{alloc}}_{it} = \max\{\mathrm{surrender}_{it} - \mathrm{allocation}_{it},\, 0\},
\label{eq:quant_dalloc}
\end{align}
and report bank-adjusted, cumulative-surplus, and lagged-surplus alternatives in Appendix~\ref{subsec:quant_appendix}.

Throughout this section, we use the one-shot known-shortfall specification of Proposition~\ref{prop:endog_access}. We back out a firm-specific access intensity $\hat{\lambda}_i$ from observed pre-surrender trading frequency. Under the model, the probability that operator $i$ receives no discretionary access opportunity over a cycle of length $\tau$ is $e^{-\lambda_i \tau}$, so a descriptive shrinkage estimator inverts this relation using each operator's empirical frequency of no-trade years.\footnote{Because observed no-trade can also reflect little motive to rebalance, $\hat{\lambda}_i$ is best read as an index of effective pre-surrender market access among positive-shortfall operators rather than as a structural access parameter for every operator-year.} The model-implied residual terminal demand is then
\begin{align}
D^{\mathrm{model}}_{A,t} = \sum_i d_{it}\, e^{-\hat{\lambda}_i \tau},
\label{eq:quant_DA}
\end{align}
and we calibrate the surrender-window return-impact sensitivity parameter $\varphi$ from the relation between observed April return premia $\pi_{A,t}$ and $D^{\mathrm{model}}_{A,t}$. Because the model imposes a nonnegative compliance-pressure premium, the headline calibration uses $\pi_{A,t}=\max\{r_{A,t},0\}$, the nonnegative part of the signed April log return. We refer to this object as the calibration premium. Over the 2009--2022 calibration cycles, the period with bankable allowances, the signed April log return averages 6.9\% (standard deviation 16.3 percentage points) and is positive in 11 of 14 years, and the calibration premium averages 10.3\%. The signed return and non-April-adjusted and residualized variants are reported in the appendix. Implied normalized access-cost indices $\hat{\tilde{\psi}}_i$ are recovered from the access-choice first-order condition $S_i\tau e^{-\lambda_i\tau}=\tilde{\psi}_i\lambda_i$ only for operator-cycles with $d_{it}>0$, $\hat{\lambda}_i>0$, and $\pi_{A,t}>0$. Zero-shortfall and nonpositive-premium cycles are omitted rather than assigned $\hat{\tilde{\psi}}_{it}=0$. This restriction leaves 37{,}910 operator-cycle observations spanning 7{,}367 operators. We winsorize the recovered cycle-level costs at the pooled 1st and 99th percentiles within each sample, then take each firm's median across valid cycles.

\subsection{Headline Distribution and Model Fit}

The headline sample uses 95{,}526 operator-year observations for 10{,}707 operators over the 2009--2022 compliance cycles (obligations for 2008--2021 emissions), the period with bankable allowances, where a year is defined as the May-to-April compliance cycle. Appendix~\ref{subsec:quant_appendix} reports the sample including the Phase I pilot cycles. Weighting each operator-cycle equally, the descriptive estimator implies a median $\hat{\lambda}_i$ of $0.693$, with a 10th-to-90th percentile range $[0.182,\,1.792]$. With the cycle normalized to unit length, the median estimate corresponds to roughly one pre-surrender trading opportunity every 1.4 compliance years. Equivalently, the Poisson benchmark implies a $50.0\%$ probability of at least one such opportunity in a given compliance year. Table~\ref{tab:quant_core_fit} reports the access-intensity distribution together with the main validation moments. The pre-surrender no-trade share is directly disciplined. Persistence, April volume, and aggregate terminal demand are untargeted. The model matches the pre-surrender no-trade share (data $0.501$, model $0.506$).\footnote{This pre-surrender no-trade share differs from the roughly 37\% annual no-trade fact documented in Figure~\ref{fig:trading-frequency}. The annual measure counts an operator as trading if it makes any market trade during the year, including purchases in April, whereas the quantitative model's Poisson access object is the probability of no discretionary market access before surrender. Operators who wait until April count as traders in the annual Figure~\ref{fig:trading-frequency} measure but as no pre-surrender trade in the model-aligned moment.} The model also matches the one-year persistence of non-trading reasonably well (data $0.667$, model $0.618$).

\begin{table}[!t]
\centering
\caption{Headline access-intensity distribution and core model fit\label{tab:quant_core_fit}}
\begin{tabular}{lcc}
\toprule
Moment & Data/Value & Model \\
\midrule
\multicolumn{3}{l}{\textit{Panel A: Distribution of }$\hat{\lambda}_i$\textit{ and implied no-access probabilities}} \\
\midrule
$\hat{\lambda}_i$ p10 & 0.182 & \\
$\hat{\lambda}_i$ p25 & 0.357 & \\
Median $\hat{\lambda}_i$ & 0.693 & \\
Mean $\hat{\lambda}_i$ & 0.851 & \\
$\hat{\lambda}_i$ p75 & 1.145 & \\
$\hat{\lambda}_i$ p90 & 1.792 & \\
Mean no-access probability $e^{-\hat{\lambda}_i\tau}$ & 0.514 & \\
Median no-access probability & 0.500 & \\
\midrule
\multicolumn{3}{l}{\textit{Panel B: Validation moments}} \\
\midrule
Pre-surrender no-trade share & 0.501 & 0.506 \\
One-year no-trade persistence & 0.667 & 0.618 \\
Median pre-surrender trade count & 0.000 & 0.000 \\
Share of annual volume in April purchases & 0.239 & 0.121 \\
Corr$(D^{\mathrm{model}}_{A,t},\,D^{\mathrm{data}}_{A,t})$ & 0.819 & \\
Slope of $D^{\mathrm{data}}_{A,t}$ on $D^{\mathrm{model}}_{A,t}$ & 1.276 & \\
$R^2$ of terminal-demand validation & 0.671 & \\
\bottomrule
\end{tabular}
\par
\medskip
\begin{minipage}{0.93\textwidth}
\footnotesize
\emph{Notes:} Headline sample: 2009--2022 compliance cycles. Panel A weights each of 95{,}526 operator-cycles equally (10{,}707 operators). Panel B's first four rows use the 90{,}079 cycles with compliance records (10{,}241 operators). Persistence uses consecutive eligible cycles. The last three rows compare yearly aggregates. The pre-surrender no-trade share directly disciplines access intensity. The count row compares medians of realized counts: its model entry is the exact median of an auxiliary Poisson-count mixture with means $\hat{\lambda}_i\tau$, using the same descriptive intensities and equal cycle weights. Its zero median follows from the fitted no-trade probability exceeding one half. The remaining rows are untargeted.
\end{minipage}
\end{table}

The constant-intensity benchmark is calibrated on whether an operator trades before surrender, not on when, and it under-predicts the share of annual volume that occurs in April, which points to access intensity rising around verification. Still, model-implied terminal demand tracks the level of observed April purchases across allocation regimes, which is the variation that anchors the calibration of $\hat{\varphi}$ and the counterfactual changes. The level mismatch on the April-share moment makes the baseline euro measure of the April premium paid by delayed buyers a conservative lower bound, and likely under-states the gains from a market-design intervention that staggers the deadline.

\begin{figure}[tbp]
\centering
\includegraphics[width=0.78\textwidth]{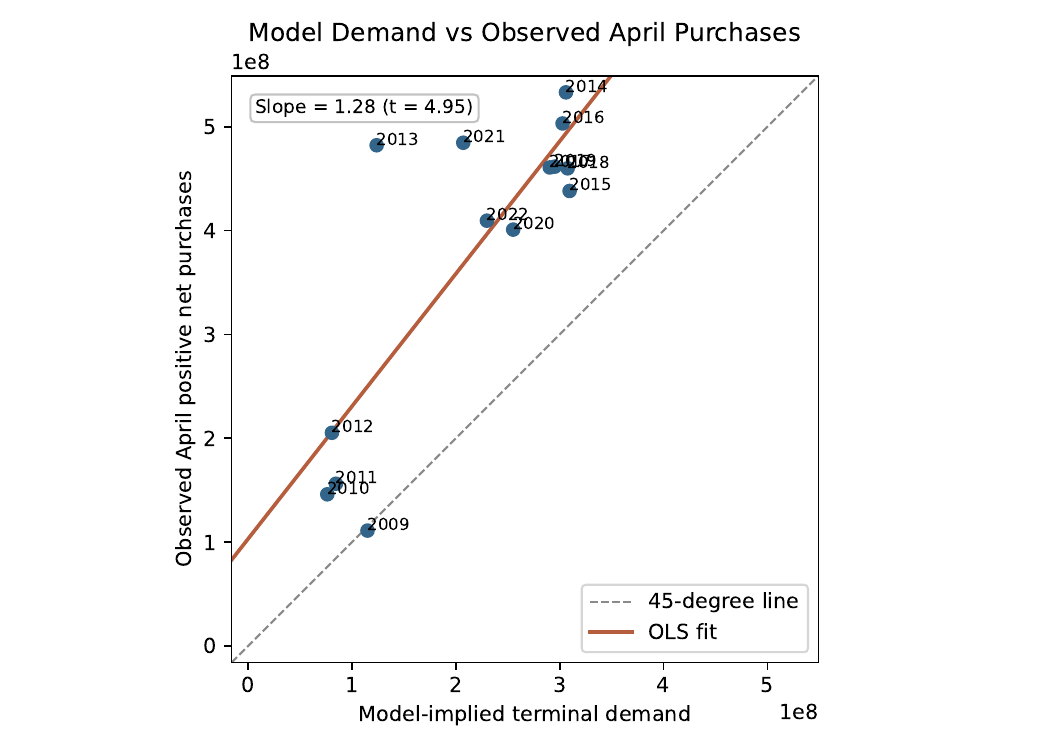}
\caption{Model-implied terminal demand and observed April net operator purchases, by year.\label{fig:quant_terminal_demand}}
\par\medskip
\begin{minipage}{0.95\textwidth}
\footnotesize \emph{Notes:} Headline 2009--2022 sample. The 45-degree line marks one-to-one correspondence. The fitted line is the OLS regression of observed on model-implied terminal demand.
\end{minipage}
\end{figure}

The terminal-demand validation is the central diagnostic for the calibration of $\varphi$, since the return-impact relation is meaningful only if the model-implied demand object tracks observed surrender-month buying pressure. In the headline sample, the correlation between $D^{\mathrm{model}}_{A,t}$ and observed April net operator purchases is $0.819$, with $R^2=0.671$ in a yearly regression of observed on model-implied terminal demand. The fitted slope of $1.28$ indicates that the model understates the magnitude of observed April buying in levels, consistent with the April-share under-prediction noted above. The fit is driven by the level of terminal demand across allocation regimes: the shift toward auctioning at the start of Phase III raises both the aggregate shortfall and April purchases, and the model tracks that shift, whereas within phases model-implied demand varies little from year to year.\footnote{Demeaning both series by trading phase before correlating them gives a within-phase correlation of $-0.07$ over the fourteen cycles and $0.25$ excluding the phase-final Aprils of 2013 and 2021 discussed in Appendix~\ref{subsec:phase_final_borrowing}. Excluding those Aprils, the raw correlation is $0.96$. The incidence regressions of Section~\ref{subsec:emp_pressure} include year fixed effects and therefore do not rely on this year-level variation.} The year-level comparison therefore validates the level of terminal demand. The access mechanism itself is tested in the cross-section by the incidence regressions of Section~\ref{subsec:emp_pressure}, which use about 41{,}000 operator-cycles. The largest deviation is April 2013, the last surrender of Phase II, when observed purchases exceed model-implied demand by the most: allowances of the new phase could not be used for that obligation and the 2013 free allocation had not yet been issued, so operators lost the option, available within a phase, of covering the previous year's obligation with the current year's allocation. Appendix~\ref{subsec:phase_final_borrowing} shows that April purchases exceed model-implied demand at all three phase-final surrenders, that the excess is concentrated among operators whose free allocation covers most of their obligation, and that the incidence estimates of Section~\ref{subsec:emp_pressure} are unaffected. Figure~\ref{fig:quant_terminal_demand} reports the year-by-year scatter.

\subsection{Calibrated Return Impact}

We calibrate $\varphi$ from the relation $\pi_{A,t} = \varphi\,D^{\mathrm{model}}_{A,t} + \varepsilon_t$ using the headline nonnegative April premium $\pi_{A,t}=\max\{r_{A,t},0\}$. The baseline counterfactuals use the all-year nonnegative least-squares (NNLS) estimate $\hat{\varphi}=4.06\times 10^{-10}$, which equals OLS through the origin. This calibration uses all 14 years, with the three nonpositive realizations set to zero by the model-aligned definition. As a sensitivity comparison, the median annual premium-to-demand ratio among the 11 positive-premium years is $5.65\times 10^{-10}$. The order of magnitude is stable across signed, non-April-adjusted, and residualized premium definitions and across the sample-period robustness checks reported in Table~\ref{tab:quant_phi} in the Online Appendix. Leave-one-out diagnostics on the headline nonnegative-April OLS specification produce a tight range of $[3.78\times 10^{-10},\,4.57\times 10^{-10}]$. As an external check, a weekly regression of futures returns on net registry inflows to operators (Appendix~\ref{subsec:daily_april_evidence}) implies a return impact of about $3.4\times 10^{-10}$ per allowance, the same order of magnitude as $\hat{\varphi}$. $\hat{\varphi}$ is pinned down by the level of April premia relative to model-implied terminal demand rather than by their year-to-year co-movement, which is weak (Appendix~\ref{subsec:quant_appendix}). We therefore treat it as a calibrated sensitivity parameter and emphasize relative counterfactual changes over euro levels.

\subsection{Counterfactuals}

We use the calibrated objects to solve the equilibrium fixed point $\pi_A=\varphi\,D_A(\pi_A)$ under alternative scenarios. The counterfactual sample contains 44{,}872 positive-shortfall operator-cycles for the 7{,}367 operators with a recovered firm-level access cost. Unlike cost recovery, it does not require the cycle's observed premium to be positive. Because each operator's chosen $\lambda_i^*$ depends on the equilibrium premium, the counterfactuals incorporate the feedback emphasized in Proposition~\ref{prop:endog_access}: a lower premium reduces the private value of early access and partially reverses the direct effect of any policy change. The baseline fixed-point premium of $0.068$ is lower than the mean nonnegative April calibration premium of $0.103$ because the counterfactual exercise isolates the compliance-demand component captured by $\pi_A=\varphi D_A(\pi_A)$ using a single calibrated $\hat{\varphi}$ and model-implied terminal demand, rather than the full signed April return including residual shocks.

Table~\ref{tab:quant_cf} reports the headline counterfactuals. Halving surrender-window return impact lowers the equilibrium premium by $37.7\%$ rather than the $50\%$ that a static return-impact model would predict, because residual terminal demand $D_A$ rises by $24.6\%$ once operators internalize the lower premium. The realized elasticity is $\log(0.623)/\log(0.5)\approx 0.68$, consistent with Corollary~\ref{cor:phi_elasticity}: the model-implied $|D_A'(\pi_A)|$ at the calibrated equilibrium absorbs roughly $30\%$ of any direct change in $\varphi$. Halving access costs lowers the premium by $20.8\%$ and the April premium paid by delayed buyers by $37.5\%$, working through both lower $D_A$ and a lower equilibrium premium. The largest effects come from staggering the surrender deadline across $K$ size-balanced groups: $K=2$ lowers the premium by $36.0\%$ and the April premium paid by delayed buyers by $22.0\%$, while $K=4$ lowers the premium by $58.4\%$ and the April premium paid by delayed buyers by $40.7\%$. Splitting terminal demand into two size-balanced surrender windows reduces the order-flow pressure faced in each window and therefore resembles a reduction in effective surrender-window return impact. This explains why the $K=2$ staggered-deadline counterfactual is numerically close to the half-$\varphi$ counterfactual. The non-linear gain from $K=2$ to $K=4$ reflects the convex absorption-cost technology combined with the endogenous-access feedback.

The joint half-$\varphi$ and half-$\tilde\psi$ row in Table~\ref{tab:quant_cf} shows that when both frictions are reduced at once, the equilibrium premium falls by exactly $50\%$ and the April premium paid by delayed buyers also falls by $50\%$, as established analytically by the scale invariance of Corollary~\ref{cor:scale_invariance}. Under the joint rescaling, the premium and access costs fall in proportion, leaving the benefit-cost ratio of access, optimal intensities, and residual demand unchanged. A separate sensitivity comparison, not reported in the table, illustrates the amplification channel. Doubling $\tilde\psi$---making market access more costly---raises the equilibrium premium to $0.085$, and from that higher-friction starting point halving $\varphi$ lowers the premium to $0.052$. The implied finite-change elasticity is about $0.70$, slightly above the $0.68$ obtained in the baseline calibration. This numerical amplification result in the heterogeneous economy is in the direction established analytically by Corollary~\ref{cor:proportional_amplification} under proportional access costs.

\begin{table}[tbp]
\centering
\caption{Headline counterfactuals from the quantitative access model\label{tab:quant_cf}}
\small
\resizebox{\textwidth}{!}{%
\begin{tabular}{l r r r r r}
\toprule
Scenario & Premium & $D_A$ change & April premium paid (\euro\ mn) & Premium-paid change & Absorption cost change \\
\midrule
Baseline & 0.068 & -- & 236.6 & -- & -- \\
Half $\varphi$ & 0.042 & $+24.6\%$ & 184.6 & $-22.0\%$ & $-22.0\%$ \\
Half normalized access cost ($\tilde{\psi}$) & 0.054 & $-20.8\%$ & 147.9 & $-37.5\%$ & $-37.5\%$ \\
Half $\varphi$ and half $\tilde{\psi}$ jointly & 0.034 & $0.0\%$ & 118.3 & $-50.0\%$ & $-50.0\%$ \\
Stagger $K=2$ & 0.043 & $+24.6\%$ & 184.7 & $-22.0\%$ & $-22.0\%$ \\
Stagger $K=4$ & 0.028 & $+52.7\%$ & 140.3 & $-40.7\%$ & $-40.7\%$ \\
\bottomrule
\end{tabular}
}
\par
\medskip
\begin{minipage}{0.95\textwidth}
\footnotesize
\emph{Notes:} The baseline row reports the level of the equilibrium premium and the April premium paid by delayed buyers in millions of euros. Counterfactual rows report changes relative to the baseline. Premium is the equilibrium fixed-point solution to $\pi_A=\varphi\,D_A(\pi_A)$. $\varphi$ is the surrender-window return-impact parameter. $D_A$ is aggregate residual net deadline buying pressure under the counterfactual, not gross intermediation volume. The April premium paid by delayed buyers is $P_e\pi_A D_A$ at the within-cycle early price level $P_e$. Absorption cost is $\tfrac{1}{2}P_e\pi_A D_A$. $\tilde{\psi}$ denotes the access-cost shifter used in the quantitative exercise. Under the one-shot access specification, the staggered-deadline counterfactual splits operators into $K$ equally sized surrender groups and solves the fixed point separately for each group.
\end{minipage}
\end{table}

\begin{figure}[tbp]
\centering
\includegraphics[width=0.78\textwidth]{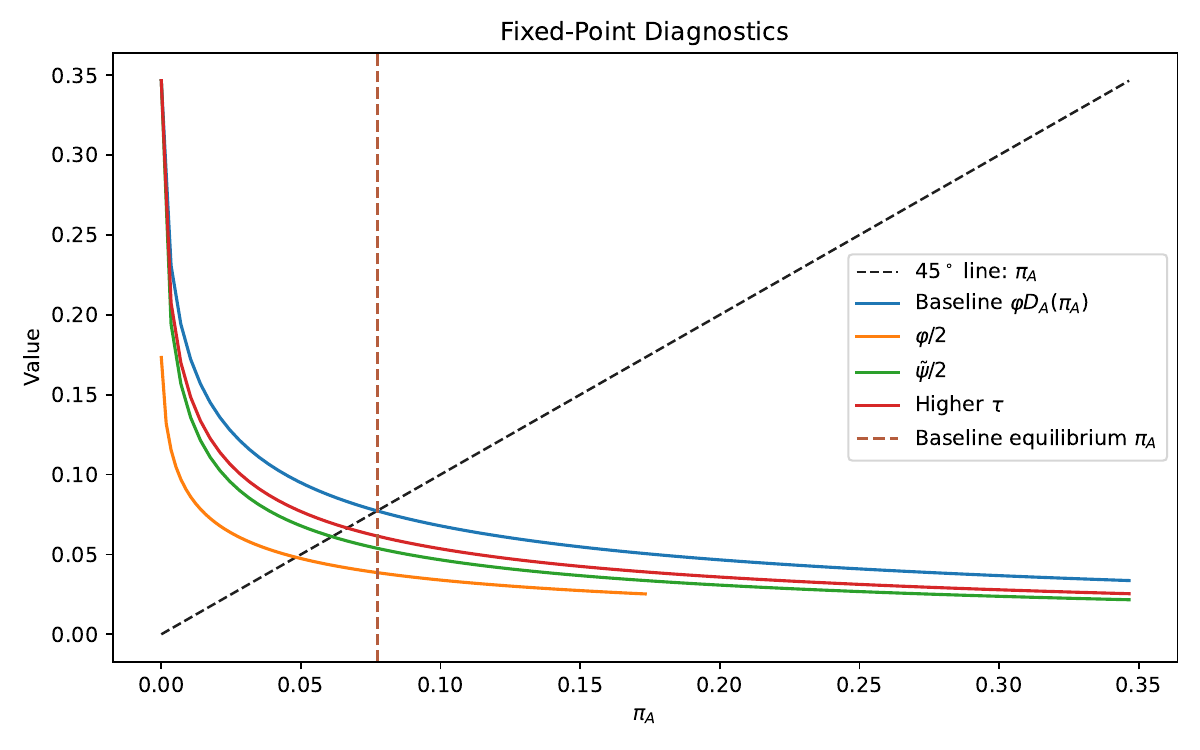}
\caption{Fixed-point logic behind the quantitative counterfactuals.\label{fig:quant_fixed_point}}
\par\medskip
\begin{minipage}{0.95\textwidth}
\footnotesize \emph{Notes:} The 45-degree line plots the equilibrium identity $\pi_A=\pi_A$. The curves plot $\varphi\,D_A(\pi_A)$ under the baseline calibration and under three counterfactuals. Equilibrium premia are the intersections of each curve with the 45-degree line. Halving $\varphi$ shifts the price-pressure curve down but leaves $D_A(\pi_A)$ unchanged at any given premium. At the new, lower equilibrium premium, access declines and residual demand rises, generating less than a one-for-one fall in $\pi_A$.
\end{minipage}
\end{figure}

The counterfactuals are partial-equilibrium: they hold measured shortfalls, recovered access-cost parameters, and early-trading prices fixed while re-solving access intensities and the equilibrium premium. Aggregate scarcity, the allowance bank, and the abatement decision are unchanged. The staggered scenarios assume non-overlapping congestion windows, renewed intermediary capacity in each window, unchanged access horizons, and no cross-window arbitrage or funding feedback. The counterfactuals compare prices and payments across market designs within the model. They are not welfare calculations. Splitting compliance demand across four equal groups reduces the implied April premium paid by delayed buyers by roughly $41\%$, with no change in aggregate emissions or the cap.

\section{Testing Additional Model Predictions\label{sec:empirical}}

Section~\ref{sec:facts} documented the three headline patterns that motivate the model. We now go beyond these patterns to examine the model's mechanisms, additional empirical implications, and main alternative explanations. Subsection~\ref{subsec:emp_slow} studies whether matched free allocations or banking can explain persistent non-trading. Subsection~\ref{subsec:emp_pressure} tests whether prior trading inactivity predicts surrender-month demand conditional on compliance shortfalls and connects surrender-month trading to the delayed-buyer premium. Subsections~\ref{subsec:emp_predictability}--\ref{subsec:emp_frequent} report the full return-predictability specifications, examine alternative explanations, and test whether the predictive content is concentrated in flow from frequent operators.

\subsection{Prediction 1: Slow Participation\label{subsec:emp_slow}}

Section~\ref{subsec:fact_participation} documented the prevalence and persistence of non-trading. The model's participation mechanism generates persistent gaps between desired and actual allowance holdings, implying delayed trading and non-participation for operators with high normalized access costs $\tilde{\psi}_i$ or low value of access $S_i$. To assess this mechanism, we ask whether two alternatives---free allocations that closely match surrender obligations and allowance banking---can account for these patterns.

\begin{figure}[!t]
\caption{Allocated vs Surrendered EUAs\label{fig:allocated_vs_surrendered}}

\medskip{}

{\footnotesize~This figure plots the relationship between total allocated and surrendered allowances for operators with no trade in a given year. Each hexagon represents the number of operator-year observations falling within a bin of allocated and surrendered EUA values, as indicated by the color bar. Both axes are in log scale. The red dashed line denotes the 45-degree line. Data is from February 2005 to September 2022.}

\medskip{}

\centering{}\includegraphics[scale=0.6]{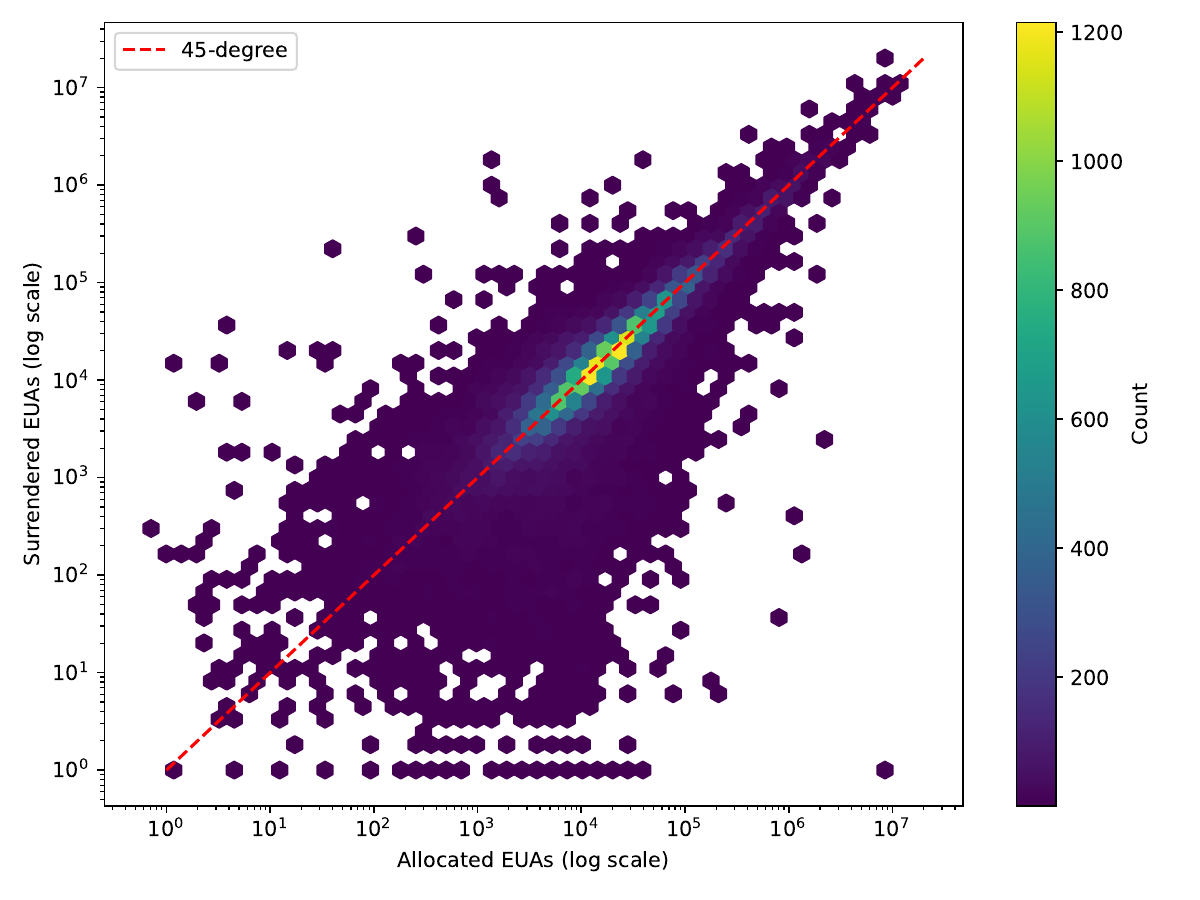}
\end{figure}

One possible justification for the large fraction of non-trading firms illustrated in Figure \ref{fig:trading-frequency} is that they receive a number of free allowances roughly equal to the amount they are required to surrender. However, the data does not support this explanation. Among these firms, the 5th and 25th percentiles of the allocated-to-surrendered ratio are 0 and 0.75, respectively, while the 75th and 95th percentiles are 1.37 and 4.93. This suggests that many non-trading firms face substantial imbalances in the years they do not trade. Figure \ref{fig:allocated_vs_surrendered} plots the relationship between total allocated and surrendered EUAs, for non-trading operators, and confirms this pattern, with a large number of operator-year observations falling above or below the 45-degree line.

A second possible justification for the large fraction of non-trading firms is that they save---or bank---excess allowances \citep{martino2013back, cantillon2023market}, perhaps in anticipation of higher future prices and a reduction in the number of freely allocated allowances. One way to test this hypothesis is to examine whether the absence of trading in a given year predicts continued inactivity in subsequent years. If banking were the main reason for non-trading, we would expect that firms eventually return to the market, and firms that bank more are more likely to engage in trading in the future. Thus, no trade in one year should not systematically and positively predict no trade in future years.

Table~\ref{tab:Test_NoTrade} presents panel regression results where the dependent variable is an operator-level dummy indicating no trade in year $t + h$ (for $h = 1,\dots,5$), and the main explanatory variable is a no-trade dummy in year $t$. Panel A reports estimates for the full sample (2005--2022), while Panel B focuses on the post-2013 period, covering Phase III and the beginning of Phase IV.\footnote{Banking was not allowed between Phase I and Phase II, so unused Phase I allowances expired at the end of 2007. The Phase III/IV transition was different: existing Phase III allowances remained valid into Phase IV, although newly issued Phase IV allowances could not be used for 2020 compliance in April 2021. Yet, in 2007 and 2020---the final years of Phases I and III---approximately 60\% and 25\% of operators, respectively, did not trade, figures that are very similar to adjacent years within those phases (e.g., 53\% in 2006 and 27\% in 2019). By contrast, in 2012---the final year of Phase II, when banking to Phase III was permitted---the non-trading share was only slightly different, at 36\%.}

 Across all horizons from $t+1$ to $t+5$, the estimated coefficients are positive and statistically significant at the 1\% level, in both samples and regardless of specification. These estimates can be interpreted as changes in probability. For instance, in the full sample baseline model, operators that did not trade in year $t$ are, on average, 40 percentage points more likely to also not trade in year $t+1$, compared to those that did trade.

Overall, the high prevalence of non-trading firms, together with its persistence, is consistent with the first model prediction: a non-trivial subset of operators chooses low effective access intensity $\lambda_i^*$ because of high normalized access costs $\tilde{\psi}_i$ or low value of access $S_i$, so that expected compliance gaps remain unresolved and trading is systematically delayed. The gradual decline in non-trading rates over time points to a fall in effective access costs and hence higher chosen $\lambda_i^*$, plausibly driven by market development and policy changes across phases, but a substantial share of firms remains effectively inactive even in the current phase.

{\footnotesize{}}
\begin{table}[tbp]
{\footnotesize{}\caption{No trade predictability\label{tab:Test_NoTrade}}
}{\footnotesize\par}

{\footnotesize{}\medskip{}
}{\footnotesize\par}

{\footnotesize~This table reports estimates from panel regressions testing whether the absence of trading by an operator in year $t$ predicts continued non-trading behavior in subsequent years ($t+1$ to $t+5$). The dependent variable is a no-trade dummy in future years, and the main regressor is a no-trade dummy in year $t$. Panel A presents results for the full sample (2005--2022), while Panel B restricts the sample to Phase III and early Phase IV (2013--2022), covering the mature market and the Phase III/IV transition and excluding the non-bankable Phase I break. Both baseline specifications and models with year fixed effects are reported. Standard errors are two-way clustered by operator and year. $t$-statistics are reported in parentheses. Clustering by operator alone yields $t$-statistics above 30 in every column. {*}, {*}{*}, and {*}{*}{*} denote significance
at the 10\%, 5\%, and 1\% levels, respectively.}

{\footnotesize{}\medskip{}
}{\footnotesize\par}
\centering{}%
\begin{tabular}{lccccc}
\toprule
\multicolumn{6}{c}{Panel A: Full sample}\tabularnewline
 & $t+1$ & $t+2$ & $t+3$ & $t+4$ & $t+5$ \tabularnewline
\midrule
\multicolumn{6}{l}{Baseline}\tabularnewline
no trade & 0.402*** & 0.322*** & 0.257*** & 0.217*** & 0.196*** \tabularnewline
 & (14.437) & (14.118) & (13.801) & (13.780) & (16.395) \tabularnewline
\tabularnewline[-1ex]
$R^2$ & 0.167 & 0.109 & 0.071 & 0.052 & 0.043 \tabularnewline
\midrule
\multicolumn{6}{l}{Year Fixed Effects}\tabularnewline
no trade & 0.394*** & 0.315*** & 0.267*** & 0.228*** & 0.203*** \tabularnewline
 & (13.245) & (13.620) & (14.499) & (14.656) & (17.797) \tabularnewline
\tabularnewline[-1ex]
$R^2$ & 0.188 & 0.129 & 0.095 & 0.077 & 0.068 \tabularnewline
\midrule
\multicolumn{6}{c}{Panel B: Phases III and IV}\tabularnewline 
 & $t+1$ & $t+2$ & $t+3$ & $t+4$ & $t+5$ \tabularnewline
\midrule
\multicolumn{6}{l}{Baseline} \tabularnewline
no trade & 0.435*** & 0.360*** & 0.302*** & 0.252*** & 0.219*** \tabularnewline
 & (11.783) & (13.501) & (15.900) & (10.527) & (12.840) \tabularnewline
\tabularnewline[-1ex]
$R^2$ & 0.189 & 0.133 & 0.096 & 0.070 & 0.055\tabularnewline
\midrule
\multicolumn{6}{l}{Year Fixed Effects} \tabularnewline
no trade & 0.432*** & 0.361*** & 0.316*** & 0.271*** & 0.234*** \tabularnewline
 & (11.686) & (14.123) & (20.487) & (14.618) & (21.632) \tabularnewline
\tabularnewline[-1ex]
$R^2$ & 0.207 & 0.161 & 0.136 & 0.107 & 0.080 \tabularnewline
\bottomrule
\end{tabular}
\end{table}
{\footnotesize\par}

\subsection{Prediction 2: Surrender-Month Demand Pressure\label{subsec:emp_pressure}}

Section~\ref{subsec:fact_pressure} documented that operator purchases and allowance returns both peak in April. We now test the model's implication for the incidence of this demand across operators with different effective market access. In the model, when desired compliance demand rises as surrender approaches and participation is gradual, unresolved aggregate demand $G_t$ translates---through limited intermediary capacity $\phi>0$---into predictable surrender-month price pressure. We note that April returns may also in part reflect news about scarcity and not entirely compliance-related price pressure. The premium does not imply unbounded arbitrage: profitable positions are bounded by finite aggregate residual demand $D_A$, and arbitrageurs face risk from end-of-March verification uncertainty and operators' endogenous banking choices. The daily evidence in Appendix~\ref{subsec:daily_april_evidence} pins down the timing: the return and the registry inflows to operators both arrive through the surrender month, concentrated after verified emissions are published rather than in the last days before the deadline. The pressure builds as compliance purchases reach the market and is worked off gradually: no reversal is visible in the ten trading days after the deadline.

The incidence prediction is that, conditional on the size of an operator's compliance shortfall, April demand should be greater when that operator has a higher probability of reaching the surrender month without earlier market access. We estimate this probability using only trading histories completed before the outcome cycle. Specifically, for operators observed in at least two prior cycles, $\widehat q_{it}=(N^0_{i,<t}+1/2)/(N_{i,<t}+1)$, where $N^0_{i,<t}$ counts prior cycles with no pre-surrender market trade. We then estimate operator-cycle PPML regressions of positive net April purchases on the log shortfall and $\log\widehat q_{it}$, together with operator size, year fixed effects, and country fixed effects. The model predicts a positive coefficient on $\log\widehat q_{it}$: holding the operator's shortfall and common conditions in year $t$ fixed, effective access affects terminal demand.

\begin{table}[!t]
\centering
\caption{Incidence of Shortfalls and Surrender-Month Demand\label{tab:incidence_terminal_demand}}
\small
\begin{tabular}{lcccc}
\toprule
 & \multicolumn{2}{c}{Full sample} & \multicolumn{2}{c}{2014--2022} \\
\cmidrule(lr){2-3}\cmidrule(lr){4-5}
 & (1) & (2) & (3) & (4) \\
\midrule
Log shortfall & 0.587*** & 0.567*** & 0.684*** & 0.652*** \\
 & (6.42) & (7.27) & (5.87) & (6.83) \\
Log predicted no-access probability & 0.259*** & 0.263*** & 0.255*** & 0.262*** \\
 & (3.90) & (3.92) & (3.97) & (4.02) \\
Log surrender obligation & 0.616*** & 0.629*** & 0.526*** & 0.553*** \\
 & (6.31) & (7.62) & (4.11) & (5.27) \\
\midrule
Shortfall measure & Allocation & Bank-adjusted & Allocation & Bank-adjusted \\
Year fixed effects & Yes & Yes & Yes & Yes \\
Country fixed effects & Yes & Yes & Yes & Yes \\
Observations & 41,171 & 27,737 & 34,503 & 22,774 \\
Operators & 7,094 & 6,536 & 6,225 & 5,657 \\
\bottomrule
\end{tabular}
\par
\medskip
\begin{minipage}{0.95\textwidth}
\footnotesize
\emph{Notes:} This table reports operator-cycle PPML regressions. The full sample covers the 2006--2022 compliance cycles. The dependent variable is the positive part of operator $i$'s net April purchases in millions of allowances, including zeros. The allocation shortfall is $\max\{\mathrm{surrender}_{it}-\mathrm{allocation}_{it},0\}$. The bank-adjusted measure additionally subtracts the lagged allowance bank. The predicted no-access probability is estimated only from cycles preceding $t$ as $(N^0_{i,<t}+1/2)/(N_{i,<t}+1)$, where $N^0_{i,<t}$ counts prior cycles with no pre-surrender market trade. At least two prior cycles are required. All specifications include year and country fixed effects. $t$-statistics based on standard errors two-way clustered by operator and year are reported in parentheses. *, **, and *** denote significance at the 10\%, 5\%, and 1\% levels, respectively.
\end{minipage}
\end{table}

Table~\ref{tab:incidence_terminal_demand} supports this incidence prediction. In the full sample, the coefficient on the predicted no-access probability is $0.259$ ($t=3.90$), so a 10\% higher predicted probability of no pre-surrender access is associated with approximately 2.6\% greater April purchases, conditional on the shortfall and operator size. The estimate is $0.263$ with the bank-adjusted shortfall and $0.255$--$0.262$ over the 2014--2022 cycles, whose obligations fall in Phases III and IV. Under the one-shot specification of Section~\ref{sec:quant}, expected April demand is $d_{it}\,e^{-\lambda_i\tau}$, so the benchmark elasticities with respect to the shortfall and to the no-access probability are both one. The estimates have the predicted signs but are below unity, which is consistent with measurement error in the allocation-based shortfall proxy, partial rather than full closure of the gap upon access, and the use of banked allowances at surrender. Because $\widehat q_{it}$ is constructed only from prior-cycle trading and the access estimator is not fitted to April purchase quantities, this provides a cross-sectional validation of the mechanism that the calibration does not target: shortfall size matters, but its incidence across operators with different effective access also predicts surrender-month demand.

To quantify the implied April premium paid by regulated firms that purchase emission allowances when prices are systematically high during the surrender month, we multiply the net amount purchased by regulated firms in each April by the allowance price at the time and by the post-2008 average April log return of 8.6\%, and sum across years. Using the average April return as a benchmark, a back-of-the-envelope calculation puts the implied premium paid by delayed buyers at about \euro5.7 billion, or about 1.8\% of the euro value of their total traded volume.\footnote{The magnitude depends on the premium convention. Using each year's realized April return instead of the post-2008 average gives about \euro7.5 billion. Using the 10.3\% nonnegative calibration premium of Section~\ref{subsec:quant_construction} gives about \euro6.7 billion.} As emphasized in Section~\ref{subsec:private_costs}, this is a premium paid to counterparties, not a measure of aggregate resource costs.

\subsection{Prediction 3: Operator Flow Imbalance and Return Predictability\label{subsec:emp_predictability}}

Beyond the aggregate predictability documented in Section~\ref{subsec:fact_predictability}, the model predicts that informative flow should be concentrated among operators with more frequent market access. We assess this mechanism by examining the return-predictability specifications and alternative explanations below, and by comparing frequent and infrequent operators in Subsection~\ref{subsec:emp_frequent}. A small number of regulated firms account for a disproportionate share of trading: these are precisely the operators whose higher chosen access intensity $\lambda_i^*$---reflecting low normalized access costs $\tilde{\psi}_i$ or large benefits of early trading---allows them, in the model, to act on private signals when they arrive. If so, allowance prices should rise following net purchases by these firms and fall following net sales.

Section~\ref{subsec:fact_predictability} introduced \emph{Operator Flow Imbalance} and summarized its positive relation with future returns. In the spirit of the model, the measure is a monthly proxy for the informative-order-flow component $M_t^{\text{info}}$. \citet{KalaitzoglouIbrahim2013} find that order flow in EUA futures carries information at the intraday level. Our registry-based measure identifies which market participants' flow is informative, at lower frequencies and longer horizons. Table~\ref{tab:Predictability} reports estimates across horizons for the full sample in Panel A and for the sample beginning in January 2008 in Panel B. The latter excludes the pilot Phase I (2005--2007), whose allowances could not be banked into Phase II, while retaining the entire second trading phase. The results are consistent with active operators trading on information not yet impounded in prices. Intermediary inventory rebalancing, the main alternative, is examined next and in Appendix~\ref{subsec:inventory_predictability}.

We find no robust evidence that intermediary inventory reversal accounts for the positive return predictability associated with operator flow.\footnote{Appendix~\ref{subsec:inventory_predictability} constructs a person-account position proxy and adds it to the return specification, which is otherwise left unchanged. The apparent negative predictive coefficient on that proxy appears only in its raw cumulative level, which is statistically indistinguishable from a unit-root process, and it disappears under any stationary transformation of the series. Because the level is unbalanced with the return regression, the coefficients estimated alongside it, including those on \emph{Operator Flow Imbalance}, are not reliably identified.}

Because trading activity clusters around compliance deadlines, we include month fixed effects to control for seasonality. In the full sample, all coefficients remain positive and significant. In economic terms, \emph{Operator Flow Imbalance} has a full-sample mean of 0.03 and a standard deviation of 0.55, so the month-fixed-effect estimates in Panel A imply that a one-standard-deviation increase is associated with cumulative returns higher by about 5.7, 18.8, and 72 percentage points over the following one, three, and twelve months, respectively. For comparison, the standard deviation of monthly log returns is 25 percentage points.

The estimates in Panel B are smaller but remain positive and statistically significant at the 5\% level or better at every horizon when month fixed effects are included. For example, the three-month coefficient is 0.153 ($t=3.21$), compared with 0.342 in the full sample, and the twelve-month coefficient is 0.521 ($t=3.79$), compared with 1.316. The predictive relation is thus not a pilot-phase artifact. Untabulated results for the more recent but shorter sample beginning with Phase III in 2013 show weaker predictability: the point estimates are smaller and estimated imprecisely, consistent with informational frictions becoming less pronounced as the market matured.

Simulation-based long-horizon inference and year-omission diagnostics (Appendix~\ref{subsec:predictability_robustness}) confirm the results: Stambaugh-type bias is at most 3\% of the coefficients, the full-sample estimates remain significant under Valkanov-style critical values at every horizon and the post-Phase-I estimates from two months onward, and every coefficient stays positive when any single year is omitted. Appendix~\ref{subsec:public_pressure_predictability} also separates operator-flow predictability from public compliance pressure. The coefficient on \emph{Operator Flow Imbalance} remains positive and significant through eight months when cycle-specific aggregate shortfalls enter for forecast windows containing April, and through six months at the 5\% level (eight months at the 10\% level) when all April-crossing windows are excluded. At twelve months, every forecast window contains an April surrender date and the coefficient is no longer significant once cycle-specific shortfalls are controlled for, so the informational attribution is less sharp at the longest horizon. Appendix~\ref{subsec:dual_account_ofi} addresses overlap in account roles. Restricting the measure to transactions between entities that hold only operator accounts and entities that hold only person accounts removes 61.3\% of gross bilateral volume, yet Operator Flow Imbalance remains positive and significant at every horizon, and joint regressions retain a significant frequent-trader coefficient.

To gauge the economic magnitude of these patterns, we value each month's net operator purchases from person holding accounts at that month's allowance price and mark them to market three months later at the realized cumulative return.\footnote{That is, $\sum_t N_t P_t\,(e^{r_{t\to t+3}}-1)$, where $N_t$ is net operator purchases from person holding accounts in month $t$, $P_t$ is the allowance price, and $r_{t\to t+3}$ is the cumulative log return over the following three months. The share divides this sum by the euro value of gross bilateral operator--person volume. The figure increases with the holding horizon: about \euro5.5 billion at one month and about \euro25 billion at six months.} The resulting cumulative ex post trading performance is about \euro13 billion over our sample, about 3.8\% of the euro value of gross bilateral operator--person volume. This is an ex post valuation, not an estimate of realized trading profits.

{\footnotesize{}}
\begin{table}[tbp]
{\footnotesize{}\caption{Time-Series Return Predictability\label{tab:Predictability}}
}{\footnotesize\par}

{\footnotesize{}\medskip{}
}{\footnotesize\par}

{\footnotesize~This table reports time-series EUA return predictability tests. Panel A uses the full sample from 2005 through September 2022. Panel B restricts forecast-origin months to January 2008 onward, excluding Phase I (2005--2007) and retaining Phases II--IV. Both panels report baseline regressions and regressions with calendar-month fixed effects. \emph{Operator Flow Imbalance} is the net flow from person holding accounts to operator holding accounts divided by total bilateral volume between the groups in each month. Its full-sample mean and standard deviation are 0.03 and 0.55. The dependent variable is the cumulative log return over the following $h$ months from daily ICE EUA December futures settlement prices. The full-sample one-month standard deviation is 0.25. Only complete forecast windows ending by September 2022 are included, so observations decline with the horizon.
Standard errors are Newey-West~\citep{newey1987simple} adjusted with $h-1$ lags to account for overlapping returns. $t$-statistics are
reported in parentheses. {*}, {*}{*}, and {*}{*}{*} denote significance
at the 10\%, 5\%, and 1\% levels, respectively.}

{\footnotesize{}\medskip{}
}{\footnotesize\par}
\centering{}%
\resizebox{\textwidth}{!}{%
\begin{tabular}{lccccccc}
\toprule
\multicolumn{8}{c}{Panel A: Full sample (2005--September 2022)}\\
 & $r_{t\to t+1}$ & $r_{t\to t+2}$ & $r_{t\to t+3}$ & $r_{t\to t+4}$ & $r_{t\to t+6}$ & $r_{t\to t+8}$ & $r_{t\to t+12}$\\
\midrule
\multicolumn{8}{l}{\textit{Baseline}}\\
\emph{Op. Flow Imb.} & 0.064$^{**}$ & 0.162$^{***}$ & 0.247$^{***}$ & 0.348$^{***}$ & 0.505$^{***}$ & 0.637$^{***}$ & 0.947$^{**}$\\
 & (2.158) & (3.032) & (3.042) & (3.185) & (2.954) & (2.817) & (2.559)\\
$R^2$ & 0.019 & 0.060 & 0.080 & 0.106 & 0.118 & 0.115 & 0.131\\
\addlinespace
\multicolumn{8}{l}{\textit{Month Fixed Effects}}\\
\emph{Op. Flow Imb.} & 0.104$^{***}$ & 0.217$^{***}$ & 0.342$^{***}$ & 0.447$^{***}$ & 0.628$^{***}$ & 0.811$^{***}$ & 1.316$^{***}$\\
 & (2.648) & (3.180) & (3.258) & (3.099) & (2.778) & (2.626) & (2.617)\\
$R^2$ & 0.090 & 0.108 & 0.142 & 0.151 & 0.146 & 0.142 & 0.183\\
Observations & 209 & 208 & 207 & 206 & 204 & 202 & 198\\
\midrule
\multicolumn{8}{c}{Panel B: January 2008--September 2022}\\
 & $r_{t\to t+1}$ & $r_{t\to t+2}$ & $r_{t\to t+3}$ & $r_{t\to t+4}$ & $r_{t\to t+6}$ & $r_{t\to t+8}$ & $r_{t\to t+12}$\\
\midrule
\multicolumn{8}{l}{\textit{Baseline}}\\
\emph{Op. Flow Imb.} & 0.030 & 0.072$^{**}$ & 0.101$^{**}$ & 0.164$^{***}$ & 0.219$^{***}$ & 0.297$^{***}$ & 0.358$^{***}$\\
 & (1.555) & (2.419) & (2.403) & (3.095) & (2.772) & (3.425) & (3.832)\\
$R^2$ & 0.011 & 0.040 & 0.053 & 0.105 & 0.115 & 0.146 & 0.132\\
\addlinespace
\multicolumn{8}{l}{\textit{Month Fixed Effects}}\\
\emph{Op. Flow Imb.} & 0.049$^{**}$ & 0.096$^{***}$ & 0.153$^{***}$ & 0.199$^{***}$ & 0.278$^{***}$ & 0.372$^{***}$ & 0.521$^{***}$\\
 & (1.991) & (2.760) & (3.208) & (3.247) & (2.897) & (3.326) & (3.790)\\
$R^2$ & 0.109 & 0.099 & 0.147 & 0.152 & 0.165 & 0.174 & 0.189\\
Observations & 176 & 175 & 174 & 173 & 171 & 169 & 165\\
\bottomrule
\end{tabular}

}
\end{table}
{\footnotesize\par}

\subsection{Frequent versus Infrequent Traders\label{subsec:emp_frequent}}

We now test the cross-sectional implication of the model's information channel by examining flow from frequent and infrequent operators separately: higher market-access intensity $\lambda_i$ increases opportunities to act on private signals, while lower $\gamma_i$ increases the size of the order submitted upon access. The empirical mapping is reduced form, since frequent trading can reflect lower access frictions, larger compliance needs, or speculative motives, only the first two of which enter the solved access decision. The intuition is direct: an operator who rarely accesses the market has few opportunities to act on private signals, so informational advantages, even when present, generate little realized order flow.

To test this prediction, we calculate the number of trades companies engage in over the sample period. We then create two subsamples based on the median number of trades and compute \emph{Operator Flow Imbalance} separately within each subsample. We denote \emph{Operator Flow Imbalance}$^{\text{high}}$ and \emph{Operator Flow Imbalance}$^{\text{low}}$ as the imbalance for firms above and below the sample median, respectively. The split uses full-sample trading frequency: it is a cross-sectional classification, not a real-time trading rule.

Table~\ref{tab:Predictability_FrequentTraders} presents the results from time-series return predictability regressions using both \emph{Operator Flow Imbalance}$^{\text{high}}$ and \emph{Operator Flow Imbalance}$^{\text{low}}$. The coefficient on \emph{Operator Flow Imbalance}$^{\text{high}}$ is positive at every horizon, significant from the two-month horizon onward without month fixed effects and at the 5\% level or better at every horizon with them. The coefficient on \emph{Operator Flow Imbalance}$^{\text{low}}$ is never significant. With month fixed effects, operator flow from frequent traders positively and significantly predicts future returns even after controlling for flow from infrequent traders, whose coefficient is not statistically significant. This pattern is consistent with the information channel underlying the model's third prediction.

{\footnotesize{}}
\begin{table}[!t]
{\footnotesize{}\caption{Time-Series Return Predictability: Frequent Traders\label{tab:Predictability_FrequentTraders}}
}{\footnotesize\par}

{\footnotesize{}\medskip{}
}{\footnotesize\par}

{\footnotesize~This table reports the results of time-series EUA return predictability tests for frequent and infrequent operators. \emph{Operator Flow Imbalance}$^{\text{low}}$ and \emph{Operator Flow Imbalance}$^{\text{high}}$ are defined as the net flow from person holding accounts to operator holding accounts divided by total bilateral transaction volume between the two groups, computed separately for operators below and above the sample median number of transactions over the full sample. The low and high measures have standard deviations of 0.52 and 0.55, respectively.
Standard errors are Newey-West~\citep{newey1987simple} adjusted with $n$ lags, where $n$ is the number of overlapping periods. $t$-statistics are reported in parentheses. {*}, {*}{*}, and {*}{*}{*} denote significance at the 10\%, 5\%, and 1\% levels, respectively.}

{\footnotesize{}\medskip{}
}{\footnotesize\par}
\centering{}%
\resizebox{\textwidth}{!}{%
\begin{tabular}{lccccccc}
\toprule
& $r_{t\to t+1}$ & $r_{t\to t+2}$ & $r_{t\to t+3}$ & $r_{t\to t+4}$ & $r_{t\to t+6}$ & $r_{t\to t+8}$ & $r_{t\to t+12}$\tabularnewline
\midrule
\multicolumn{8}{l}{Baseline} \tabularnewline
\emph{Op. Flow Imb.}$^{\text{low}}$ & 0.012 & -0.009 & 0.007 & 0.004 & 0.078 & 0.143 & 0.225\tabularnewline
 & (0.332) & (-0.160) & (0.074) & (0.029)& (0.424) & (0.600) & (0.683)\tabularnewline
\emph{Op. Flow Imb.}$^{\text{high}}$ & 0.057 & 0.168{*}{*} & 0.242{*}{*} & 0.344{*}{*} & 0.454{*}{*} & 0.545{*}{*}{*} & 0.806{*}{*}{*}\tabularnewline
 & (1.333) & (2.227) & (2.180) & (2.356) & (2.329) & (2.593) & (2.924)\tabularnewline
$R^2$ & 0.020 & 0.060 & 0.081 & 0.106 & 0.119 & 0.118 & 0.135\tabularnewline
\midrule
\multicolumn{8}{l}{Month Fixed Effects} \tabularnewline
\emph{Op. Flow Imb.}$^{\text{low}}$ & 0.011 & -0.007 & 0.012 & 0.008 & 0.082 & 0.217 & 0.461\tabularnewline
 & (0.248) & (-0.083) & (0.094) & (0.046) & (0.361) & (0.794) & (1.163)\tabularnewline
\emph{Op. Flow Imb.}$^{\text{high}}$ & 0.097{*}{*} & 0.219{*}{*}{*} & 0.331{*}{*}{*} & 0.437{*}{*}{*} & 0.575{*}{*}{*} & 0.684{*}{*}{*} & 1.052{*}{*}{*}\tabularnewline
 & (2.116) & (2.637) & (2.638) & (2.708) & (2.634) & (2.701) & (2.921)\tabularnewline
$R^2$ & 0.090 & 0.108 & 0.142 & 0.151 & 0.147 & 0.147 & 0.194\tabularnewline
\bottomrule
\end{tabular}}
\end{table}
{\footnotesize\par}

\section{Conclusions}

This paper studies how a major regulated asset market---the EU ETS---absorbs predictable compliance demand and information. Using the universe of EU ETS registry transactions from 2005 through September 2022, we document three facts that are difficult to reconcile with a frictionless allowance market. A substantial fraction of regulated firms do not trade in a given year. Many operators concentrate purchases in the April surrender month, when prices are systematically high. Operator flow imbalance positively predicts future allowance returns, and the predictive content comes from frequent traders. We organize these facts around a model of trading frictions in which operators choose how often to access the market at a convex attention cost, intermediaries supply immediacy while bearing inventory-deviation costs, and a subset of operators trade on private information. The solved access-intermediation equilibrium generates delayed trading and surrender-month price pressure, while the reduced-form information component links access to informative flow. When intermediary capacity changes, operators' access choices partially offset the direct effect, so the equilibrium price reaction is smaller than a static demand-pressure model would predict. The two frictions also interact non-additively. In the proportional-cost benchmark, higher access costs weaken the dampening, so the equilibrium response to intermediary return impact is amplified relative to a lower-access-cost economy. The baseline calibration exhibits the same direction. Halving intermediary return impact alone lowers the equilibrium premium by about $38\%$, halving access costs alone lowers it by about $21\%$, but halving both frictions jointly lowers it by exactly $50\%$. Model-implied terminal demand has a correlation of $0.82$ with observed April compliance purchases at the year level. Staggering surrender deadlines across four equally sized groups reduces the surrender-month premium by about $58\%$ and the implied April premium paid by delayed buyers by about $41\%$, with no change in aggregate emissions or the cap. The mechanism is not specific to carbon: it operates wherever a designed deadline concentrates demand that limited intermediation must absorb, and the EU ETS is a laboratory in which the regulator can move the deadline.

\begin{singlespace}
\setlength{\bibsep}{3pt}
\bibliographystyle{aer}
\bibliography{bib_emissions}

@article{LipseyLancaster1956,
  author  = {Lipsey, R. G. and Lancaster, Kelvin},
  title   = {The General Theory of Second Best},
  journal = {Review of Economic Studies},
  year    = {1956},
  volume  = {24},
  number  = {1},
  pages   = {11--32},
  doi     = {10.2307/2296233}
}

@article{GrossmanStiglitz1980,
  author  = {Grossman, Sanford J. and Stiglitz, Joseph E.},
  title   = {On the Impossibility of Informationally Efficient Markets},
  journal = {American Economic Review},
  year    = {1980},
  volume  = {70},
  number  = {3},
  pages   = {393--408}
}

@unpublished{abrell_database,
	author = {Abrell, Jan},
	note = {Available at Euets.info},
	title = {Database for the European Union Transaction Log},
	year = {2023}}

@article{Duffie2010,
  author  = {Duffie, Darrell},
  title   = {Presidential Address: Asset Price Dynamics with Slow-Moving Capital},
  journal = {Journal of Finance},
  year    = {2010},
  volume  = {65},
  number  = {4},
  pages   = {1237--1267},
  doi     = {10.1111/j.1540-6261.2010.01569.x}
}

@article{DuffieGarleanuPedersen2007,
  author  = {Duffie, Darrell and G{\^a}rleanu, Nicolae and Pedersen, Lasse Heje},
  title   = {Valuation in Over-the-Counter Markets},
  journal = {Review of Financial Studies},
  year    = {2007},
  volume  = {20},
  number  = {6},
  pages   = {1865--1900},
  doi     = {10.1093/rfs/hhm037}
}

@article{BrunnermeierPedersen2009,
  author  = {Brunnermeier, Markus K. and Pedersen, Lasse Heje},
  title   = {Market Liquidity and Funding Liquidity},
  journal = {Review of Financial Studies},
  year    = {2009},
  volume  = {22},
  number  = {6},
  pages   = {2201--2238},
  doi     = {10.1093/rfs/hhn098}
}

@article{Rubin1996,
  author  = {Rubin, Jonathan D.},
  title   = {A Model of Intertemporal Emission Trading, Banking, and Borrowing},
  journal = {Journal of Environmental Economics and Management},
  year    = {1996},
  volume  = {31},
  number  = {3},
  pages   = {269--286},
  doi     = {10.1006/jeem.1996.0044}
}

@article{GlostenMilgrom1985,
  author  = {Glosten, Lawrence R. and Milgrom, Paul R.},
  title   = {Bid, Ask and Transaction Prices in a Specialist Market with Heterogeneously Informed Traders},
  journal = {Journal of Financial Economics},
  year    = {1985},
  volume  = {14},
  number  = {1},
  pages   = {71--100},
  doi     = {10.1016/0304-405X(85)90044-3}
}

@article{nordhaus1992optimal,
	author = {Nordhaus, William D},
	journal = {Science},
	number = {5086},
	pages = {1315--1319},
	publisher = {American Association for the Advancement of Science},
	title = {An optimal transition path for controlling greenhouse gases},
	volume = {258},
	year = {1992}}

@article{montgomery1972markets,
	author = {Montgomery, W David},
	journal = {Journal of Economic Theory},
	number = {3},
	pages = {395--418},
	publisher = {Elsevier},
	title = {Markets in licenses and efficient pollution control programs},
	volume = {5},
	year = {1972}}

@article{jaraite2010transaction,
	author = {Jarait{\.e}, J{\=u}rat{\.e} and Convery, Frank and Di Maria, Corrado},
	journal = {Climate Policy},
	number = {2},
	pages = {190--215},
	publisher = {Taylor \& Francis},
	title = {Transaction costs for firms in the EU ETS: lessons from Ireland},
	volume = {10},
	year = {2010}}

@article{schmalensee2013so2,
	author = {Schmalensee, Richard and Stavins, Robert N},
	journal = {Journal of Economic Perspectives},
	number = {1},
	pages = {103--122},
	publisher = {American Economic Association},
	title = {The SO2 allowance trading system: The ironic history of a grand policy experiment},
	volume = {27},
	year = {2013}}

@article{goulder2013markets,
	author = {Goulder, Lawrence H},
	journal = {Journal of Economic Perspectives},
	number = {1},
	pages = {87--102},
	publisher = {American Economic Association},
	title = {Markets for pollution allowances: what are the (new) lessons?},
	volume = {27},
	year = {2013}}

@article{harstad2010trading,
	author = {Harstad, B{\aa}rd and Eskeland, Gunnar S},
	journal = {Journal of Public Economics},
	number = {9-10},
	pages = {749--760},
	publisher = {Elsevier},
	title = {Trading for the future: Signaling in permit markets},
	volume = {94},
	year = {2010}}

@techreport{kanzig2023unequal,
	author = {K{\"a}nzig, Diego R},
	institution = {National Bureau of Economic Research},
	title = {The unequal economic consequences of carbon pricing},
	year = {2023}}

@book{dales1968pollution,
	author = {Dales, John Harkness},
	publisher = {University of Toronto Press},
	title = {Pollution, property, and prices},
	year = {1968}}

@Article{valkanov2003long,
  author    = {Valkanov, Rossen},
  title     = {Long-horizon regressions: theoretical results and applications},
  journal   = {Journal of Financial Economics},
  year      = {2003},
  volume    = {68},
  number    = {2},
  pages     = {201--232},
  publisher = {Elsevier},
}

@Article{stambaugh1999predictive,
  author    = {Stambaugh, Robert F},
  title     = {Predictive regressions},
  journal   = {Journal of Financial Economics},
  year      = {1999},
  volume    = {54},
  number    = {3},
  pages     = {375--421},
  publisher = {Elsevier},
}

@Article{newey1987simple,
  author  = {Newey, Whitney and West, Kenneth},
  title   = {A simple, positive semi-definite, heteroskedasticity and autocorrelation consistent covariance matrix},
  journal = {Econometrica},
  year    = {1987},
  volume  = {55},
  number  = {3},
  pages   = {703--708},
}

@article{acharya2013limits,
  author    = {Acharya, Viral V and Lochstoer, Lars A and Ramadorai, Tarun},
  title     = {Limits to arbitrage and hedging: Evidence from commodity markets},
  journal   = {Journal of Financial Economics},
  year      = {2013},
  volume    = {109},
  number    = {2},
  pages     = {441--465},
  publisher = {Elsevier},
}

@article{garleanu2009demand,
  author    = {G{\^a}rleanu, Nicolae and Pedersen, Lasse Heje and Poteshman, Allen M},
  title     = {Demand-based option pricing},
  journal   = {Review of Financial Studies},
  year      = {2009},
  volume    = {22},
  number    = {10},
  pages     = {4259--4299},
  publisher = {Oxford University Press},
}

@article{hendershott2014price,
  author    = {Hendershott, Terrence and Menkveld, Albert J},
  title     = {Price pressures},
  journal   = {Journal of Financial Economics},
  year      = {2014},
  volume    = {114},
  number    = {3},
  pages     = {405--423},
  publisher = {Elsevier},
}

@article{subrahmanyam1991risk,
  author    = {Subrahmanyam, Avanidhar},
  title     = {Risk aversion, market liquidity, and price efficiency},
  journal   = {Review of Financial Studies},
  year      = {1991},
  volume    = {4},
  number    = {3},
  pages     = {417--441},
  publisher = {Oxford University Press},
}

@article{gabaix2015international,
  author    = {Gabaix, Xavier and Maggiori, Matteo},
  title     = {International liquidity and exchange rate dynamics},
  journal   = {The Quarterly Journal of Economics},
  year      = {2015},
  volume    = {130},
  number    = {3},
  pages     = {1369--1420},
  publisher = {MIT Press},
}

@incollection{cantillon2023market,
  author    = {Cantillon, Estelle and Slechten, Aur\'elie},
  title     = {Market design for the environment},
  booktitle = {Handbook of the Economics of Corporate Finance},
  year      = {2023},
}

@article{colmer2025does,
  author    = {Colmer, Jonathan and Martin, Ralf and Mu{\^u}ls, Mirabelle and Wagner, Ulrich J.},
  title     = {Does Pricing Carbon Mitigate Climate Change? Firm-Level Evidence from the European Union Emissions Trading System},
  journal   = {Review of Economic Studies},
  year      = {2025},
  volume    = {92},
  number    = {3},
  pages     = {1625--1660}
}

@article{martino2013back,
  author    = {Betz, Regina and Schmidt, Tobias S},
  title     = {Transfer patterns in Phase I of the European Union Emissions Trading System: a first reality check based on cluster analysis},
  journal   = {Climate Policy},
  year      = {2016},
  volume    = {16},
  number    = {4},
  pages     = {474--495},
  publisher = {Taylor \& Francis},
}

@unpublished{BiaisHombertSchmidtWeill2026,
  author = {Biais, Bruno and Hombert, Johan and Schmidt, Daniel and Weill, Pierre-Olivier},
  title  = {Cap and Trade with Imperfect Hedging},
  year   = {2026},
  note   = {Working paper, February 26, 2026. HEC Research Papers Series No. 1574. Available at SSRN},
  doi    = {10.2139/ssrn.5308210},
  url    = {https://ssrn.com/abstract=5308210}
}

@article{BiaisLandier2026,
  author  = {Biais, Bruno and Landier, Augustin},
  title   = {Emission Caps and Investment in Green Technologies},
  journal = {Review of Financial Studies},
  year    = {2026},
  note    = {Forthcoming},
  doi     = {10.2139/ssrn.4100087},
  url     = {https://ssrn.com/abstract=4100087}
}

@article{Pedersen2026CarbonPricing,
  author  = {Pedersen, Lasse Heje},
  title   = {Carbon Pricing versus Green Finance},
  journal = {Journal of Finance},
  year    = {2026},
  volume  = {81},
  number  = {2},
  pages   = {561--602},
  doi     = {10.1111/jofi.70022}
}

@article{BarrageNordhaus2024,
  author  = {Barrage, Lint and Nordhaus, William D.},
  title   = {Policies, Projections, and the Social Cost of Carbon: Results from the {DICE}-2023 Model},
  journal = {Proceedings of the National Academy of Sciences},
  year    = {2024},
  volume  = {121},
  number  = {13},
  pages   = {e2312030121},
  doi     = {10.1073/pnas.2312030121}
}

@article{Barrage2020,
  author  = {Barrage, Lint},
  title   = {Optimal Dynamic Carbon Taxes in a Climate--Economy Model with Distortionary Fiscal Policy},
  journal = {The Review of Economic Studies},
  year    = {2020},
  volume  = {87},
  number  = {1},
  pages   = {1--39},
  doi     = {10.1093/restud/rdz055}
}

@article{Kubler2025,
  author  = {K{\"u}bler, Felix},
  title   = {Incomplete Financial Markets, the Social Cost of Carbon and Constrained Efficient Carbon Pricing},
  journal = {Journal of Economic Theory},
  year    = {2025},
  volume  = {230},
  pages   = {106105},
  doi     = {10.1016/j.jet.2025.106105}
}

@article{FoliniFriedlKueblerScheidegger2025,
  author  = {Folini, Doris and Friedl, Aleksandra and K{\"u}bler, Felix and Scheidegger, Simon},
  title   = {The Climate in Climate Economics},
  journal = {The Review of Economic Studies},
  year    = {2025},
  volume  = {92},
  number  = {1},
  pages   = {299--338},
  month   = {01},
  doi     = {10.1093/restud/rdae011}
}

@article{PedersenFitzgibbonsPomorski2021,
  author  = {Pedersen, Lasse Heje and Fitzgibbons, Shaun and Pomorski, Lukasz},
  title   = {Responsible Investing: The {ESG}-Efficient Frontier},
  journal = {Journal of Financial Economics},
  year    = {2021},
  volume  = {142},
  number  = {2},
  pages   = {572--597},
  doi     = {10.1016/j.jfineco.2020.11.001}
}

@article{SautnerVanLentVilkovZhang2023,
  author  = {Sautner, Zacharias and Van Lent, Laurence and Vilkov, Grigory and Zhang, Ruishen},
  title   = {Firm-Level Climate Change Exposure},
  journal = {Journal of Finance},
  year    = {2023},
  volume  = {78},
  number  = {3},
  pages   = {1449--1498},
  doi     = {10.1111/jofi.13219}
}

@article{BoltonKacperczyk2023,
  author  = {Bolton, Patrick and Kacperczyk, Marcin},
  title   = {Global Pricing of Carbon-Transition Risk},
  journal = {Journal of Finance},
  year    = {2023},
  volume  = {78},
  number  = {6},
  pages   = {3677--3754},
  doi     = {10.1111/jofi.13272}
}

@article{Weitzman1974,
  author  = {Weitzman, Martin L.},
  title   = {Prices vs. Quantities},
  journal = {Review of Economic Studies},
  year    = {1974},
  volume  = {41},
  number  = {4},
  pages   = {477--491},
  doi     = {10.2307/2296698}
}

@article{GolosovHasslerKrusellTsyvinski2014,
  author  = {Golosov, Mikhail and Hassler, John and Krusell, Per and Tsyvinski, Aleh},
  title   = {Optimal Taxes on Fossil Fuel in General Equilibrium},
  journal = {Econometrica},
  year    = {2014},
  volume  = {82},
  number  = {1},
  pages   = {41--88},
  doi     = {10.3982/ECTA10217}
}

@article{PastorStambaughTaylor2021,
  author  = {P{\'a}stor, {\v L}ubo{\v s} and Stambaugh, Robert F. and Taylor, Lucian A.},
  title   = {Sustainable Investing in Equilibrium},
  journal = {Journal of Financial Economics},
  year    = {2021},
  volume  = {142},
  number  = {2},
  pages   = {550--571},
  doi     = {10.1016/j.jfineco.2020.12.011}
}

@unpublished{AcemogluAghionBarrageHemous2024,
  author = {Acemoglu, Daron and Aghion, Philippe and Barrage, Lint and H{\'e}mous, David},
  title  = {Climate Change, Directed Innovation, and Energy Transition: The Long-run Consequences of the Shale Gas Revolution},
  year   = {2024},
  note   = {Working paper}
}

@article{Kyle1985,
  author  = {Kyle, Albert S.},
  title   = {Continuous Auctions and Insider Trading},
  journal = {Econometrica},
  year    = {1985},
  volume  = {53},
  number  = {6},
  pages   = {1315--1335},
  doi     = {10.2307/1913210}
}

@article{GiglioKellyStroebel2021,
  author  = {Giglio, Stefano and Kelly, Bryan and Stroebel, Johannes},
  title   = {Climate Finance},
  journal = {Annual Review of Financial Economics},
  year    = {2021},
  volume  = {13},
  pages   = {15--36},
  doi     = {10.1146/annurev-financial-102620-103311}
}

@article{EngleGiglioKellyLeeStroebel2020,
  author  = {Engle, Robert F. and Giglio, Stefano and Kelly, Bryan and Lee, Heebum and Stroebel, Johannes},
  title   = {Hedging Climate Change News},
  journal = {Review of Financial Studies},
  year    = {2020},
  volume  = {33},
  number  = {3},
  pages   = {1184--1216},
  doi     = {10.1093/rfs/hhz072}
}

@article{FernandezVillaverdeGillinghamScheidegger2025,
  author  = {Fern{\'a}ndez-Villaverde, Jes{\'u}s and Gillingham, Kenneth T. and Scheidegger, Simon},
  title   = {Climate Change Through the Lens of Macroeconomic Modeling},
  journal = {Annual Review of Economics},
  year    = {2025},
  volume  = {17},
  pages   = {125--150},
  doi     = {10.1146/annurev-economics-091124-045357}
}

@article{croce2025green,
  title   = {Green Coins},
  author  = {Croce, Mariano Max Massimiliano and Guinez, Nicolas and Inzunza M{\'e}ndez, Alejandra and Nguyen, Thien T. and Tebaldi, Claudio},
  journal = {Available at SSRN 5139356},
  year    = {2025}
}

@article{HongWangYang2023,
  author  = {Hong, Harrison and Wang, Neng and Yang, Jinqiang},
  title   = {Mitigating Disaster Risks in the Age of Climate Change},
  journal = {Econometrica},
  year    = {2023},
  volume  = {91},
  number  = {5},
  pages   = {1763--1802},
  doi     = {10.3982/ECTA20442}
}

@article{BierbrauerPolbornRitterrathWeizsacker2026,
  author = {Bierbrauer, Felix J. and Polborn, Mattias and Ritterrath, Marten and Weizs{\"a}cker, Georg},
  title  = {Morals and the Political Economy of Corrective Taxes},
  year   = {2026},
  note   = {Working Paper}
}

@article{GrossmanMiller1988,
  author  = {Grossman, Sanford J. and Miller, Merton H.},
  title   = {Liquidity and Market Structure},
  journal = {Journal of Finance},
  year    = {1988},
  volume  = {43},
  number  = {3},
  pages   = {617--633}
}

@article{LouYanZhang2013,
  author  = {Lou, Dong and Yan, Hongjun and Zhang, Jinfan},
  title   = {Anticipated and Repeated Shocks in Liquid Markets},
  journal = {Review of Financial Studies},
  year    = {2013},
  volume  = {26},
  number  = {8},
  pages   = {1891--1912}
}

@article{DuTepperVerdelhan2018,
  author  = {Du, Wenxin and Tepper, Alexander and Verdelhan, Adrien},
  title   = {Deviations from Covered Interest Rate Parity},
  journal = {Journal of Finance},
  year    = {2018},
  volume  = {73},
  number  = {3},
  pages   = {915--957}
}

@article{HartzmarkSolomon2025,
  author  = {Hartzmark, Samuel M. and Solomon, David H.},
  title   = {Market-Wide Predictable Price Pressure},
  journal = {American Economic Review},
  year    = {2025},
  volume  = {115},
  number  = {9},
  pages   = {3171--3213}
}

@article{AbelEberlyPanageas2007,
  author  = {Abel, Andrew B. and Eberly, Janice C. and Panageas, Stavros},
  title   = {Optimal Inattention to the Stock Market},
  journal = {American Economic Review},
  year    = {2007},
  volume  = {97},
  number  = {2},
  pages   = {244--249}
}

@article{AbelEberlyPanageas2013,
  author  = {Abel, Andrew B. and Eberly, Janice C. and Panageas, Stavros},
  title   = {Optimal Inattention to the Stock Market with Information Costs and Transactions Costs},
  journal = {Econometrica},
  year    = {2013},
  volume  = {81},
  number  = {4},
  pages   = {1455--1481}
}

@article{AlvarezGuisoLippi2012,
  author  = {Alvarez, Fernando and Guiso, Luigi and Lippi, Francesco},
  title   = {Durable Consumption and Asset Management with Transaction and Observation Costs},
  journal = {American Economic Review},
  year    = {2012},
  volume  = {102},
  number  = {5},
  pages   = {2272--2300}
}

@article{DuffieSun1990,
  author  = {Duffie, Darrell and Sun, {Tong-sheng}},
  title   = {Transactions Costs and Portfolio Choice in a Discrete-Continuous-Time Setting},
  journal = {Journal of Economic Dynamics and Control},
  year    = {1990},
  volume  = {14},
  number  = {1},
  pages   = {35--51}
}

@article{BertaGautheratGun2017,
  author  = {Berta, Nathalie and Gautherat, Emmanuelle and Gun, Ozgur},
  title   = {Transactions in the {European} Carbon Market: A Bubble of Compliance in a Whirlpool of Speculation},
  journal = {Cambridge Journal of Economics},
  year    = {2017},
  volume  = {41},
  number  = {2},
  pages   = {575--593}
}

@article{CludiusBetz2020,
  author  = {Cludius, Johanna and Betz, Regina},
  title   = {The Role of Banks in {EU} Emissions Trading},
  journal = {The Energy Journal},
  year    = {2020},
  volume  = {41},
  number  = {2},
  pages   = {275--300}
}

@article{QueminPahle2023,
  author  = {Quemin, Simon and Pahle, Michael},
  title   = {Financials Threaten to Undermine the Functioning of Emissions Markets},
  journal = {Nature Climate Change},
  year    = {2023},
  volume  = {13},
  number  = {1},
  pages   = {22--31}
}

@article{AbrellCludiusLehmannSchleichBetz2022,
  author  = {Abrell, Jan and Cludius, Johanna and Lehmann, Sascha and Schleich, Joachim and Betz, Regina},
  title   = {Corporate Emissions-Trading Behaviour During the First Decade of the {EU ETS}},
  journal = {Environmental and Resource Economics},
  year    = {2022},
  volume  = {83},
  number  = {1},
  pages   = {47--83}
}

@article{EllermanTrotignon2009,
  author  = {Ellerman, A. Denny and Trotignon, Rapha{\"e}l},
  title   = {Cross Border Trading and Borrowing in the {EU ETS}},
  journal = {The Energy Journal},
  year    = {2009},
  volume  = {30},
  number  = {Special Issue 2},
  pages   = {53--78}
}

@article{HitzemannUhrigHomburg2018,
  author  = {Hitzemann, Steffen and Uhrig-Homburg, Marliese},
  title   = {Equilibrium Price Dynamics of Emission Permits},
  journal = {Journal of Financial and Quantitative Analysis},
  year    = {2018},
  volume  = {53},
  number  = {4},
  pages   = {1653--1678}
}

@article{HitzemannUhrigHomburgEhrhart2015,
  author  = {Hitzemann, Steffen and Uhrig-Homburg, Marliese and Ehrhart, Karl-Martin},
  title   = {Emission Permits and the Announcement of Realized Emissions: Price Impact, Trading Volume, and Volatilities},
  journal = {Energy Economics},
  year    = {2015},
  volume  = {51},
  pages   = {560--569}
}

@article{KalaitzoglouIbrahim2013,
  author  = {Kalaitzoglou, Iordanis and Ibrahim, Boulis Maher},
  title   = {Does Order Flow in the {European} Carbon Futures Market Reveal Information?},
  journal = {Journal of Financial Markets},
  year    = {2013},
  volume  = {16},
  number  = {3},
  pages   = {604--635}
}

@article{Stoll1978,
  author  = {Stoll, Hans R.},
  title   = {The Supply of Dealer Services in Securities Markets},
  journal = {Journal of Finance},
  year    = {1978},
  volume  = {33},
  number  = {4},
  pages   = {1133--1151}
}

@article{HoStoll1981,
  author  = {Ho, Thomas and Stoll, Hans R.},
  title   = {Optimal Dealer Pricing under Transactions and Return Uncertainty},
  journal = {Journal of Financial Economics},
  year    = {1981},
  volume  = {9},
  number  = {1},
  pages   = {47--73}
}

@article{RamadoraiZeni2024,
  author  = {Ramadorai, Tarun and Zeni, Federica},
  title   = {Climate Regulation and Emissions Abatement: Theory and Evidence from Firms' Disclosures},
  journal = {Management Science},
  year    = {2024},
  volume  = {70},
  number  = {12},
  pages   = {8366--8385},
  doi     = {10.1287/mnsc.2021.00482}
}
\end{singlespace}

\clearpage

\appendix

\setcounter{section}{0} \renewcommand{\thesection}{\Alph{section}} \setcounter{table}{0} \renewcommand{\thetable}{A.\arabic{table}} \setcounter{figure}{0} \renewcommand{\thefigure}{A.\arabic{figure}} \setcounter{equation}{0} \renewcommand{\theequation}{A.\arabic{equation}}
\setcounter{page}{1}

\section{Online Appendix}\label{sec:online_appendix}

\subsection{Market Overview}

The EU ETS is the first large carbon market in the world. The system
covers more than 13,000 facilities spanning diverse industrial sectors.
Functioning as a cap-and-trade system, the EU ETS allocates emission
permits, commonly known as European Union Allowances (EUAs), to regulated
companies. Regulated companies are required to report their annual
carbon emission amount to the regulatory authorities and also must
surrender a corresponding amount of permits. 

The EU ETS was launched in 2005 as a cornerstone of the EU policy
to reduce greenhouse gas emissions. The evolution of the EU ETS is
organized in various ``phases''. The first trading phase (Phase
I) goes from 2005 to 2007. This is a pilot phase
that lays the groundwork for subsequent phases by establishing trading
rules and procedures. The second trading phase (Phase II) goes from 2008 to 2012. Building upon Phase I, the EU ETS expanded
its scope and refined its mechanisms, including efforts to standardize
the allocation rules and improve the pricing and functionality of
the trading system. The third phase (Phase III) goes
from 2013 to 2020, while the current phase (Phase IV) started in 2021
and will end in 2030.

During our sample period, the EU ETS has a fixed verifying
and surrendering schedule each year. In particular, each year operators
must submit an emission report. The emissions data for a given year
must be verified by an accredited verifier by the end of March of
each year. Then, operators must surrender the equivalent number of
allowances to match the installation's emissions by the end of April.
In other words, April is the surrender month for all regulated firms under the EU ETS. For each ton of emissions where
no allowance is surrendered on time, there is a penalty of \euro100,
in addition to the cost of surrendering allowances. The names of penalized
operators are also made public.

\subsection{Data}

The EU ETS individual trading and compliance data are publicly available
and are from the European Union Transaction Log (EUTL)~\citep{abrell_database}.\footnote{The data used in the paper is publicly available through the EUTL website (\href{https://ec.europa.eu/clima/ets/}{https://ec.europa.eu/clima/ets/}). We downloaded the data using the tools available at \href{https://www.euets.info}{https://www.euets.info}.} The EUTL serves
as the primary platform for reporting and monitoring within the EU
ETS. This tool enables the European Commission to transparently disclose
data regarding compliance by regulated entities, the participation
of stakeholders within the system, and the transactions conducted
among these participants. Our sample contains the universe of transactions
and associated compliance information from the EUTL from February 2005 to September 2022,
covering the
first three phases of the EU ETS and part of the current fourth phase. The EUTL data that
we use in this paper contain a number of components, including compliance,
account, and transaction information, which we describe separately
below.

\paragraph*{Compliance Information}

Compliance within the EU ETS occurs at the level of installations. The installations have the obligation to surrender emission allowances
for the verified amounts. The EUTL data contain detailed information
about the identities of the installations as well as the compliance
information. The compliance information contains the allocated amount
and the surrendered amount for each installation in each year. In
the sample, the compliance rate of installations is close to 100\%,
and therefore, non-compliance is unlikely to be an issue in our analysis. 

\paragraph*{Account Information}

There are three basic account types in the EUTL
~\citep{abrell_database}: operator
holding accounts, person holding accounts, and administrative accounts.
An operator holding account is the account through which an installation---the regulated
participant---receives, transfers, and surrenders emission allowances. Non-regulated
participants (or financial intermediaries) can participate in the EU ETS by using person holding
accounts to trade emission allowances. Regulators must use administrative
accounts to allocate and receive surrendered emission allowances. Accounts are held by account holders or companies. Sometimes, a company may have multiple accounts, including operator holding accounts and/or person holding accounts.

\paragraph*{Transaction Information}

The EUTL records all emission allowance transactions between accounts.
Transactions can happen between any two accounts, including operator,
person, and administrative accounts. Reallocations of emission allowances
within the same firm are recorded as transactions in the EUTL. Receiving
and surrendering emission allowances from and to administrative accounts
are also recorded as transactions. In this paper, we exclude all EUTL recorded transactions involving administrative accounts, as our focus is on actual market activity---particularly transactions involving regulated entities---rather than allowance allocation or surrender. However, we retain information on surrender activity and the identities of firms that only interact with administrative accounts. Furthermore, since we aggregate transaction information at the company level, transactions recorded
within the same company in the EUTL always net out to zero. 

The composition and behavior of different account types in the EU ETS highlight important market dynamics. While person holding accounts are fewer in number than operator holding accounts, they account for the majority of trading value in the market. Their activity has increased over time, particularly from Phase I to Phase III. In contrast, operators typically engage in relatively few transactions, most often with intermediaries rather than other operators. The next subsection reports account-type evolution and transaction volumes, followed by operator-level trading statistics.

\subsection{EU ETS Transaction Information}\label{sec:transactions_detail}

\begin{figure}[H]
\caption{EUA Buy and Sell Transactions by Holding Type\label{fig:EUA-Buy-and-Sell}}

\medskip{}

{\footnotesize~This figure presents the evolution in the total euro amount (top panels)
and transaction counts (bottom panel) of EUA sold and purchased by
holding type. Data is from February 2005 to September 2022.}

\medskip{}

\centering{}\includegraphics[scale=0.5]{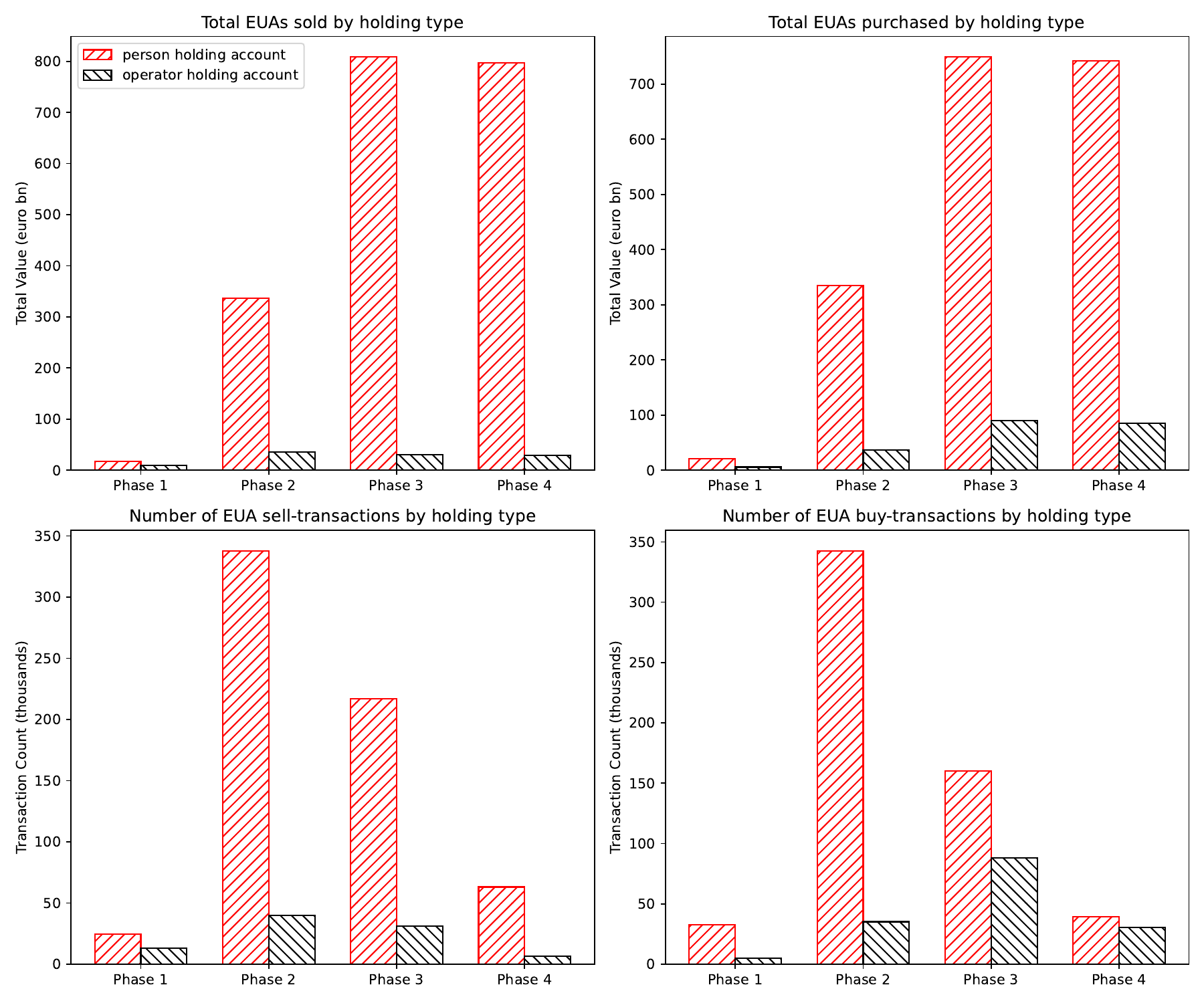}
\end{figure}

The number of active person account holders was
270 at the onset of the EU ETS. This number increased,
peaking at 1,453 in 2009, and then generally declined,
reaching 635 in the first nine months of 2022. In contrast, the
number of active operator account holders started at 3,770,
reached a peak of 7,806 in 2013, and was 5,455 in the first nine months
of 2022. While the number of person account holders is a fraction of the number of operator account holders,
person holding accounts represent the largest share of the total value
of EUAs sold and purchased, as illustrated in Figure \ref{fig:EUA-Buy-and-Sell}.
Furthermore, the figure shows that the total value of EUAs sold and
purchased by person holding accounts has been increasing from Phase
I to Phase III of the EU ETS, reaching approximately \euro808 billion (65 billion EUAs) sold and \euro749 billion (60 billion EUAs) purchased in Phase III. Operator
holding accounts, instead, account for approximately \euro30 billion
in EUAs sold (2.7 billion EUAs) and \euro90 billion in EUAs purchased (7.5 billion EUAs) in Phase III.\footnote{Throughout the paper, EUA prices and returns are based on the daily settlement price of the ICE EUA December futures contract (front December, continuous series), which is available with two-decimal precision from April 2005; the euro value of transfers uses the most recent settlement price on or before the transfer date. Values for Phase IV correspond to 2021 through September 2022.}

Table \ref{tab:Descriptive-Statistics-of-Operators-Trading} provides descriptive statistics on operator trading throughout the entire sample period, as well as across the three distinct phases of the EU ETS market and the first months of the current Phase IV. The statistics reflect the number of transactions---both sales and purchases---rather than the volume of EUAs traded. Individual transactions may correspond to very different quantities of EUAs, with some involving small transfers and others representing large blocks of allowances.

Panel A shows that, on average, an operator engaged in approximately 8 sell transactions and 14 buy transactions over the entire sample. The average share of transactions involving sales to other operators was 20\%, while purchases from other operators accounted for 16\%. Overall, the average operator conducted around 22 transactions in total, with 19\% involving another operator as the counterparty. These proportions are relatively stable across the first three market phases. Panel B highlights that the distribution of transaction counts is right-skewed, indicating that median operators trade significantly less than the average. Specifically, the median operator engaged in just 3 sell and 5 buy transactions, typically interacting with intermediaries rather than with other operators---a pattern that holds across all phases of the market.

\begin{table}[tbp]
\caption{Operator Trading in the EU ETS Market\label{tab:Descriptive-Statistics-of-Operators-Trading}}
{\footnotesize{}\medskip{}
}{\footnotesize\par}

{\footnotesize~This table shows the fraction of trades operators engage with other
operators. Panel A reports the average numbers of sales, buys, and
total trades as well as the fractions of sales, buys, and total trades
with other operators. Panel B reports the median numbers of sales,
buys, and total trades as well as the fractions of sales, buys, and
total trades with other operators.}

{\footnotesize{}\medskip{}
}{\footnotesize\par}
\centering{}%
\begin{tabular}{ccccccc}
\toprule
 & Sales & To Operators (\%) & Buys & From Operators (\%) & Total Trades & With Operators (\%)\tabularnewline
\midrule
& \multicolumn{6}{c}{Panel A: Averages}\tabularnewline
Overall & 8.11 & 0.20 & 14.30 & 0.16 & 22.41 & 0.19\\
Phase 1 & 3.51 & 0.13 & 1.41 & 0.16 & 4.91 & 0.22\\
Phase 2 & 6.05 & 0.17 & 5.27 & 0.19 & 11.32 & 0.22\\
Phase 3 & 3.46 & 0.20 & 9.82 & 0.14 & 13.27 & 0.20\\
Phase 4 & 1.15 & 0.16 & 5.43 & 0.15 & 6.58 & 0.21\\
\midrule
 & Sales & To Operators (\%) & Buys & From Operators (\%) & Total Trades & With Operators (\%)\tabularnewline
\midrule
& \multicolumn{6}{c}{Panel B: Medians}\tabularnewline
Overall & 3 & 0 & 5 & 0 & 10 & 0\\
Phase 1 & 1 & 0 & 0 & 0 & 2 & 0\\
Phase 2 & 3 & 0 & 2 & 0 & 6 & 0\\
Phase 3 & 1 & 0 & 4 & 0 & 6 & 0\\
Phase 4 & 0 & 0 & 2 & 0 & 3 & 0\\
\bottomrule
\end{tabular}
\end{table}

\subsection{Further results in terms of emission allowance return predictability\label{subsec:predictability_robustness}}

We refresh long-horizon inference for both samples in Table~\ref{tab:Predictability}, using the same monthly futures returns, forecast windows, and baseline and calendar-month specifications. Overlapping cumulative returns and persistent predictors can distort conventional inference~\citep{valkanov2003long}, while correlated return and predictor innovations can bias estimated predictive slopes~\citep{stambaugh1999predictive}. We use a calibrated AR(1) model for Operator Flow Imbalance, allowing its innovation to be contemporaneously correlated with the one-month return disturbance and retaining calendar-month terms when appropriate. Under the no-predictability null, the lagged-flow return coefficient is set to zero. For each sample and specification, we run 49{,}999 replications with Gaussian innovation pairs and separately with resampled empirical residual pairs. Each replication reconstructs the overlapping future-return windows and computes both $t_{\mathrm{OLS}}/\sqrt{T_h}$, where $T_h$ is the number of complete $h$-month forecasts, and the paper's HAC statistic with $h-1$ lags. The resulting two-sided Monte Carlo $p$-values are calibrated finite-sample diagnostics, not a uniformly valid near-unit-root test; the simulations condition on estimated parameters and assume innovation pairs are independent over time.

With calendar-month effects, the Valkanov-style scaled-OLS check rejects no predictability at 5\% at every reported full-sample horizon. In the sample beginning in 2008, it rejects from two months onward, but the one-month $p$-value is 0.057--0.059. Simulating the HAC statistic gives the same post-Phase-I qualification: the one-month $p$-value is 0.060--0.063 and the remaining horizons reject at 5\%. In the full sample, the simulated HAC check rejects through eight months; the twelve-month result is borderline, with $p$-values of 0.044 and 0.050 for resampled and Gaussian innovations, respectively. The estimated AR(1) coefficients with month effects are 0.72 and 0.67 in the full and post-Phase-I samples. Stambaugh-type Monte Carlo bias, computed relative to the known predictive slopes in the calibrated alternative, is small but not uniformly below 1\%: its absolute magnitude is at most about 3\% of the observed coefficient across the samples, horizons, specifications, and innovation schemes. We leave the headline coefficients unchanged.

We also omit each calendar year in turn, either removing forecast origins in that year or removing all observations whose origin or future-return window touches the year. Returns are constructed before omission, and HAC lag distances retain the original monthly calendar. All coefficients remain positive in both samples and specifications. With month effects, full-sample significance survives every year omission at the reported horizons; after excluding Phase I, the one- and two-month results are sensitive to the omitted year, while significance survives from three months onward. Magnitudes remain sensitive: under the stricter window-omission check, the full-sample twelve-month coefficient ranges from 0.529 to 1.621, compared with 1.316 in Table~\ref{tab:Predictability}. These checks do not overturn the separate finding that estimates beginning in Phase III are weaker and imprecise.

\subsection{Daily Evidence on the Surrender-Month Return\label{subsec:daily_april_evidence}}

\begin{figure}[H]
\centering
\includegraphics[width=\textwidth]{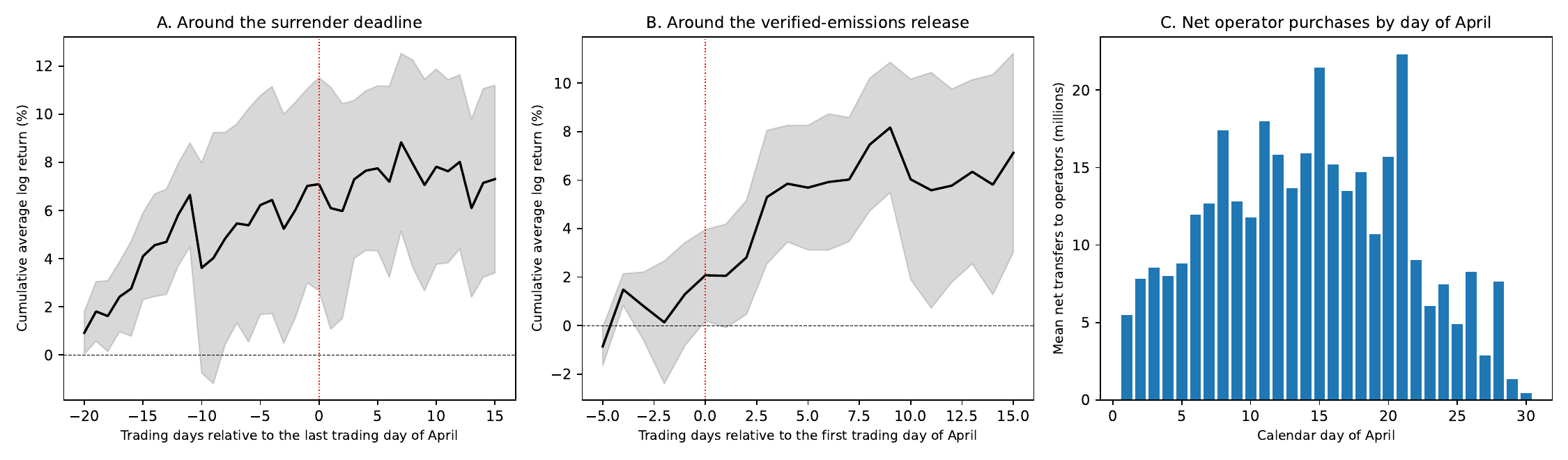}
\caption{Daily evidence on the surrender-month return, 2009--2022.\label{fig:daily_april_evidence}}
\par\medskip
\begin{minipage}{0.95\textwidth}
\footnotesize \emph{Notes:} Panels~A and B plot the average across the fourteen surrender years of the cumulative daily log return of the ICE EUA December futures contract around the last trading day of April (Panel~A) and the first trading day of April, when verified emissions for the previous year are published (Panel~B); the shaded band is plus or minus one standard error of the cumulative average across years. Panel~C plots the average net transfer of allowances from person holding accounts to operator holding accounts by calendar day of April, in millions of allowances.
\end{minipage}
\end{figure}

The monthly evidence in Section~\ref{subsec:fact_pressure} shows that April returns are high on average but cannot say when within the month the return accrues, whether it reverses after the deadline, or how it relates to the timing of compliance purchases. This appendix uses daily settlement prices of the ICE EUA December futures contract and daily registry transfers over the fourteen surrender years 2009--2022, the period with bankable allowances and a liquid futures market. Figure~\ref{fig:daily_april_evidence} plots the average cumulative log return around the surrender deadline (Panel~A) and around the publication of verified emissions on 1 April (Panel~B), and the average net transfer of allowances from person to operator accounts by calendar day in April (Panel~C). Table~\ref{tab:daily_april_evidence} reports the corresponding window statistics and weekly flow--return regressions. \citet{HitzemannUhrigHomburgEhrhart2015} document price and volume reactions to the annual publication of verified emissions in exchange data.

Three patterns stand out. First, the April return accrues in the first half of the month. The average cumulative return over trading days $-20$ to $-11$ before the deadline, roughly April 1--15, is 6.7\% (standard error across years 2.2\%, positive in 10 of 14 years), and the return over the five trading days after the release of verified emissions is 4.4\% (standard error 1.4\%, positive in 11 of 14 years). Over the last two weeks before the deadline the average return is 0.4\% with a standard error of 3.6\%, and over the last week 1.7\% with a standard error of 1.3\%. Second, the premium does not reverse in the ten trading days after the deadline: the average return over that window is 0.7\%, with a median of $-1.9$\%, so the pressure is worked off gradually rather than unwound within days. Third, compliance purchases reach operator accounts throughout April rather than in the last days: 53\% of April net transfers to operators occur in the first fifteen days and 13\% after April 23, so the timing of inflows matches the timing of the return.

Panel B of the table relates non-overlapping weekly futures returns to weekly net registry inflows to operators, scaled by the trailing 52-week average of weekly gross bilateral operator--person volume. The contemporaneous coefficient is 0.0065 ($t=2.4$): a week in which net inflows equal one week of average gross bilateral volume is associated with a return about 0.65 percentage points higher. The interaction with an April indicator is negative and insignificant, so the impact is not larger in the surrender month, and the following week's and the following four weeks' returns do not load negatively on the inflow. Dividing the coefficient by the average scaling volume gives a return impact of about $3.4\times10^{-10}$ per allowance of net inflow, the same order of magnitude as the surrender-window return-impact parameter $\hat{\varphi}=4.1\times10^{-10}$ calibrated in Section~\ref{sec:quant} from fourteen annual April premia. The two estimates use different data and different windows, so we read the agreement as an order-of-magnitude check rather than as an independent identification of $\varphi$; registry transfers lag the underlying trades by settlement, which argues for the weekly rather than daily aggregation and attenuates the estimate.

Taken together, the daily evidence supports the price-pressure interpretation of the surrender-month premium and refines its timing. Once verified emissions are known, operators buy throughout April, intermediaries absorb a month-long wave of demand, and prices are elevated while the wave lasts; the absence of a reversal in the ten trading days after the deadline indicates that intermediaries work off the resulting positions gradually rather than within days. Scarcity information released with verification can contribute to the April return, and fourteen years of daily data cannot separate a slow unwinding from a permanent component, which is why we treat $\hat{\varphi}$ as a calibrated sensitivity rather than a structural estimate of transitory price impact.

\begin{table}[H]
\centering
\caption{Daily Evidence on the Surrender-Month Return, 2009--2022\label{tab:daily_april_evidence}}
\footnotesize
\resizebox{\textwidth}{!}{%
\begin{tabular}{lrrrrr}
\toprule
\multicolumn{6}{l}{Panel A: Cumulative futures log returns over event windows (percent), across fourteen surrender years} \\
Window & Mean & S.e. & $t$ & Median & Positive years \\
\midrule
Deadline, days $-20$ to $-11$ (about April 1--15) & 6.65 & 2.15 & 3.09 & 5.60 & 10/14 \\
Deadline, days $-10$ to $0$ (last two weeks) & 0.44 & 3.57 & 0.12 & 2.71 & 10/14 \\
Deadline, days $-5$ to $0$ (last week) & 1.71 & 1.26 & 1.36 & 2.94 & 9/14 \\
Post-deadline, days $+1$ to $+10$ & 0.73 & 1.94 & 0.37 & -1.86 & 6/14 \\
Release window, days $0$ to $+5$ & 4.40 & 1.42 & 3.09 & 3.29 & 11/14 \\
Pre-release, days $-5$ to $-1$ & 1.29 & 2.12 & 0.61 & 2.95 & 10/14 \\
\midrule
\multicolumn{6}{l}{Panel B: Weekly futures returns and operator inflow, January 2009--September 2022} \\
 & Inflow & Inflow $\times$ April & April & $R^2$ & Weeks \\
\midrule
Weekly return on weekly operator inflow & 0.0065 (2.43) &  &  & 0.016 & 691 \\
With April interaction & 0.0099 (2.89) & -0.0061 (-1.00) & 0.0023 (0.21) & 0.019 & 691 \\
Next-week return on inflow & 0.0024 (1.22) &  &  & 0.002 & 690 \\
Four-week-ahead return on inflow & 0.0021 (0.47) &  &  & 0.000 & 687 \\
\bottomrule
\end{tabular}
}
\par
\smallskip
\begin{minipage}{0.95\textwidth}
\footnotesize
\emph{Notes:} Daily log returns are computed from ICE EUA December-futures settlement prices (continuous series). Panel A sums daily log returns over each window in each of the fourteen surrender years 2009--2022 and reports the mean, its standard error across years, the $t$-statistic, the median, and the number of years with a positive window return. Event day 0 is the last trading day on or before April 30 for the deadline windows and the first trading day on or after April 1, when verified emissions for the previous year are published, for the release window. The monthly April log return over the same years averages 6.9\% (median 10.0\%, positive in 11 of 14 years). Panel B regresses non-overlapping weekly (Monday--Friday) futures log returns on net registry transfers from person to operator accounts in the same week, scaled by the trailing 52-week average of weekly gross bilateral operator--person volume, with Newey--West (4 lags) $t$-statistics in parentheses; April is an indicator for weeks with at least three April trading days; the reversal rows use the following week's return and the cumulative return over the following four weeks. The contemporaneous coefficient implies a return impact of about 3.4e-10 per allowance of net inflow, the same order of magnitude as the calibrated surrender-window return-impact parameter of Section~\ref{sec:quant} (4.1e-10 per allowance).
\end{minipage}
\end{table}

\subsection{Public Compliance Pressure and Return Predictability\label{subsec:public_pressure_predictability}}

Operator Flow Imbalance may contain both informative trading and publicly predictable compliance demand. Calendar-month fixed effects absorb average seasonality, but they do not absorb cycle-specific variation in aggregate shortfalls. We therefore construct two annual measures from the compliance records. For cycle $y$, aggregate positive shortfall is
\begin{align}
Q_y=\sum_i\max\{\text{surrender}_{iy}-\text{allocation}_{iy},0\},
\qquad
q_y=\frac{Q_y}{\sum_i\text{surrender}_{iy}},
\label{eq:appendix_public_pressure}
\end{align}
where $q_y$ is the corresponding shortfall share. These realized quantities provide conservative ex post measures of public compliance pressure. We map May through December of year $y-1$ and January through April of year $y$ to the same compliance cycle, and $Q_y$ refers to the obligation surrendered in April of year $y$ (emissions of year $y-1$).

For each forecast horizon $h$, let $A_{t,h}$ indicate that the cumulative-return window from $t+1$ through $t+h$ contains April. We augment the baseline regression with either standardized $\log(1+Q_y)\times A_{t,h}$ or standardized $q_y\times A_{t,h}$. This targets the proposed channel directly: cycle-specific compliance pressure is allowed to explain returns when the future window contains the surrender month, without turning the annual control into an unrestricted phase proxy in all other months. We also re-estimate the baseline after removing every forecast window for which $A_{t,h}=1$.

Table~\ref{tab:public_compliance_pressure} reports the results. With log aggregate shortfall in Panel B and with the scale-free shortfall share in Panel C, the Operator Flow Imbalance coefficient remains positive and statistically significant through eight months but not at twelve months. Panel D excludes April-crossing windows entirely; the coefficient remains positive and significant at the 5\% level through six months and at the 10\% level at eight months, with magnitudes close to the baseline. A twelve-month estimate is unavailable in that panel because every twelve-month window necessarily contains April.

Panel E reports the unrestricted alternatives, in which the annual controls enter in every month rather than only when the forecast window contains April, together with the limiting case of compliance-cycle fixed effects. With the unrestricted log-shortfall control, the Operator Flow Imbalance coefficient remains significant at one and three months but not beyond; with the unrestricted shortfall share it is significant only at three months and only at the 10\% level; and with compliance-cycle fixed effects it is positive at every horizon but significant only at three months (5\% level) and six months (10\% level). We report these specifications for completeness, but they are less informative about the channel at issue, for two reasons. First, they test a broader hypothesis than the one raised: the concern is that operator flow proxies for publicly predictable compliance demand when the forecast window contains a surrender month, whereas an unrestricted annual control also absorbs variation in months whose forecast window contains no April. Second, and more importantly, they leave little variation with which to identify any monthly predictor. The sample contains seventeen compliance cycles with a shortfall measure, and the baseline already includes calendar-month fixed effects, so an annual control entering in every month is close to a compliance-cycle fixed effect. The final row makes that limiting case explicit: with both calendar-month and cycle fixed effects, Operator Flow Imbalance is identified only from within-cycle deviations that are also orthogonal to the average seasonal pattern. The April-window restriction in Panels B and C is therefore a deliberate feature of the test rather than an incidental choice, and Panel D provides the complementary check that does not rely on it.

Taken together, public surrender-month pressure does not account for the shorter-horizon return predictability in the specifications that isolate it, while the twelve-month relation is not separately identified from cycle-specific compliance pressure.

\begin{table}[tbp]
\centering
\caption{Public Compliance Pressure and Operator Flow Predictability}
\label{tab:public_compliance_pressure}
\small
\begin{tabular}{lccccc}
\toprule
Forecast horizon (months) & 1 & 3 & 6 & 8 & 12 \\
\midrule
\multicolumn{6}{l}{Panel A: Baseline} \\
Operator Flow Imbalance & 0.104*** & 0.342*** & 0.628*** & 0.811*** & 1.316*** \\
 & (2.65) & (3.26) & (2.78) & (2.63) & (2.62) \\
Observations & 209 & 207 & 204 & 202 & 198 \\
\midrule
\multicolumn{6}{l}{Panel B: Aggregate shortfall in April-crossing windows} \\
Operator Flow Imbalance & 0.092** & 0.336*** & 0.636** & 0.784** & 1.033 \\
 & (2.36) & (3.04) & (2.47) & (1.99) & (1.42) \\
Log aggregate shortfall $\times$ April window & 0.779* & 0.198 & -0.133 & 0.309 & 0.222 \\
 & (1.69) & (0.34) & (-0.14) & (0.22) & (1.10) \\
Observations & 209 & 207 & 204 & 202 & 197 \\
\midrule
\multicolumn{6}{l}{Panel C: Shortfall share in April-crossing windows} \\
Operator Flow Imbalance & 0.091** & 0.325*** & 0.541*** & 0.553** & 0.535 \\
 & (2.36) & (3.10) & (2.62) & (2.11) & (1.34) \\
Shortfall share $\times$ April window & 0.059 & 0.045 & 0.122 & 0.253** & 0.514*** \\
 & (1.61) & (0.74) & (1.16) & (2.17) & (3.02) \\
Observations & 209 & 207 & 204 & 202 & 197 \\
\midrule
\multicolumn{6}{l}{Panel D: Forecast windows that do not cross April} \\
Operator Flow Imbalance & 0.088** & 0.306*** & 0.529** & 0.460* & -- \\
 & (2.26) & (2.68) & (2.36) & (1.73) & -- \\
Observations & 192 & 156 & 102 & 68 & -- \\
\midrule
\multicolumn{6}{l}{Panel E: Unrestricted annual controls} \\
\emph{OFI}, with log aggregate shortfall & 0.102** & 0.361*** & 0.548 & 0.708 & 1.033 \\
 & (1.97) & (2.69) & (1.61) & (1.56) & (1.42) \\
\emph{OFI}, with shortfall share & 0.048 & 0.190* & 0.200 & 0.233 & 0.535 \\
 & (0.99) & (1.88) & (0.89) & (0.89) & (1.34) \\
\emph{OFI}, with compliance-cycle fixed effects & 0.096 & 0.282** & 0.272* & 0.168 & 0.147 \\
 & (1.30) & (2.02) & (1.71) & (1.22) & (1.03) \\
Observations & 204 & 204 & 203 & 201 & 197 \\
\bottomrule
\end{tabular}
\begin{minipage}{0.97\textwidth}
\footnotesize\vspace{0.15cm}
\textit{Notes:} This table reports predictive regressions of future cumulative EUA log returns on Operator Flow Imbalance. All specifications include calendar-month fixed effects and use overlapping-horizon Newey--West standard errors. Aggregate shortfall is the sum across operators of positive surrender obligations net of free allocations in a compliance cycle; the shortfall share divides this quantity by aggregate surrender obligations. The controls in Panels B and C are standardized and enter only when the forecast window includes the cycle's April surrender month. Panel D removes all such forecast windows; a twelve-month estimate is therefore unavailable. Panel E instead lets the annual controls enter in every month rather than only in April-crossing windows, and its final row replaces them with compliance-cycle fixed effects; the reported observation counts are those of the two shortfall rows, while the cycle-fixed-effect row uses the Panel A sample. Aggregate shortfalls refer to the obligation surrendered in the cycle's April. $t$-statistics are in parentheses. *, **, and *** denote significance at the 10\%, 5\%, and 1\% levels, respectively.
\end{minipage}
\end{table}

\subsection{Dual-Account Entities and Operator Flow Predictability\label{subsec:dual_account_ofi}}

The distinction between operator and person accounts is an account classification, not necessarily a complete separation of economic entities. A regulated firm can hold an operator account for compliance and a person account for trading, so transactions involving such entities may blur the operator--intermediary interpretation of Operator Flow Imbalance. We identify an entity as dual role if a common account-holder identifier or a valid company registration number is linked to both account types. The registry contains 761 dual-role account-holder identifiers and 815 dual-role registration numbers; the latter represent 7.0\% of registration numbers associated with operator accounts.

We reconstruct monthly Operator Flow Imbalance from the same transactions and retain the paper's normalization by gross bilateral operator--person volume. Panel B of Table~\ref{tab:dual_account_ofi} excludes dual-role entities when they appear as person-account counterparties. Panels C and D impose the stricter restriction that neither side of a transaction can be dual role. This pure-role restriction removes 34.0\% of bilateral transactions and 61.3\% of gross bilateral volume, so it is a demanding robustness check rather than a marginal sample adjustment. All return regressions otherwise preserve the baseline specification, including calendar-month fixed effects, overlapping-horizon Newey--West inference, and the original full-sample operator-frequency classification.

The result is not driven by dual-role entities. Relative to the existing coefficients reproduced in Panel A, the pure-role estimates in Panel C are close in magnitude and remain statistically significant at every horizon from one to twelve months. Panel D also preserves the cross-sectional prediction: pure-role flow from frequent operators is positive and significant at every horizon, whereas flow from infrequent operators is never significant. Thus, neither the aggregate predictability result nor its concentration among frequent operators is an artifact of regulated entities also operating through person accounts.

\begin{table}[tbp]
\centering
\caption{Dual-Account Entities and Operator Flow Predictability}
\label{tab:dual_account_ofi}
\small
\begin{tabular}{lccccc}
\toprule
Forecast horizon (months) & 1 & 3 & 6 & 8 & 12 \\
\midrule
\multicolumn{6}{l}{Panel A: Existing measure} \\
Operator Flow Imbalance & 0.104*** & 0.342*** & 0.628*** & 0.811*** & 1.316*** \\
 & (2.65) & (3.26) & (2.78) & (2.63) & (2.62) \\
Observations & 209 & 207 & 204 & 202 & 198 \\
\midrule
\multicolumn{6}{l}{Panel B: Exclude dual-role person-account counterparties} \\
Operator Flow Imbalance & 0.091** & 0.306*** & 0.585*** & 0.758*** & 1.224*** \\
 & (2.54) & (3.30) & (2.88) & (2.76) & (2.67) \\
Observations & 209 & 207 & 204 & 202 & 198 \\
\midrule
\multicolumn{6}{l}{Panel C: Pure-role transactions} \\
Operator Flow Imbalance & 0.093*** & 0.280*** & 0.597*** & 0.771*** & 1.227*** \\
 & (3.00) & (3.34) & (2.85) & (2.75) & (2.59) \\
Observations & 209 & 207 & 204 & 202 & 198 \\
\midrule
\multicolumn{6}{l}{Panel D: Pure-role transactions by operator trading frequency} \\
Infrequent-operator flow & -0.016 & 0.041 & 0.042 & 0.145 & 0.346 \\
 & (-0.32) & (0.40) & (0.22) & (0.58) & (0.94) \\
Frequent-operator flow & 0.101** & 0.255*** & 0.563*** & 0.671*** & 0.994*** \\
 & (2.08) & (2.66) & (2.84) & (2.97) & (2.95) \\
Observations & 209 & 207 & 204 & 202 & 198 \\
\bottomrule
\end{tabular}
\begin{minipage}{0.97\textwidth}
\footnotesize\vspace{0.15cm}
\textit{Notes:} This table reports predictive regressions of future cumulative EUA log returns on Operator Flow Imbalance with calendar-month fixed effects and overlapping-horizon Newey--West standard errors. Panel A reproduces the existing measure. Panel B excludes person-account counterparties whose entity also holds an operator account. Panels C and D retain only transactions for which neither side is a dual-role entity. Dual role is identified by a common account-holder ID or registration number; 815 registration numbers, about 7.0\% of registrations with operator accounts, are dual role. The strict exclusion removes 61.3\% of gross bilateral operator--person volume. Panel D retains the paper's original full-sample operator-frequency classification. $t$-statistics are in parentheses. *, **, and *** denote significance at the 10\%, 5\%, and 1\% levels, respectively.
\end{minipage}
\end{table}

\subsection{Intermediary Position Proxies and Return Predictability\label{subsec:inventory_predictability}}

We examine whether intermediary inventory reversal can account for the positive return predictability associated with Operator Flow Imbalance. The registry records transfers rather than historical month-end account balances, so it does not provide a direct measure of intermediary inventory. Moreover, within bilateral operator--person transfers, the aggregate change in person-account positions is mechanically the negative of aggregate operator net flow. Those bilateral flows therefore cannot separately identify an inventory channel after controlling for Operator Flow Imbalance.

For this diagnostic, we construct a broader position proxy from all recorded external transfers involving pure-person account holders, defined as holders with a person account but no operator account. We cumulate their net acquisitions from February 2005, including transfers involving administrative accounts. The resulting series is a proxy rather than a balance measure because opening positions are unobserved. It is also extremely persistent: in the return-regression sample its estimated AR(1) coefficient is 1.008, and an augmented Dickey--Fuller test with a constant has a $p$-value of 0.99.

We adopt a holdings convention throughout, so the proxy rises when pure-person account holders accumulate allowances. This fixes the sign that each hypothesis predicts. Under an inventory-reversal interpretation, an intermediary sector carrying a larger accumulated position must eventually unwind it, and the associated selling pressure should forecast lower subsequent returns; the competing channel therefore predicts a \emph{negative} coefficient on the position proxy. Under the interpretation in the main text, operator flow is informative about future scarcity, and the position proxy should add little once Operator Flow Imbalance is included, so the predicted coefficient is approximately zero.

Table~\ref{tab:inventory_predictability} keeps the return specification unchanged: future cumulative allowance returns are regressed jointly on Operator Flow Imbalance and the standardized position proxy, with calendar-month fixed effects and overlapping-horizon Newey--West standard errors. Panel A uses the cumulative position level. Its coefficient is negative at every horizon, and one-sided persistence-adjusted $p$-values based on the Valkanov-style simulation are 0.021 and 0.040 at six and twelve months. Panel A should nonetheless be read as a diagnostic rather than as an estimate. The position level is statistically indistinguishable from a unit-root process, and including it makes the regression unbalanced, so every coefficient in the panel is unreliable. The Operator Flow Imbalance coefficients illustrate the point: they fall to well under half their magnitude elsewhere in the table and lose significance at the 5\% level at three of the four horizons, even though the same predictor is significant in Table~\ref{tab:Predictability} and in Panels B through D here. The panel is therefore uninformative about both variables, and it motivates the stationary specifications that follow rather than providing evidence for the inventory channel.

Panels B and C remove, respectively, a linear trend and phase means from the position series before estimation. The adjusted position coefficients are small and insignificant at every horizon, while the Operator Flow Imbalance coefficients recover the magnitudes and significance reported in Table~\ref{tab:Predictability}. Removing this low-frequency variation is a substantive choice rather than a statistical convenience. Inventory effects in the tradition of \cite{GrossmanMiller1988} operate at daily to monthly frequencies, over which intermediaries absorb and then unwind order imbalances. A multi-year trend in cumulative registry positions is far more plausibly attributable to the growth of the market itself, including entry of new person accounts, rising traded volume, and the secular increase in allowance prices, than to inventory pressure at the horizons the theory concerns. Panel D uses the monthly position change and likewise finds no negative relation, the only significant coefficient being positive; because a change is a flow rather than a stock, it is scaled by trailing twelve-month gross pure-person-account flows so that a given change is measured relative to contemporaneous market activity rather than in absolute allowances.

Taken together, the negative relation between the position proxy and future returns appears only in the raw, near-unit-root level and does not survive any stationary transformation of the series. We therefore cannot distinguish it from low-frequency growth in market size, and we find no evidence of inventory reversal at the horizons over which the mechanism is usually understood to operate. This conclusion is necessarily limited by the absence of direct intermediary balance data, and we interpret the exercise as a competing-channel diagnostic rather than as a structural inventory test.

\begin{table}[tbp]
\caption{Intermediary Position Proxies and Return Predictability\label{tab:inventory_predictability}}

\medskip{}

{\footnotesize This table reports joint predictive regressions of future cumulative EUA log returns on \emph{Operator Flow Imbalance} and alternative pure-person-account position proxies. Both predictors are standardized within each estimation sample. All specifications retain the paper's calendar-month fixed effects and use Newey--West standard errors with $h-1$ lags at horizon $h$; $t$-statistics are reported in parentheses. The cumulative position level is the sum of recorded net external acquisitions by holders with person accounts but no operator accounts from February 2005 onward. The linear-detrended and phase-adjusted proxies are residuals from projecting that level on, respectively, a linear time trend and EU ETS phase indicators before running the unchanged return regression. The monthly-change proxy is scaled by trailing twelve-month gross pure-person-account flows before standardization. The registry does not report opening balances, so these variables are position proxies rather than direct intermediary inventories. All proxies use a holdings convention, so the variable increases when pure-person account holders accumulate allowances; an inventory-reversal channel predicts a negative coefficient. The position level in Panel A is statistically indistinguishable from a unit-root process, so the coefficients in that panel, including those on \emph{Operator Flow Imbalance}, are not reliably estimated.}

\medskip{}

\centering
\begin{tabular}{lcccc}
\toprule
 & $r_{t\to t+1}$ & $r_{t\to t+3}$ & $r_{t\to t+6}$ & $r_{t\to t+12}$\\
\midrule
\multicolumn{5}{l}{Panel A: Cumulative position level}\\
\emph{Op. Flow Imb.} & 0.032 & 0.117 & 0.162 & 0.363\\
 & (1.361) & (2.024) & (1.498) & (1.671)\\
Position proxy & -0.031 & -0.091 & -0.233 & -0.453\\
 & (-2.138) & (-2.545) & (-2.917) & (-2.802)\\
$R^2$ & 0.098 & 0.163 & 0.194 & 0.236\\
$N$ & 209 & 207 & 204 & 198\\
\midrule
\multicolumn{5}{l}{Panel B: Linear-detrended cumulative position}\\
\emph{Op. Flow Imb.} & 0.055 & 0.181 & 0.328 & 0.673\\
 & (2.652) & (3.312) & (2.849) & (2.795)\\
Position proxy & 0.018 & 0.064 & 0.146 & 0.408\\
 & (1.080) & (1.340) & (1.207) & (1.411)\\
$R^2$ & 0.095 & 0.160 & 0.179 & 0.261\\
$N$ & 209 & 207 & 204 & 198\\
\midrule
\multicolumn{5}{l}{Panel C: Phase-adjusted cumulative position}\\
\emph{Op. Flow Imb.} & 0.057 & 0.187 & 0.343 & 0.724\\
 & (2.653) & (3.224) & (2.752) & (2.585)\\
Position proxy & 0.003 & -0.001 & -0.007 & -0.015\\
 & (0.285) & (-0.050) & (-0.175) & (-0.151)\\
$R^2$ & 0.090 & 0.142 & 0.147 & 0.183\\
$N$ & 209 & 207 & 204 & 198\\
\midrule
\multicolumn{5}{l}{Panel D: Monthly position change}\\
\emph{Op. Flow Imb.} & 0.068 & 0.221 & 0.399 & 0.793\\
 & (2.903) & (3.663) & (3.090) & (2.580)\\
Position proxy & 0.008 & 0.044 & 0.075 & 0.050\\
 & (0.929) & (1.665) & (2.221) & (0.693)\\
$R^2$ & 0.099 & 0.165 & 0.170 & 0.201\\
$N$ & 206 & 204 & 201 & 195\\
\bottomrule
\end{tabular}
\end{table}

\subsection{Continuous-Time Model: Derivations and Proofs\label{subsec:ct_appendix}}

This subsection provides the derivations and proofs underlying the
continuous-time model in Section~\ref{sec:ctmodel}. We follow the
notation of the main text: calendar time is continuous with a
recurring annual compliance cycle, $s_t$ denotes time remaining until
the next surrender date, and $X_t=(B_t,z_t,s_t)$ is the public state.
The aggregate bank $B_t$ evolves deterministically with aggregate net
compliance use, and the public scarcity state $z_t$ follows a
mean-reverting diffusion
\begin{align}
dz_t = -\kappa_z(z_t-\bar z)\, dt + \sigma_z\, dZ_t,
\label{eq:ap_zt}
\end{align}
where $\kappa_z \geq 0$ and $\bar z$ is the long-run mean. The
specification is stylized: any affine state dynamics delivers the
same conclusions.

\subsubsection*{Affine Benchmark Pricing}

In the frictionless benchmark, intermediary capacity is unlimited,
participation is continuous, and information is public. Consider a
pricing functional $P_t^*=P^*(X_t)$ and conjecture the affine form
\begin{align}
P^*(X_t) = A_0(s_t) + A_B(s_t)\, B_t + A_z(s_t)\, z_t.
\label{eq:ap_affine}
\end{align}
Absent frictions, allowance prices satisfy a no-arbitrage condition
relating the instantaneous discount rate to the drift of $P^*$ plus
any expected compliance-driven dividend flow. Substituting
\eqref{eq:ap_affine} and matching coefficients on $B_t$, $z_t$, and
the constant yields a system of ordinary differential equations in
the within-cycle calendar variable $s_t$ for $A_0(s_t)$, $A_B(s_t)$,
and $A_z(s_t)$, with periodic boundary conditions imposed by the
recurrence of the compliance cycle. Closed-form solutions obtain
under constant aggregate compliance demand and mean-reverting $z_t$;
in general, the coefficients are smooth functions of $s_t$. Further
details are standard in affine asset pricing and omitted.

\subsubsection*{Endogenous Market Access\label{subsec:ap_endogenous}}

This subsection proves Proposition~\ref{prop:endog_access} for the one-shot known-shortfall specialization used in the quantitative section. Consider a compliance cycle with remaining horizon $\tau$. Operator $i$ has expected positive shortfall $d_i$ and takes the expected surrender-month return premium $\pi_A$ as given. If the operator receives at least one discretionary access opportunity before surrender, it can avoid buying the shortfall at the surrender premium. Normalizing the per-cycle access cost by the early-cycle price level, the value of access is $S_i = d_i\,\pi_A$ and the normalized access-cost parameter is $\tilde{\psi}_i\equiv \psi_i/P_e$. Choosing access intensity $\lambda_i\geq 0$ delivers access probability $1-e^{-\lambda_i\tau}$ and convex per-cycle cost $\tilde{\psi}_i\lambda_i^2/2$. The operator's problem is
\begin{align}
\max_{\lambda_i\geq 0}
\left\{ S_i\!\left(1-e^{-\lambda_i\tau}\right) - \frac{\tilde{\psi}_i}{2}\lambda_i^2 \right\},
\label{eq:ap_access_problem}
\end{align}
which is strictly concave in $\lambda_i$ because its second derivative is $-S_i\tau^2 e^{-\lambda_i\tau}-\tilde{\psi}_i<0$. The first-order condition is therefore sufficient and given by $S_i\,\tau\, e^{-\lambda_i\tau} = \tilde{\psi}_i\,\lambda_i$. Multiplying through by $\tau e^{\lambda_i\tau}$ gives $(\lambda_i\tau)\, e^{\lambda_i\tau} = S_i\tau^2/\tilde{\psi}_i$, so
\begin{align}
\lambda_i^* = \frac{1}{\tau}\, W\!\left(\frac{S_i\,\tau^2}{\tilde{\psi}_i}\right),
\label{eq:ap_lambdaStar}
\end{align}
where $W(\cdot)$ denotes the Lambert $W$ function on $\mathbb{R}_+$, that is, the inverse of $y\mapsto ye^y$ for nonnegative arguments. Since $W$ is increasing on $\mathbb{R}_+$, $\lambda_i^*$ is increasing in $S_i = d_i\pi_A$ and decreasing in $\tilde{\psi}_i$. The probability of receiving no discretionary access opportunity before surrender is $e^{-\lambda_i^*\tau}$, so expected residual terminal demand for operator $i$ is
\begin{align}
R_{iA} = d_i\, e^{-\lambda_i^*\tau}.
\label{eq:ap_RiA}
\end{align}
Aggregating across operators,
\begin{align}
D_A(\pi_A) = \sum_i d_i\, e^{-\lambda_i^*(\pi_A)\,\tau}.
\label{eq:ap_DA}
\end{align}
The equilibrium surrender-month return premium then satisfies the fixed-point condition
\begin{align}
\pi_A = \varphi\, D_A(\pi_A) = \varphi\,\sum_i d_i\, e^{-\lambda_i^*(\pi_A)\,\tau}.
\label{eq:ap_eqm}
\end{align}
The right-hand side of \eqref{eq:ap_eqm} is continuous and strictly decreasing in $\pi_A$ whenever $\sum_i d_i>0$: a higher expected premium raises every $\lambda_i^*(\pi_A)$ for which $d_i>0$ and lowers each corresponding term $d_i e^{-\lambda_i^*(\pi_A)\tau}$. At $\pi_A=0$, the right-hand side equals $\varphi\sum_i d_i>0$. As $\pi_A\to\infty$, each $\lambda_i^*(\pi_A)\to\infty$, so each term $d_i e^{-\lambda_i^*(\pi_A)\tau}\to 0$ and therefore the right-hand side converges to zero. The left-hand side is increasing in $\pi_A$ and equals zero at the origin. Hence there exists a unique positive fixed point whenever $\sum_i d_i>0$, so the equilibrium surrender-month return premium is a uniquely determined object rather than an assumed primitive. This establishes Proposition~\ref{prop:endog_access}.

To establish Corollary~\ref{cor:phi_elasticity}, define $F(\pi,\varphi)\equiv\pi-\varphi\,D_A(\pi)$ so that the equilibrium is the implicit solution to $F(\pi_A,\varphi)=0$. Differentiating the fixed point yields, by the implicit function theorem,
\begin{align}
\frac{d\pi_A}{d\varphi} = \frac{D_A(\pi_A)}{1-\varphi\, D_A'(\pi_A)}.
\label{eq:ap_dDeltadphi}
\end{align}
The differentiability of $D_A(\cdot)$ on $\mathbb{R}_+$ follows from differentiability of the Lambert $W$ function on $\mathbb{R}_+$. Because $\lambda_i^*(\pi_A)$ is strictly increasing in $\pi_A$ for each $i$ with $d_i>0$, the function $D_A(\pi_A)=\sum_i d_i e^{-\lambda_i^*(\pi_A)\tau}$ is strictly decreasing in $\pi_A$, that is, $D_A'(\pi_A)<0$ whenever $\sum_i d_i>0$. Hence $1-\varphi\,D_A'(\pi_A)>1$. At the equilibrium, $\varphi\,D_A(\pi_A)=\pi_A$, so dividing both sides of \eqref{eq:ap_dDeltadphi} by $\pi_A/\varphi$ gives
\begin{align}
\frac{d\log \pi_A}{d\log \varphi} = \frac{1}{1-\varphi\,D_A'(\pi_A)} = \frac{1}{1+\varphi\,|D_A'(\pi_A)|},
\label{eq:ap_phi_elasticity}
\end{align}
which lies strictly in $(0,1)$ whenever $\sum_i d_i>0$. The elasticity approaches one in the static-access limit $|D_A'(\pi_A)|\to 0$, which obtains as normalized access costs become uniformly large: at any equilibrium $0<\pi_A\leq\varphi D_0$ with $D_0\equiv\sum_i d_i$, so as $\tilde{\psi}_i\to\infty$ for all $i$, $\chi_i\equiv d_i\pi_A\tau^2/\tilde\psi_i\leq d_i\varphi D_0\tau^2/\tilde{\psi}_i\to 0$, hence $W_i\to 0$ and $\lambda_i^*\to 0$; moreover $\pi_A\geq\varphi D_0e^{-\max_i W_i}\to\varphi D_0>0$, so by the expression for $D_A'$ derived below, $|D_A'(\pi_A)|=\pi_A^{-1}\sum_i d_ie^{-W_i}W_i/(1+W_i)\to 0$. This establishes Corollaries~\ref{cor:phi_elasticity} and~\ref{cor:friction_interaction}.

To establish Corollary~\ref{cor:elasticity_bound}, we compute $|D_A'(\pi_A)|$ explicitly. The chain rule gives $dW_i/d\pi_A = W'(\chi_i)\,dx_i/d\pi_A$, where $\chi_i\equiv d_i\pi_A\tau^2/\tilde\psi_i$ and $dx_i/d\pi_A=\chi_i/\pi_A$. Differentiating the defining identity $W(x)e^{W(x)}=x$ yields $W'(x)=e^{-W(x)}/(1+W(x))=W(x)/[x(1+W(x))]$, where the second equality uses the exact identity $e^{-W(x)}=W(x)/x$. Combining,
\begin{align}
\frac{dW_i}{d\pi_A} = \frac{W_i}{(1+W_i)\,\pi_A}.
\label{eq:ap_dWi}
\end{align}
Differentiating $D_A(\pi_A)=\sum_i d_i e^{-W_i}$ then yields
\begin{align}
D_A'(\pi_A) = -\frac{1}{\pi_A}\sum_i d_i e^{-W_i}\,\frac{W_i}{1+W_i}.
\label{eq:ap_DA_prime}
\end{align}
Substituting into \eqref{eq:ap_phi_elasticity} and using $\varphi/\pi_A=1/D_A(\pi_A)$ at the equilibrium gives
\begin{align}
\frac{d\log\pi_A}{d\log\varphi} = \frac{1}{1+\omega}, \qquad \omega = \sum_i \varrho_i\,\frac{W_i}{1+W_i},
\label{eq:ap_omega}
\end{align}
where $\varrho_i\equiv d_i e^{-W_i}/\sum_j d_j e^{-W_j}$ is operator $i$'s share of aggregate residual terminal demand. Since $W_i/(1+W_i)\in(0,1)$ pointwise whenever $d_i,\pi_A,\tau,\tilde\psi_i>0$, $\omega$ is a convex combination of values in $(0,1)$ and therefore $\omega\in(0,1)$. Hence the elasticity is bounded strictly in $(1/2,1)$. For the limit, let $\tilde\psi_i=\epsilon c_i$ with $c_i>0$ fixed and $\epsilon\to 0$. Using $d_ie^{-W_i}=\tilde\psi_iW_i/(\pi_A\tau^2)$, the fixed point reads $\pi_A^2\tau^2/\varphi=\sum_j\tilde\psi_jW_j$. If some $\chi_i$ remained bounded along a sequence $\epsilon\to 0$, all $W_j$ would be bounded (since $\chi_j=(d_j/c_j)(c_i/d_i)\chi_i$), hence $\pi_A=O(\epsilon^{1/2})$ and $\chi_i=d_i\pi_A\tau^2/(\epsilon c_i)=\Omega(\epsilon^{-1/2})\to\infty$, a contradiction. Hence $W_i\to\infty$ for each $i$ with $d_i>0$, $W_i/(1+W_i)\to 1$, $\omega\to 1$, and the elasticity converges to $1/2$. This establishes Corollary~\ref{cor:elasticity_bound}.

To establish Corollary~\ref{cor:scale_invariance}, suppose $(\varphi,\tilde\psi_i)\mapsto(k\varphi,\,k\tilde\psi_i)$ for all $i$ with $k>0$, and conjecture that the new equilibrium premium is $\pi_A^{\text{new}}=k\pi_A$, where $\pi_A$ is the equilibrium under the original parameters. Under the scaled parameters, the Lambert-$W$ argument evaluated at the conjectured equilibrium is
\begin{align}
\frac{d_i\,\pi_A^{\text{new}}\,\tau^2}{k\tilde\psi_i}=\frac{d_i\,(k\pi_A)\,\tau^2}{k\tilde\psi_i}=\frac{d_i\,\pi_A\,\tau^2}{\tilde\psi_i},
\label{eq:ap_scale_invariance_arg}
\end{align}
identical to the original Lambert-$W$ argument. Hence each scaled access intensity $\lambda_i^{*,\text{new}}(\pi_A^{\text{new}})=\lambda_i^*(\pi_A)$, and the scaled residual terminal demand $D_A^{\text{new}}(\pi_A^{\text{new}})=D_A(\pi_A)$. The scaled fixed-point condition becomes
\begin{align}
\pi_A^{\text{new}}=k\varphi\,D_A^{\text{new}}(\pi_A^{\text{new}})=k\varphi\,D_A(\pi_A)=k\pi_A,
\label{eq:ap_scale_invariance_fp}
\end{align}
where the last equality uses the original fixed point $\pi_A=\varphi D_A(\pi_A)$. The conjecture is verified, and uniqueness from Proposition~\ref{prop:endog_access} guarantees this is the equilibrium of the scaled system. This establishes Corollary~\ref{cor:scale_invariance}.

To establish Corollary~\ref{cor:proportional_amplification}, impose $\tilde\psi_i=a d_i$ for every positive-shortfall operator. The optimal-access rule then gives a common $w\equiv\lambda_i^*\tau=W(\pi_A\tau^2/a)$, so $D_A=D_0e^{-w}$, where $D_0=\sum_i d_i>0$. Combining the defining identity of $W$ with the equilibrium condition yields
\begin{align}
w e^w=\frac{\pi_A\tau^2}{a},\qquad
\pi_A=\varphi D_0e^{-w}
\quad\Longrightarrow\quad
w e^{2w}=\frac{\varphi D_0\tau^2}{a}.
\label{eq:ap_proportional_fixed_point}
\end{align}
Hence $w=\tfrac{1}{2}W(2\varphi D_0\tau^2/a)$. Holding $D_0$ and $\tau$ fixed, log-differentiation gives
\begin{align}
\frac{\partial w}{\partial\log\varphi}
=\frac{w}{1+2w},\qquad
\frac{\partial w}{\partial\log a}
=-\frac{w}{1+2w}.
\label{eq:ap_proportional_w_derivatives}
\end{align}
Since $\log\pi_A=\log\varphi+\log D_0-w$, the premium elasticity is $\varepsilon_\varphi=1-w/(1+2w)=(1+w)/(1+2w)$. Differentiating this elasticity with respect to $\log a$ gives
\begin{align}
\frac{\partial\varepsilon_\varphi}{\partial\log a}
=-\frac{1}{(1+2w)^2}\frac{\partial w}{\partial\log a}
=\frac{w}{(1+2w)^3}>0.
\label{eq:ap_proportional_amplification}
\end{align}
This proves Corollary~\ref{cor:proportional_amplification}. The log-change elasticity for any fixed proportional reduction in $\varphi$ is the average of $\varepsilon_\varphi$ over the corresponding log-$\varphi$ interval, so it too increases with $a$. The restriction is sufficient rather than necessary: it fixes residual-demand shares at $\varrho_i=d_i/D_0$, whereas under unrestricted heterogeneity those shares change with access costs and can overturn monotonicity.

\subsubsection*{Aggregation of Compliance Demand}

Let $c_{it}$ denote operator $i$'s compliance allowance holdings and
$h_{it}^*$ target compliance holdings. Under the endogenous Poisson access specification
\eqref{eq:ct_access}, the compliance gap $g_{it}=h_{it}^*-c_{it}$
satisfies the reduced-form adjustment equation in the main text,
\begin{align}
E_t[dg_{it}] = E_t[dh_{it}^*] - \lambda_i^*\, g_{it}\, dt.
\label{eq:ap_gap}
\end{align}
Speculative positions are tracked separately from $c_{it}$ in
Section~\ref{subsec:ct_info} of the main text and do not enter
\eqref{eq:ap_gap}. Write $E_t[dh_{it}^*]=a_i(s_t,X_t)\,dt$, where
$a_i(s_t,X_t)$ captures how desired compliance demand rises as the
surrender date approaches ($s_t\to 0$). Consistent with the discussion in Section~\ref{sec:ctmodel}, we treat $h_{it}^*$ as a stochastic target with deterministic seasonal drift, idiosyncratic firm-level innovations, and a public-state component that loads on $z_t$; this stochastic structure ensures that operator-level gaps are not fully predictable away from access shocks and gives the private-signal component in Section~\ref{subsec:ct_info} a non-trivial object to be informative about. Under mild integrability conditions, starting from date $t_0$, the conditional expected gap satisfies
\begin{align}
E_{t_0}[g_{it}] = e^{-\lambda_i^*\,(t-t_0)}\, g_{i,t_0} + \int_{t_0}^{t}
e^{-\lambda_i^*\,(t-u)}\, E_{t_0}\!\left[a_i(s_u,X_u)\right] du.
\label{eq:ap_gap_sol}
\end{align}
Aggregating across operators yields
\begin{align}
E_t[dG_t] = A(s_t,X_t)\, dt - \sum_i \lambda_i^*\, g_{it}\, dt = A(s_t,X_t)\, dt - \bar\lambda_t^*\, G_t\, dt,
\qquad
G_t = \sum_i g_{it},
\label{eq:ap_G}
\end{align}
with $A(s_t,X_t)=\sum_i a_i(s_t,X_t)$ and $\bar\lambda_t^* \equiv \sum_i \lambda_i^* g_{it}/G_t$ the gap-weighted average of cycle-level chosen access intensities. Equation~\eqref{eq:ap_G} is the
aggregate analog of \eqref{eq:ap_gap} and is the law of motion for
unresolved aggregate compliance demand. This richer dynamic object motivates the general seasonal discussion in the main text, but it is not the exact object calibrated in Section~\ref{sec:quant}, which instead uses the one-shot specialization $D_A(\pi_A)=\sum_i d_i e^{-\lambda_i^*(\pi_A)\tau}$. A closed-form expression for
$E[G_t]$ under a simple seasonal profile and a locally constant $\bar\lambda_t^*$ is derived below.

\subsubsection*{Intermediary Price Impact}

We give the intermediary supply relation a short-window inventory
interpretation. An intermediary enters the trading window with allowance
inventory $I_{t^-}$, has target inventory $\bar I_t$, and assigns continuation
value $v_t$ to each allowance retained after the window. If it supplies
$M_t^s$ allowances to operators, its post-trade inventory is
$I_t=I_{t^-}-M_t^s$. Normalizing the short window to unit length, intermediary
supply solves
\begin{align}
\max_{M_t^s}\quad
(P_t-v_t)M_t^s
-\frac{\phi}{2}\left(I_{t^-}-M_t^s-\bar I_t\right)^2,
\label{eq:ap_intermediary_problem}
\end{align}
where $\phi>0$ indexes the cost of deviating from target inventory and hence
limited intermediary capacity. The objective is strictly concave, and its
first-order condition is
\begin{align}
P_t-v_t=\phi\left(M_t^s+\bar I_t-I_{t^-}\right).
\label{eq:ap_intermediary_foc}
\end{align}
We use a short-window, fast-reset benchmark in which the intermediary enters
at its target, $I_{t^-}=\bar I_t$. Over the short surrender window, the
continuation value of a retained allowance is pinned by the intermediary's
ability to transact at the benchmark outside the congested window. Because
the window is short relative to the compliance cycle, discounting and drift
in $P_t^*$ are second order; accordingly, we set $v_t=P_t^*(X_t)$ in this
local benchmark. Equation~\eqref{eq:ap_intermediary_foc} then becomes
$P_t-P_t^*(X_t)=\phi M_t^s$. Market clearing at operator demand
$M_t^s=M_t$ yields the equilibrium pricing relation
\begin{align}
P_t = P_t^*(X_t) + \phi\, M_t,
\label{eq:ap_price}
\end{align}
which is equation~\eqref{eq:ct_price} of the main text. Operators and
intermediaries transact at the same actual price $P_t$; $P_t^*(X_t)$ is a
counterfactual frictionless benchmark, not a second transaction price. The
intermediary does not choose operator order flow; rather, $\phi$ is the local
slope of the dealers' inverse supply curve, and \eqref{eq:ap_price} obtains by
clearing. The
decomposition $M_t=M_t^{\text{comp}}+M_t^{\text{info}}$ follows by
splitting operator flow into a compliance-related component and an
informational component. In the positive-shortfall buyer specification, $M_t^{\text{comp}}\geq 0$, while $M_t^{\text{info}}\in\mathbb{R}$ is signed. For the compliance component, under full reset upon access (that is, $\alpha_i=1$), the expected compliance order-flow rate is
$E_t[M_t^{\text{comp}}]=\sum_i \lambda_i^* g_{it}=\bar\lambda_t^* G_t$ stated in~\eqref{eq:ct_Mcomp} of the main text; $\bar\lambda_t^*$ is the gap-weighted access intensity and inherits the comparative statics of the underlying $\lambda_i^*$ in $S_i=d_i\pi_A$ and $\tilde{\psi}_i$.

\subsubsection*{Informed Demand}

Let $\mathcal{F}_t=\sigma(X_t)$ denote the public information set
generated by the public state $X_t$, and let
$\mathcal{I}_{it}\supseteq\mathcal{F}_t$ denote operator $i$'s
information set, including any private signal. Define the
informational advantage $\mu_{it}$ in expected-price-change units as in
\eqref{eq:ct_mu}. Upon access, operator $i\in\mathcal{I}$ submits a
speculative order $q_{it}^{\text{info}}$ that solves
$\max_{q}\{\mu_{it}q-\frac{\gamma_i}{2}q^2\}$, which yields the
closed-form rule
$q_{it}^{\text{info}}=\mu_{it}/\gamma_i$. Because access arrivals are
Poisson with chosen intensity $\lambda_i^*$, the expected contribution of
operator $i$ to the informative order-flow rate is $\lambda_i^*\,
\mu_{it}/\gamma_i$ (each access re-establishes the speculative position from zero, so the order equals the position). Aggregating in expected-rate form across informed
operators,
\begin{align}
M_t^{\text{info}}
= \sum_{i\in\mathcal{I}} \lambda_i^*\,
\frac{\mu_{it}}{\gamma_i},
\label{eq:ap_info}
\end{align}
which avoids conditioning on a time-varying active set of operators
and makes information incorporation a conditional expected-flow
object. Because access is intermittent, information is impounded into
prices gradually through the arrival of market-access opportunities,
and a higher chosen $\lambda_i^*$ implies faster and more informative flow.

\subsubsection*{Derivation of Prediction 1 (Slow Participation)}

From \eqref{eq:ap_gap_sol}, $E_{t_0}[g_{it}]$ is decreasing in $\lambda_i^*$
pointwise in $t$ when $g_{i,t_0}\geq 0$ and $a_i(\cdot)\geq 0$, strictly if either is strictly positive on a set of positive measure; in particular, a lower chosen intensity $\lambda_i^*$ then implies a larger expected compliance gap at every point in the compliance cycle. By \eqref{eq:ap_lambdaStar}, $\lambda_i^*$ is increasing in $S_i = d_i\pi_A$ and decreasing in $\tilde{\psi}_i$, so operators with high normalized attention or treasury costs $\tilde{\psi}_i$ or small expected shortfalls $d_i$ optimally choose lower access intensity. Because a realized trade at $i$
requires a Poisson access arrival, the probability that operator $i$
trades in any interval of length $\Delta$ equals
$1-e^{-\lambda_i^* \Delta}$, which is monotonically increasing in
$\lambda_i^*$. Consequently, operators with lower chosen $\lambda_i^*$ trade less
often and have larger unresolved compliance gaps. Because
$a_i(s_t,X_t)$ is larger for smaller $s_t$ (the surrender date is
approaching), conditional on eventual trading the expected share of
purchases realized near surrender is also larger for operators with
lower $\lambda_i^*$. This establishes Prediction 1. $\hfill\square$

\subsubsection*{Derivation of Prediction 2 (Seasonal Price Pressure)}

Combining \eqref{eq:ap_price} with the decomposition
$M_t=M_t^{\text{comp}}+M_t^{\text{info}}$, and using the full-reset access-weighted aggregation $E_t[M_t^{\text{comp}}]=\bar\lambda_t^* G_t$, where $\bar\lambda_t^* \equiv \sum_i \lambda_i^* g_{it}/G_t$ is the gap-weighted chosen access intensity,
\begin{align}
E_t[P_t - P_t^*(X_t)] = \phi\,\bar\lambda_t^*\, G_t + \phi\, E_t[M_t^{\text{info}}].
\label{eq:ap_pressure}
\end{align}
Equation \eqref{eq:ap_G} implies that $E_{t_0}[G_t]$ is a weighted integral of the past path of $A$, increasing in $t$ whenever $A$ is nondecreasing within the cycle and $G$ starts below $A/\bar\lambda^*$; it is therefore largest just before surrender. Hence $E_t[P_t - P_t^*(X_t)]$ is strictly positive near surrender whenever $\phi>0$, $\bar\lambda_t^*>0$, $G_t>0$, and $E_t[M_t^{\text{info}}]\geq 0$. Comparative statics in $\varphi$ are immediate from the fixed-point relation \eqref{eq:ap_eqm}. Comparative statics in access costs follow from \eqref{eq:ap_lambdaStar}: higher normalized access costs $\tilde{\psi}_i$ lower each $\lambda_i^*$, raise expected residual terminal demand $D_A$, and---through the equilibrium fixed point \eqref{eq:ap_eqm}---raise $\pi_A$.
Comparative statics in scarcity and the allowance bank follow from
the dependence of $A(s_t,X_t)$ on $z_t$ and $B_t$: specifically,
tighter cap ($z_t$ high) or a smaller bank ($B_t$ low) raises desired
compliance demand $\sum_i a_i(s_t,X_t)$, and hence raises $G_t$ near
surrender. This establishes Prediction 2. $\hfill\square$

\subsubsection*{Derivation of Prediction 3 (Informative Operator Flow)}

Under the maintained positive linear signal-loading condition in
\eqref{eq:ap_signal_slope}, informative operator flow forecasts future
benchmark-price innovations. The expected contribution of
operator $i$ to $M_t^{\text{info}}$ is $\lambda_i^*\mu_{it}/\gamma_i$,
so that informative order flow generated by operator $i$ scales
linearly with its chosen access intensity. Holding signal quality and $\gamma_i$ fixed, higher $\lambda_i^*$ generates more informative flow. Here the intensity is determined by compliance savings $S_i=d_i\pi_A$, not by speculative gains.
Cross-sectional comparative statics in $\gamma_i$ follow from
$q_{it}^{\text{info}}=\mu_{it}/\gamma_i$. This establishes the benchmark-price-innovation form of Prediction 3; the actual-return extension follows from \eqref{eq:ap_actual_price}--\eqref{eq:ap_actual_return} below under the maintained reduced-form condition that contemporaneous price impact under-reveals the private signal. $\hfill\square$

The empirical tests use actual allowance returns, not benchmark-price innovations. Actual prices satisfy
\begin{align}
P_t = P_t^*(X_t) + \phi M_t,
\label{eq:ap_actual_price}
\end{align}
so that
\begin{align}
P_{t+\Delta} - P_t = \left(P_{t+\Delta}^* - P_t^*\right) + \phi\left(M_{t+\Delta} - M_t\right).
\label{eq:ap_actual_return}
\end{align}
The reduced-form return-predictability implication requires that the signal content of $M_t^{\text{info}}$ is not fully incorporated into the contemporaneous price. In particular, if
\begin{align}
E_t\!\left[P_{t+\Delta}^* - P_t^* \mid \mathcal{F}_t, M_t^{\text{info}}\right] - E_t\!\left[P_{t+\Delta}^* - P_t^* \mid \mathcal{F}_t\right] = \beta\, M_t^{\text{info}},
\label{eq:ap_signal_slope}
\end{align}
and the expected reversal of contemporaneous price impact is smaller than the signal-revelation component, then $E_t[P_{t+\Delta}-P_t \mid M_t^{\text{info}}]$ is increasing in $M_t^{\text{info}}$. In the special case in which neither $E_t[M_{t+\Delta}\mid M_t^{\text{info}}]$ nor $E_t[M_t^{\text{comp}}\mid M_t^{\text{info}}]$ loads on $M_t^{\text{info}}$, the condition for positive predictability of actual returns is exactly $\beta>\phi$; more generally, the required condition is that contemporaneous price impact under-reveals the private signal relative to the subsequent adjustment in the benchmark price. We do not derive this condition from a market-maker filtering or Kyle-style equilibrium; we treat it as a maintained reduced-form assumption. Positivity of $\beta$ itself is not an additional assumption: by the definition of $\mu_{it}$ and iterated expectations, $\mathrm{Cov}\big(P_{t+\Delta}^*-P_t^*-E_t[P_{t+\Delta}^*-P_t^*\mid\mathcal{F}_t],\,M_t^{\text{info}}\big)=\sum_{i\in\mathcal{I}}\lambda_i^*E[\mu_{it}^2]/\gamma_i>0$ when the private components are cross-sectionally orthogonal; only $\beta>\phi$ is maintained.

The frequent-trader split used in the empirical section should therefore also be read in reduced-form terms. In a richer extension, the value of market access could include both compliance and speculative components, for example $S_i=d_i\pi_A+V_i^{\mathrm{info}}$ in return units, so that frequent trading could reflect lower access frictions, higher informational value of access, or both. We do not solve or estimate that extension here; the access intensities and fixed point in the paper depend only on compliance savings.

\subsubsection*{April Premium Paid by Delayed Buyers and Real Frictions}

This subsection clarifies how the empirical measure of the April premium paid by delayed buyers differs from aggregate resource costs. The empirical calculation measures a financial premium paid in April by delayed buyers; aggregate resource costs depend on the underlying participation, adjustment, and immediacy-provision frictions.

A simple quadratic benchmark illustrates the resource-cost components associated with deviations from the frictionless allocation. We do not evaluate this criterion in the calibration or solve for an optimal cap:
\begin{align}
\mathcal{W}
= E_0\int_0^{\infty} e^{-\rho t}
\Bigg[
- \sum_i \frac{1}{2}(\theta_{it}-x_{it})^2
- \sum_i \frac{\psi_i}{2}\lambda_i^2
- \frac{\phi}{2}M_t^2
\Bigg]\, dt,
\label{eq:ap_welfare}
\end{align}
where $\rho>0$ is a continuous-time discount rate. Relative to the frictionless allocation, the resource-cost components are (i) misallocation of abatement across firms and over time, (ii) real participation or attention costs, and (iii) intermediary inventory-deviation costs. The access-cost term $\sum_i \tfrac{\psi_i}{2}\lambda_i^2$ in Equation~\eqref{eq:ap_welfare} should be read as a continuous-time analogue of the per-cycle access costs used in the main model: in the quantitative implementation, access costs are imposed at the compliance-cycle level rather than as continuously paid flow costs. Under the at-target short-window benchmark, supplying $M_t$ leaves the intermediary with inventory deviation $-M_t$, so the inventory cost is exactly $\frac{\phi}{2}M_t^2$ and Equation~\eqref{eq:ap_welfare} is unchanged. The gross payment wedge $(P_t-P_t^*)M_t=\phi M_t^2$ is not itself an aggregate resource cost: under the inventory interpretation, only the quadratic inventory cost is real and the remainder is an intermediary rent or transfer. If $\phi M_t$ is instead interpreted as pure transitory price impact with no underlying inventory or risk-bearing cost, neither the payment wedge nor the quadratic term should be counted as an aggregate resource cost. The April premium paid by delayed buyers is measurable, but it is not a direct measure of aggregate resource costs because allowance payments are partly transfers across market participants.

\subsubsection*{Closed-Form Seasonal Compliance Gap and Price Premium}

This subsection displays a simple closed-form object that illustrates
Prediction 2. Let $u\in[0,T]$ denote time within the compliance cycle,
with surrender at $u=T$, and consider the piecewise-constant
aggregate desired-compliance profile
\begin{align}
A(u) =
\begin{cases}
a_0, & 0 \leq u < T-\tau_r, \\
a_0 + a_1, & T-\tau_r \leq u \leq T,
\end{cases}
\label{eq:ap_Aseason}
\end{align}
with $a_1>0$ capturing the run-up in desired compliance adjustment
near surrender and $\tau_r\in(0,T)$ the length of the surrender run-up.
Set the aggregate gap dynamics to
\begin{align}
\frac{dG(u)}{du} = A(u) - \bar\lambda\, G(u),
\label{eq:ap_Geq}
\end{align}
where $\bar\lambda$ is treated as a constant proxy for the endogenous access-adjustment rate $\bar\lambda_t^*$ for transparency in this displayed example,
and, for expositional simplicity, take $G(0)=0$. Equation
\eqref{eq:ap_Geq} is a linear first-order ODE with piecewise-constant
forcing, and its solution is
\begin{align}
G(u) = \frac{a_0}{\bar\lambda}\left(1 - e^{-\bar\lambda u}\right),
\qquad 0 \leq u < T-\tau_r,
\label{eq:ap_Gsol1}
\end{align}
and, for $T-\tau_r \leq u \leq T$,
\begin{align}
G(u) = G(T-\tau_r)\, e^{-\bar\lambda(u-(T-\tau_r))}
+ \frac{a_0+a_1}{\bar\lambda}
\left(1 - e^{-\bar\lambda(u-(T-\tau_r))}\right),
\label{eq:ap_Gsol2}
\end{align}
with $G(T-\tau_r)=\frac{a_0}{\bar\lambda}(1-e^{-\bar\lambda(T-\tau_r)})$.
Combining \eqref{eq:ap_price} with $M_t^{\text{comp}}=\bar\lambda G_t$ and
abstracting from $M_t^{\text{info}}$ in expectation delivers the
seasonal-premium representation
\begin{align}
P(u) - P^*(u) = \phi\, \bar\lambda\, G(u),
\label{eq:ap_Pdiff}
\end{align}
so that the price premium inherits the piecewise-closed-form structure
of $G(u)$.

Equations~\eqref{eq:ap_Gsol1}--\eqref{eq:ap_Pdiff} describe the pre-surrender accumulation of unresolved compliance demand within the cycle. Because surrender is mandatory, the remaining positive gap at $u=T^{-}$ is converted into terminal compliance order flow:
\begin{align}
M_T^{\text{term}} = \sum_i g_{iT-}^{+}.
\label{eq:ap_Mterm}
\end{align}
Under the same reduced-form immediacy-supply relation \eqref{eq:ap_price}, and treating the surrender date $T$ (the date $A$ of the main text, with $h_T$ in the role of $h_A$) as a windowed event of length $h_T$ over which $M_T^{\text{term}}$ and $M_T^{\text{info}}$ are stock-equivalent integrated flows,
\begin{align}
P_T - P_T^*(X_T) = \frac{\phi}{h_T}\left(M_T^{\text{term}} + M_T^{\text{info}}\right).
\label{eq:ap_PTterm}
\end{align}
Thus, the closed-form expression for $G(T^{-})$ provides the model counterpart of the surrender-month demand pressure documented in the data; the relevant object for terminal price pressure is the pre-compliance gap $G(T^{-})$, not the post-compliance value $G(T)$. Equations~\eqref{eq:ap_Gsol1}--\eqref{eq:ap_Pdiff}
illustrate Prediction 2: the expected equilibrium premium is largest near
surrender because aggregate unresolved compliance demand accumulates
most rapidly when $A(u)=a_0+a_1$, and the premium is linear in $\phi$: before the run-up it equals $\phi a_0(1-e^{-\bar\lambda u})$, and on the run-up it is bounded above by $\phi(a_0+a_1)$. For given $\phi$, the within-cycle flow premium $\phi\bar\lambda G(u)$ is increasing in $\bar\lambda$ because faster closing generates more contemporaneous order flow, whereas the terminal premium $\varphi D_A$ of Proposition~\ref{prop:endog_access} is decreasing in access intensities because it depends on residual demand.

An analogous
smoother alternative is the linear ramp
$A(u)=\alpha_0+\alpha_1(1-s_u/s_0)$ with $\alpha_1>0$ as surrender
approaches, which delivers a qualitatively identical pattern.

\subsection{Quantitative Robustness for Section~\ref{sec:quant}\label{subsec:quant_appendix}}

This appendix documents the construction details, sample-period harmonization, shortfall robustness, premium definitions, and price-impact diagnostics underlying Section~\ref{sec:quant}.

\subsubsection*{Sample harmonization}

Table~\ref{tab:quant_sample_comparison} reports operator-year counts
(one observation per operator and compliance cycle) for the headline
2009--2022 headline sample and for two robustness samples: one adding the
non-bankable Phase~I pilot cycles (2006--2022) and one restricted to cycles whose obligations fall in
Phases~III and IV.

In the headline sample, access intensities are estimated using all 95{,}526 operator-cycles; the reported distribution weights cycles equally rather than giving each operator equal weight. The no-trade, trade-count, and April-volume validation moments restrict to 90{,}079 cycles with compliance records (10{,}241 operators). Persistence uses 78{,}903 consecutive eligible cycle pairs, with 38{,}835 no-trade origin cycles defining the conditional probability. Annual terminal-demand comparisons aggregate the full panel using observed shortfalls and flows.

\begin{table}[H]
\centering
\caption{Sample harmonization for the quantitative analysis\label{tab:quant_sample_comparison}}
\begin{tabular}{p{3.0cm}p{1.8cm}rrp{4.8cm}}
\toprule
Sample & Years & N operator-years & N operators & Notes \\
\midrule
Main sample & 2009--2022 & 95{,}526 & 10{,}707 & Baseline; bankable allowances \\
Including Phase I & 2006--2022 & 112{,}127 & 10{,}859 & Adds the non-bankable pilot phase \\
Phase III onward & 2014--2022 & 62{,}977 & 9{,}416 & Mature market robustness \\
\bottomrule
\end{tabular}
\end{table}

\subsubsection*{Shortfall measures}

Table~\ref{tab:quant_shortfall} reports four alternative shortfall constructions. The allocation baseline $d^{\mathrm{alloc}}_{it}$ is the headline measure used in Section~\ref{sec:quant}; the lagged-bank measure is the preferred bank-adjusted robustness; cumulative-surplus and lagged-surplus measures are simpler fallbacks. The allocation, lagged-bank, and lagged-surplus constructions move within a band of approximately seven percent, while the cumulative-surplus construction, which nets surpluses accumulated over all prior cycles of a phase, is about a quarter lower; the share of operator-cycles with $d_{it}>0$ is meaningfully smaller under the bank-adjusted measures. Access costs are recovered only when $d_{it}>0$, $\hat\lambda_i>0$, and the nonnegative calibration premium is strictly positive. Under the allocation baseline, 37{,}910 operator-cycles for 7{,}367 operators satisfy these restrictions; zero-shortfall cycles are omitted rather than assigned a zero access cost. Cycle-level cost indices are winsorized at the pooled 1st and 99th percentiles within each sample before computing firm medians. In the headline sample, these cutoffs are 2.54 and 193{,}039, affecting 380 observations in each tail. Counterfactuals then retain the 44{,}872 positive-shortfall cycles belonging to operators with a recovered firm median, including cycles whose own observed premium is nonpositive.

\begin{table}[H]
\centering
\caption{Headline shortfall measures\label{tab:quant_shortfall}}
\small
\begin{tabular}{l r r r r l}
\toprule
Shortfall measure & Mean $d_{it}$ & Median $d_{it}$ & Share $d_{it}>0$ & Aggregate $d_{it}$ & Notes \\
\midrule
Allocation baseline & 115{,}033 & 1 & 50.0\% & 10{,}362{,}081{,}802 & Main \\
Lagged bank & 110{,}391 & 0 & 34.9\% & 9{,}943{,}889{,}025 & Preferred robustness \\
Cumulative surplus & 105{,}792 & 0 & 29.5\% & 7{,}718{,}871{,}939 & Fallback robustness \\
Lagged surplus & 112{,}851 & 0 & 46.2\% & 10{,}165{,}518{,}472 & Simple fallback \\
\bottomrule
\end{tabular}
\end{table}

\subsubsection*{April premium definitions}

Table~\ref{tab:quant_premium_summary} summarizes the four premium measures used to discipline the surrender-window return-impact calibration. The headline model-aligned object is the nonnegative April premium $\max\{r_{A,t},0\}$; the non-April-adjusted measure subtracts the average non-April monthly return; the residualized measure is the residual from a return regression on lagged returns, lagged volatility, and phase indicators; and the signed April return retains negative realizations. The unconditional signed mean is 0.069, with standard error 0.044 and a 95\% $t$-interval of $[-0.025,\,0.163]$. The nonnegative calibration object's mean is 0.103, with standard error 0.020 and a corresponding interval of $[0.059,\,0.147]$. Over the seventeen cycles including the Phase~I pilot, the signed mean is $-0.033$, because the Phase~I allowance price collapsed in April 2006 and April 2007, and the nonnegative mean is 0.089. These wide intervals motivate treating $\hat\varphi$ as a calibrated sensitivity rather than a tightly estimated structural parameter.

\begin{table}[H]
\centering
\caption{Headline April premium definitions\label{tab:quant_premium_summary}}
\begin{tabular}{l r r r r r}
\toprule
Premium measure & Mean & Median & Share positive & Std.\ dev.\ & N years \\
\midrule
Nonnegative April & 0.103 & 0.100 & 78.6\% & 0.076 & 14 \\
Non-April adjusted & 0.065 & 0.084 & 78.6\% & 0.174 & 14 \\
Residualized & 0.057 & 0.062 & 85.7\% & 0.157 & 14 \\
Signed April return & 0.069 & 0.100 & 78.6\% & 0.163 & 14 \\
\bottomrule
\end{tabular}
\end{table}

\subsubsection*{Return-impact calibration}

Table~\ref{tab:quant_phi} reports headline $\hat{\varphi}$ across premium definitions and estimation samples. The positive-only rows report the median annual premium-to-demand ratio among positive-premium years; the all-year OLS rows estimate $\pi_{A,t} = \varphi\,D^{\mathrm{model}}_{A,t}+\varepsilon_t$ without an intercept on all available years; the NNLS rows additionally impose $\varphi\geq 0$. The baseline counterfactuals use the nonnegative-April NNLS estimate, $4.06\times10^{-10}$, which equals OLS; positive-year medians provide sensitivity comparisons. Order-of-magnitude estimates are stable across specifications. The reported squared correlations between observed and fitted premia are small, indicating weak year-to-year co-movement; they are not regression $R^2$s.

\begin{table}[H]
\centering
\caption{Headline return-impact calibration across premium definitions\label{tab:quant_phi}}
\small
\resizebox{\textwidth}{!}{%
\begin{tabular}{l l r l r r}
\toprule
Premium measure & Sample & $\hat{\varphi}$ & 80\% CI & Corr$^2$ & N years \\
\midrule
Nonnegative April & Positive only & $5.65\times 10^{-10}$ &  &  & 11 \\
Nonnegative April & All years OLS & $4.06\times 10^{-10}$ & $[3.25,\,5.38]\times 10^{-10}$ & 0.0001 & 14 \\
Nonnegative April & All years NNLS & $4.06\times 10^{-10}$ & $[3.25,\,5.38]\times 10^{-10}$ & 0.0001 & 14 \\
Non-April adjusted & Positive only & $6.14\times 10^{-10}$ &  &  & 11 \\
Non-April adjusted & All years OLS & $2.79\times 10^{-10}$ & $[0.90,\,4.57]\times 10^{-10}$ & 0.0062 & 14 \\
Non-April adjusted & All years NNLS & $2.79\times 10^{-10}$ & $[0.90,\,4.57]\times 10^{-10}$ & 0.0062 & 14 \\
Residualized & Positive only & $4.14\times 10^{-10}$ &  &  & 12 \\
Residualized & All years OLS & $2.41\times 10^{-10}$ & $[0.94,\,3.96]\times 10^{-10}$ & 0.0032 & 14 \\
Residualized & All years NNLS & $2.41\times 10^{-10}$ & $[0.94,\,3.96]\times 10^{-10}$ & 0.0032 & 14 \\
\bottomrule
\end{tabular}
}
\par
\medskip
\begin{minipage}{0.95\textwidth}
\footnotesize
\emph{Notes:} All rows use the headline 2009--2022 sample and the allocation shortfall proxy $d^{\mathrm{alloc}}_{it}$. Positive-only rows report the median ratio of premium to demand in positive-premium years. OLS and NNLS use all years and no intercept; NNLS imposes $\varphi\geq0$. The nonnegative-April NNLS estimate is the baseline counterfactual calibration. The 80\% percentile intervals use the 10th and 90th percentiles of 250 bootstrap replications resampling years. Corr$^2$ is the squared Pearson correlation between observed and fitted premia, not the $R^2$ of the through-origin regression.
\end{minipage}
\end{table}

\begin{table}[H]
\centering
\caption{Leave-one-out robustness for the headline $\hat{\varphi}$\label{tab:quant_phi_loo}}
\begin{tabular}{l r r r r}
\toprule
Specification & Median LOO $\hat{\varphi}$ & Min & Max & Median Corr$^2$ \\
\midrule
Nonnegative April, OLS & $4.04\times 10^{-10}$ & $3.78\times 10^{-10}$ & $4.57\times 10^{-10}$ & 0.0042 \\
Nonnegative April, NNLS & $4.04\times 10^{-10}$ & $3.78\times 10^{-10}$ & $4.57\times 10^{-10}$ & 0.0042 \\
\bottomrule
\end{tabular}
\par
\medskip
\begin{minipage}{0.93\textwidth}
\footnotesize
\emph{Notes:} The leave-one-out exercise drops one calendar year at a time and re-estimates the headline $\varphi$ specification on the remaining 13 years. We report the median, minimum, and maximum across the 14 leave-one-out runs. The last column is the median squared Pearson correlation between observed and fitted premia, not regression $R^2$.
\end{minipage}
\end{table}

\subsection{Phase-Final Surrenders and the Within-Phase Borrowing Option\label{subsec:phase_final_borrowing}}

The quantitative model treats each compliance cycle symmetrically. One institutional feature breaks that symmetry. Within a trading phase, the free allocation for year $t$ is credited to operator accounts by the end of February of year $t$ and can be surrendered against the obligation for year $t-1$ that falls due in April $t$, so an operator can cover part of an interior obligation with allowances it has just received rather than with market purchases. At the last surrender of a phase (April 2008, April 2013, and April 2021), the allowances of the following phase cannot be used for the previous phase's obligation, and in April 2013 the 2013 allocations had in any case not yet been issued. \citet{EllermanTrotignon2009} first documented this within-phase borrowing in the Phase I compliance data. Because the model's terminal demand $D_A$ is built from the accounting shortfall of the obligation settled in each April, it does not distinguish interior from phase-final Aprils. Table~\ref{tab:phase_final_borrowing} asks whether the data do.

Panel~A regresses aggregate April purchases on model-implied terminal demand across the seventeen 2006--2022 cycles. Adding an indicator for the three phase-final Aprils raises the $R^2$ from $0.74$ to $0.88$; April purchases in those years exceed the level implied by the model by about 175 million allowances, and the slope on model-implied demand rises only slightly, from 1.45 to 1.61. The same holds with the aggregate allocation shortfall in place of model demand. Panel~B shows that the excess is not concentrated among operators with a high predicted probability of no pre-surrender access: the coefficient on the predicted no-access probability in the incidence specification of Table~\ref{tab:incidence_terminal_demand} is unaffected by the phase-final indicator or its interactions, so the incidence result does not depend on these three deadlines. The excess is instead concentrated among operators whose shortfall is small relative to their surrender obligation, that is, those whose free allocation covers most of the obligation and who therefore have the most scope to cover an interior obligation with the next allocation: in phase-final Aprils their April purchases rise by about 30\% more than those of high-ratio operators with the same shortfall, whereas the average level effect across all positive-shortfall operators is small and imprecisely estimated (a coefficient of 0.13 with a $t$-statistic of 1.5). Panel~C makes the same point without regression structure. Among positive-shortfall operators with a low shortfall ratio, the share making net April purchases roughly doubles at the phase-final April relative to interior Aprils in Phases~III and IV (from 18\% to 41\% and from 19\% to 43\%), and aggregate April purchases relative to the shortfall rise by between one half and more than double in all three phases (from 0.72 to 1.11, 0.45 to 1.00, and 0.39 to 0.90); among high-ratio operators neither measure moves systematically.

We read this as evidence that the borrowing option shifts part of the deadline demand of well-allocated operators forward within a phase and concentrates it at the phase-final surrender, consistent with the paper's broader theme that the timing of predictable compliance demand is shaped by the regulatory calendar. We do not build the option into the model: the three events are confounded (April 2008 settled Phase~I obligations with allowances that were close to worthless; April 2013 coincided with the delayed 2013 allocation and a large negative April return; April 2021 was the first Phase~IV surrender), and the largest single deviation in Figure~\ref{fig:quant_terminal_demand}, April 2013, is only partly accounted for by the indicator. The exercise is a diagnostic of what the accounting shortfall misses, not a structural extension.

\begin{table}[tbp]
\centering
\caption{Phase-Final Surrenders and the Within-Phase Borrowing Option\label{tab:phase_final_borrowing}}
\footnotesize
\begin{tabular}{lcccc}
\toprule
\multicolumn{5}{l}{Panel A: Aggregate April purchases (millions of allowances), 2006--2022 cycles} \\
 & (1) & (2) & (3) & (4) \\
\midrule
Model-implied terminal demand & 1.452*** & 1.613*** &  &  \\
 & (8.78) & (12.03) &  &  \\
Aggregate allocation shortfall &  &  & 0.393*** & 0.438*** \\
 &  &  & (7.36) & (10.97) \\
Phase-final April &  & 174.7*** &  & 175.6** \\
 &  & (2.81) &  & (2.50) \\
$R^2$ & 0.742 & 0.878 & 0.717 & 0.854 \\
Years & 17 & 17 & 17 & 17 \\
\midrule
\multicolumn{5}{l}{Panel B: Operator-cycle PPML, positive-shortfall operators} \\
 & (1) & (2) & (3) & \\
\midrule
Log shortfall & 0.587 & 0.602 & 0.432 & \\
 & (6.42) & (7.36) & (3.65) & \\
Log predicted no-access probability & 0.259 & 0.252 & 0.257 & \\
 & (3.90) & (3.84) & (3.87) & \\
Phase-final April &  & 0.125 &  & \\
 &  & (1.51) &  & \\
Low shortfall ratio &  &  & -0.383 & \\
 &  &  & (-2.45) & \\
Phase-final $\times$ low shortfall ratio &  &  & 0.276 & \\
 &  &  & (2.14) & \\
Fixed effects & Year, country & Phase, country & Year, country & \\
Observations & 41,171 & 41,171 & 41,171 & \\
\midrule
\multicolumn{5}{l}{Panel C: April buying by positive-shortfall operators, interior versus phase-final Aprils} \\
 & \multicolumn{2}{c}{Low shortfall ratio} & \multicolumn{2}{c}{High shortfall ratio} \\
\cmidrule(lr){2-3}\cmidrule(lr){4-5}
 & Interior & Phase-final & Interior & Phase-final \\
\midrule
\multicolumn{5}{l}{\emph{Share with net April purchases}} \\
\quad Phase II (April 2008) & 0.19 & 0.25 & 0.37 & 0.26 \\
\quad Phase III (April 2013) & 0.18 & 0.41 & 0.37 & 0.38 \\
\quad Phase IV (April 2021) & 0.19 & 0.43 & 0.35 & 0.49 \\
\multicolumn{5}{l}{\emph{April purchases / shortfall}} \\
\quad Phase II (April 2008) & 0.72 & 1.11 & 0.57 & 0.60 \\
\quad Phase III (April 2013) & 0.45 & 1.00 & 0.48 & 0.66 \\
\quad Phase IV (April 2021) & 0.39 & 0.90 & 0.54 & 0.53 \\
\bottomrule
\end{tabular}
\par
\smallskip
\begin{minipage}{0.95\textwidth}
\footnotesize
\emph{Notes:} Phase-final Aprils are the last surrenders of Phases I--III (April 2008, 2013, and 2021), when the following year's free allocation cannot be surrendered against the current obligation. Panel A: OLS of aggregate April positive net purchases on model-implied terminal demand (Section~\ref{sec:quant}) or on the aggregate allocation shortfall of the obligation settled in that April, with and without a phase-final indicator; heteroskedasticity-robust $t$-statistics in parentheses. Panel B: the PPML specification of Table~\ref{tab:incidence_terminal_demand} augmented with the phase-final indicator and its interaction with an indicator for a shortfall below the median ratio of shortfall to surrender obligation; $t$-statistics two-way clustered by operator and year. Panel C: share of positive-shortfall operator-cycles with positive net April purchases, and the ratio of aggregate April purchases to the aggregate shortfall, by shortfall-ratio group; the phase label is the trading phase in which the April falls. April 2008 settled Phase~I obligations with allowances that were close to worthless, and April 2013 coincided with the delayed 2013 allocation and a large negative April return. *, **, and *** denote significance at the 10\%, 5\%, and 1\% levels in Panel A.
\end{minipage}
\end{table}

\subsection{Two-Period Model Derivations}

This subsection provides the derivations underlying the simple
model introduced in Section \ref{sec:model}.

\subsubsection*{Frictionless Benchmark and Intermediation}

We first characterize the frictionless benchmark, $\phi=0$, in which the
transaction price coincides with the benchmark price $p_t^*$; this benchmark
defines $p_t^*$. Consider the two-period problem of firm $i$:
\begin{align}
\max_{x_{i0},x_{i1},m_{i0},m_{i1},b_{i1}}
\left[
-\frac{1}{2}\left(\theta_{i0}-x_{i0}\right)^2-p_0^* m_{i0}
\right]
\notag\\
+\delta E_0\left[
-\frac{1}{2}\left(\theta_{i1}-x_{i1}\right)^2-p_1^* m_{i1}
\right]
\end{align}
subject to
\begin{align}
b_{i1}=b_{i0}+q_{i0}+m_{i0}-x_{i0}, \qquad
x_{i1}=b_{i1}+q_{i1}+m_{i1}, \qquad b_{i1}\geq 0.
\end{align}
Figure~\ref{fig:model_timeline} in the main text summarizes the timing
of the two-period model and the role of intermediation.

Let $\xi_{i0}$ and $\xi_{i1}$ denote the shadow values of one
additional allowance in periods 0 and 1. The first-order conditions
for an interior solution are
\begin{align}
\theta_{i0}-x_{i0} &= \xi_{i0}, \\
\theta_{i1}-x_{i1} &= \xi_{i1}, \\
p_0^* &= \xi_{i0}, \\
p_1^* &= \xi_{i1}, \\
\xi_{i0} &= \delta E_0[\xi_{i1}].
\end{align}
Thus, in the absence of frictions, firms trade until marginal
abatement costs are equalized across firms and over time, and the
frictionless price satisfies $p_0^*=\delta E_0[p_1^*]$.

With limited intermediation, $\phi>0$, operators and intermediaries
transact at a common actual price $p_t$, which operators take as given.
The operator's static conditions are those above with $p_t$ in place of $p_t^*$:
they read $\theta_{it}-x_{it}=\xi_{it}$ and
$\xi_{it}=p_t$: operators equate marginal abatement cost to the price
they actually pay, and $p_t^*$ enters their problem only through its
effect on $p_t$. The benchmark price $p_t^*$ enters the intermediaries'
problem as the continuation value of a retained allowance, as follows.
Let $M_t=\sum_i m_{it}$ denote aggregate operator net order flow. An intermediary enters the short trading window with allowance inventory $I_{t^-}$, has target inventory $\bar I_t$, and assigns continuation value $v_t$ to a retained allowance. If it supplies $M_t^s$ allowances to operators, its post-trade inventory is $I_t=I_{t^-}-M_t^s$. Normalizing the short window to unit length, it solves
\begin{align}
\max_{M_t^s}\quad
(p_t-v_t)M_t^s
-\frac{\phi}{2}\left(I_{t^-}-M_t^s-\bar I_t\right)^2,
\label{eq:ap_tp_intermediary_problem}
\end{align}
where $\phi>0$ indexes the cost of deviating from target inventory. The objective is strictly concave, and its first-order condition is
\begin{align}
p_t-v_t=\phi\left(M_t^s+\bar I_t-I_{t^-}\right).
\label{eq:ap_tp_intermediary_foc}
\end{align}
We use the same short-window, fast-reset benchmark as in the continuous-time model: the intermediary enters at its target, $I_{t^-}=\bar I_t$, and retained allowances have continuation value $v_t=p_t^*$. The latter equality follows because the intermediary can transact at the benchmark outside the congested window; discounting and drift in the benchmark are second order over the short interval. The first-order condition therefore becomes $p_t-p_t^*=\phi M_t^s$. Market clearing at operator demand $M_t^s=M_t$ then yields the equilibrium pricing relation
\begin{align}
p_t=p_t^*+\phi M_t.
\end{align}
Operators and intermediaries transact at the same actual price $p_t$; $p_t^*$ is a counterfactual frictionless benchmark, not a second transaction price. The intermediary does not choose operator order flow; rather, $\phi$ traces the local slope of the dealers' inverse supply curve, and the equilibrium condition $p_t=p_t^*+\phi M_t$ obtains by clearing.
The two sets of conditions therefore play distinct roles: the
frictionless first-order conditions above define the benchmark price
$p_t^*$ and its no-arbitrage property $p_0^*=\delta E_0[p_1^*]$, while
the price-impact relation $p_t=p_t^*+\phi M_t$ describes how the common
transaction price departs from that benchmark when $\phi>0$, with
operators optimizing at $p_t$. Extending
the model to many periods is straightforward and yields the same Euler
equation between adjacent dates, together with a standard
transversality condition ruling out unbounded accumulation of unused
allowances.

\subsubsection*{Participation Costs and Surrender-Month Trading}

To derive the participation threshold, consider a firm expecting to be
short by $d_i$ allowances at the surrender date. If the firm becomes
active at the early trading date, it pays $p_e d_i + K_i$. If instead
it delays until April, its expected cost is $E_0[p_A] d_i$. Early
trading is therefore optimal if and only if
\begin{align}
p_e d_i + K_i \leq E_0[p_A] d_i,
\end{align}
which is equivalent to
\begin{align}
d_i\left(E_0[p_A]-p_e\right)\geq K_i.
\end{align}
At each within-cycle date $\ell\in\{e,A\}$, intermediary absorption
costs imply
\begin{align}
p_{\ell}=p_{\ell}^*+\phi M_{\ell}.
\end{align}
Hence, if delayed compliance demand is concentrated at the surrender
date, then $M_A$ is large and the April price can exceed the early
price even when the compliance wave is predictable.

\subsubsection*{Private Information and Speculative Trading}

\begin{figure}[tbp]
\caption{Within-Period Timeline with Private Information\label{fig:model_within_period}}

\medskip{}

{\footnotesize~This figure illustrates the within-period timeline from the early
trading date $e$ to the surrender month $A$ within a compliance
cycle. It highlights the arrival of private information, the
end-March verification date, the surrender date, intermediary
absorption of order imbalances, and the subsequent realization of
returns. Delayed operator demand raises the order imbalance absorbed
by intermediaries at the surrender date and can generate an April
price spike.}

\medskip{}

\centering{}\includegraphics[width=0.9\textwidth]{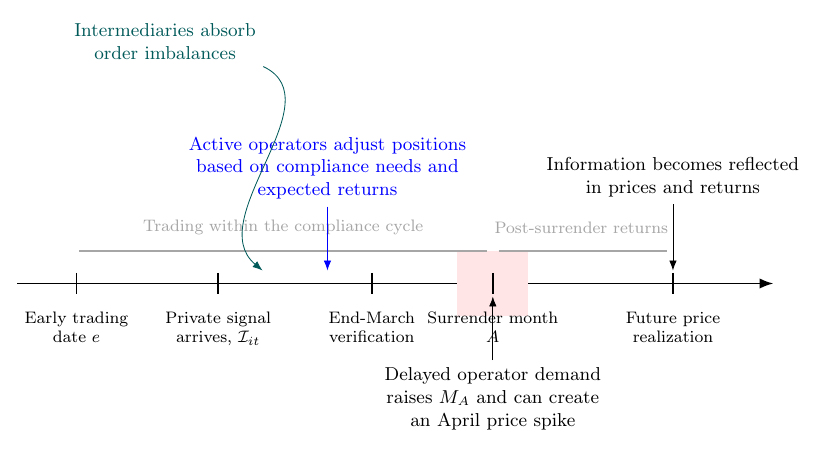}
\end{figure}

Suppose an active firm receives a private signal summarized by the
information set $\mathcal{I}_{it}$. Let $\mathcal{F}_t$ denote the
public information set. Consistent with the main text, define
\begin{align}
\mu_{it}
= E_t\!\left[p_{t+1}^*-p_t^* \mid \mathcal{I}_{it}\right]
- E_t\!\left[p_{t+1}^*-p_t^* \mid \mathcal{F}_t\right].
\end{align}
Figure~\ref{fig:model_within_period} summarizes the within-period
timeline between the early trading date $e$ and the surrender month
$A$, including private-signal arrival, verification, surrender, and
subsequent return realization. This timing is useful for interpreting
both the predictable April price pressure and the return patterns
documented in the main text.

The firm chooses a speculative inventory position $n_{it}$ to solve
\begin{align}
\max_{n_{it}}
\left\{
\mu_{it} n_{it} - \frac{\gamma_i}{2} n_{it}^2
\right\}.
\end{align}
The first-order condition is
\begin{align}
\mu_{it}-\gamma_i n_{it}=0,
\end{align}
so the optimal speculative position is
\begin{align}
n_{it}=\frac{\mu_{it}}{\gamma_i}.
\end{align}
Aggregate operator order flow can therefore predict future benchmark-price changes when a subset of firms trades on private information; for actual returns, the under-revelation condition of Section~\ref{subsec:ct_info} is required in addition.

\subsubsection*{April Premium Paid by Delayed Buyers and Participation Frictions}

This subsection gives the two-period version of the distinction
between the April premium paid by delayed buyers and underlying real frictions. The additional payment from delayed purchases is given by
\begin{align}
\text{Premium}_i^{\text{delay}} = d_i(p_A-p_e),
\end{align}
which need not coincide with aggregate resource costs because the higher April
price partly reflects a transfer to the counterparties selling
allowances. When those counterparties are intermediaries, these
payments also represent transfers from regulated firms to financial
participants outside the manufacturing sector.
Because allowance payments are partly transfers, the resource-cost decomposition below strips them out and counts only the real immediacy-provision (or order-flow absorption) cost.

A simple quadratic benchmark illustrates the resource-cost components associated with deviations from the frictionless allocation. Let $\{x_{i0}^*,x_{i1}^*\}$ denote the frictionless allocation and $\{x_{i0},x_{i1}\}$ the actual
allocation. Assume that the cap fixes aggregate emissions in each period, so that $\sum_i x_{it}=\sum_i x_{it}^*$ for $t=0,1$; with interior banking the same identity holds, because the cross term below then equals $-(p_0^*-\delta E_0[p_1^*])\sum_i\Delta x_{i0}=0$. Under the
quadratic benchmark, period-$t$ abatement costs are
\begin{align}
\frac{1}{2}\sum_i \left(\theta_{it}-x_{it}\right)^2.
\end{align}
Writing $x_{it}=x_{it}^*+\Delta x_{it}$ and using $\sum_i \Delta x_{it}=0$ together with the equalization of marginal abatement costs in the frictionless allocation, $\theta_{it}-x_{it}^*=p_t^*$ for every $i$, so that the cross term $-\sum_i(\theta_{it}-x_{it}^*)\Delta x_{it}=-p_t^*\sum_i\Delta x_{it}$ vanishes, the difference in aggregate abatement costs
between the actual and frictionless allocations is exactly
\begin{align}
\frac{1}{2}\sum_i \left(\theta_{it}-x_{it}\right)^2
- \frac{1}{2}\sum_i \left(\theta_{it}-x_{it}^*\right)^2
= \frac{1}{2}\sum_i (\Delta x_{it})^2.
\end{align}
Summing over the two periods and adding the real participation costs and immediacy-provision costs gives the resource-cost components corresponding to the loss relative to the frictionless allocation, complementing the April premium of Section~\ref{subsec:private_costs}:
\begin{align}
L
= \frac{1}{2}\sum_i (x_{i0}-x_{i0}^*)^2
+ \frac{\delta}{2} E_0\left[\sum_i (x_{i1}-x_{i1}^*)^2\right]
+ \sum_i \iota_i K_i
+ \frac{\phi}{2}\left(M_0^2+\delta E_0[M_1^2]\right),
\end{align}
where $\iota_i\in\{0,1\}$ indicates that firm $i$ pays the participation cost $K_i$ and trades early. Finally, the participation condition implies that if a firm delays
trading, then
\begin{align}
K_i \geq d_i\left(E_0[p_A]-p_e\right),
\end{align}
so, for firms with positive expected shortfalls that delay purchases, the observed April trading costs provide a lower bound on the attention or participation frictions needed to rationalize delayed trading.

\end{document}